\documentclass{article}

\usepackage{rotating, graphicx, overpic, wrapfig, amsfonts, 
  amsmath, amssymb, amsthm, color, fullpage, url, natbib, caption, 
  subcaption, authblk, color, hyperref, bm, blkarray, array, verbatim}
\usepackage[boxed]{algorithm2e}

\newcommand{\al}{\alpha}
\newcommand{\be}{\beta}

\newcommand{\ep}{\epsilon}

\renewcommand{\th}{\theta}

\newcommand{\la}{\lambda}
\newcommand{\si}{\sigma}

\newcommand{\Th}{\Theta}

\newcommand{\Si}{\Sigma}

\newcommand{\sisq}{\sigma^{2}}

\newcommand{\f}[2]{\frac{#1}{#2}}
\renewcommand{\bar}{\overline}

\newcommand{\pa}{\partial}
\newcommand{\mc}[1]{\mathcal{#1}}

\newcommand{\E}[1]{\mathbb{E}\left[#1\right]}

\newcommand{\Unif}[2]{\mbox{Unif}(#1,#2)}
\newcommand{\Norm}[2]{\mathcal{N}\left(#1,#2\right)}
\newcommand{\MVpNorm}[3]{\mathcal{N}_{#1}\left(#2,#3\right)}
\newcommand{\HMVpNorm}[3]{\mathfrak{N}_{#1}\left(#2,#3\right)}

\newcommand{\ExpD}[1]{\mbox{Exp}\left(#1\right)}

\newcommand{\PRRV}[1]{P\left[#1\right]}

\newcommand{\set}[1]{\left\{#1\right\}}

\newcommand{\lpnorm}[2]{\left|\left|#2\right|\right|_{#1}}

\newcommand{\C}{\mathbb{C}}
\newcommand{\R}{\mathbb{R}}
\newcommand{\N}{\mathbb{N}}

\renewcommand{\P}{\mathbb{P}}
\renewcommand{\S}{\mathbb{S}}

\renewcommand{\hat}{\widehat}
\renewcommand{\tilde}{\widetilde}

\newcommand{\varty}[1]{\mc{#1}}

\renewcommand{\hbar}{\mbox{------}}

\newcommand{\ma}[1]{\bm{#1}}
\newcommand{\ve}[1]{\bm{#1}}

\newcommand{\twovec}[2]{\left[\begin{matrix}#1 \\ #2 \end{matrix}\right]}
\newcommand{\twomat}[4]{\left[\begin{matrix}#1 & #3 \\ #2 & #4 \end{matrix}\right]}
\newcommand{\threevec}[3]{\left[\begin{matrix}#1 \\ #2 \\ #3\end{matrix}\right]}

\newcommand{\pkg}[1]{\textbf{\pl{#1}}}
\newcommand{\pl}[1]{{\fontfamily{\sfdefault}\selectfont #1}}

\DeclareMathOperator*{\argmin}{\arg\!\min}

\definecolor{BUgreen}{RGB}{24,144,41} 

\definecolor{mmBlue}{RGB}{0,92,141} 
\definecolor{mmRed}{RGB}{180,0,0} 
\definecolor{mmPurple}{RGB}{153,61,113} 
\definecolor{mmYellow}{RGB}{153,139,61} 
\definecolor{mmGreen}{RGB}{61,153,86} 
\definecolor{orangeRed}{RGB}{254,76,55} 

\definecolor{darkBlue}{RGB}{0,92,141}

\definecolor{darkgreen}{RGB}{0, 120, 36}
\definecolor{darkblue}{RGB}{0, 59, 114}
\definecolor{notsobrightblue}{RGB}{0, 0, 175}
\hypersetup{colorlinks,breaklinks,
            linkcolor=notsobrightblue,urlcolor=notsobrightblue,
            anchorcolor=notsobrightblue,citecolor=notsobrightblue}

\definecolor{shadecolor}{rgb}{.97, .97, .97} 
\makeatletter
 {\par\unskip\endMakeFramed%
 \at@end@of@kframe}
\makeatother

\newtheorem{defn}{Definition}

\begin{document}

\title{Stochastic Exploration of Real Varieties via Variety Distributions}
\date{\today}
\author{
David Kahle\thanks{Associate Professor, Department of Statistical Science,
Baylor University.
\href{mailto:david_kahle@baylor.edu}{david\_kahle@baylor.edu}} \\
Jonathan D. Hauenstein\thanks{Professor, Department of Applied and Computational
Mathematics and Statistics, University of Notre Dame.
\href{mailto:hauenstein@nd.edu}{hauenstein@nd.edu}}
}
\maketitle

\begin{abstract}
\noindent
Nonlinear systems of polynomial equations arise naturally in many applied
	settings, for example loglinear models on contingency tables and
	Gaussian graphical models. The solution sets to these systems over the
	reals are often positive dimensional spaces that in general may be very
	complicated yet have very nice local behavior almost everywhere.
	Standard methods in real algebraic geometry for describing positive
	dimensional real solution sets include cylindrical algebraic
	decomposition and numerical cell decomposition, both of which can be
	costly to compute in many practical applications.  In this work we
	communicate recent progress towards a Monte Carlo framework for
	exploring such real solution sets. After describing how to construct
	probability distributions whose mass focuses on a variety of interest,
	we describe how Hamiltonian Monte Carlo methods can be used to sample
	points near the variety that may then be moved to the variety using
	endgames. We conclude by showcasing trial experiments using practical
	implementations of the method in the Bayesian engine \pl{Stan}.
\end{abstract}


\section{Introduction and background}\label{sec:intro}

In this article we consider stochastic methods to explore real algebraic
varieties, solution sets of systems of algebraic equations.  Such systems arise
naturally throughout applied science, including statistical science where they
can be seen in areas such as (conditional) independence and graphical models,
regression and regression-like analyses, and the design of experiments.

As in all areas of mathematics, a bit of notation is required to orient oneself
with the relevant concepts.  Let $k$ denote a field, $k^{n}$ the $n$-dimensional
affine/coordinate vector space of $n$-tuples of elements of $k$, $k[\ve{x}] =
k[x_{1}, x_{2}, \ldots, x_{n}]$ the commutative ring of polynomials in the $n$
variables $x_{1}, \ldots, x_{n}$ with coefficients from $k$, and $g_{1}, \ldots,
g_{m} \in k[\ve{x}]$ $m$ polynomials in $k[\ve{x}]$, collectively denoted
$\ve{g} = [g_{i}]_{i=1}^{m} \in k[\ve{x}]^{m}$.  The variety $\varty{V}(\ve{g})
= \varty{V}(g_{1}, \ldots, g_{m})$ defined by the polynomials $g_{1}, \ldots,
g_{m}$, sometimes called an algebraic set, is the collection of points
$$\varty{V}(\ve{g}) \ \ = \ \ \varty{V}(g_{1}, \ldots, g_{m}) \ \ = \ \
\set{\ve{x} \in k^{n}: g_{1}(\ve{x}) = g_{2}(\ve{x}) = \cdots = g_{m}(\ve{x}) =
0} \ \ = \ \ \set{\ve{x} \in k^{n}: \ve{g}(\ve{x}) = \ve{0}_{m}}.$$ It is the
solution set to the $m \times n$ system of equations described by $g_{1},
\ldots, g_{m}$. If $m = n$, so that the number of equations and unknowns is the
same, the system is said to be square.

Polynomials can be viewed from two complimentary perspectives, first as symbolic
objects among which operations are defined and second as functions. Every
polynomial $g \in k[\ve{x}]$ is an expression of the form $g = \sum_{\bm{\al}
\in \bm{\mc{A}}}\be_{\bm{\al}}\ve{x}^{\bm{\al}}$, where $\ve{x}^{\bm{\al}} =
\prod_{i=1}^{n}x_{i}^{\al_{i}}$ is monomial notation, $\bm{\mc{A}} \subset
\N^{n}$, and $|\bm{\mc{A}}| = b < \infty$ elements so that $g$ has $b$ terms. We
refer to $\ve{x}^{\bm{\al}}$ as a monomial; that is, the product of powers of
the variables without the coefficient. As convenient, we write polynomials in
different ways: $g = \sum_{\bm{\al} \in
\bm{\mc{A}}}\be_{\bm{\al}}\ve{x}^{\bm{\al}} =
\sum\be_{\bm{\al}}\ve{x}^{\bm{\al}} = \ve{\be}'\ve{x}^{\bm{\mc{A}}} =
g(\ve{x}|\bm{\be})$, where $\bm{\be}$ and $\ve{x}^{\bm{\mc{A}}}$ have $b$
elements, the latter being the vector of monomials in $\ve{x}$ with powers given
by the elements of $\bm{\mc{A}}$, which are assumed to have some order by
convention. The (total) degree of a polynomial is the maximum sum of such powers
over the terms of the polynomial, $d = \max_{\bm{\al} \in
\bm{\mc{A}}}\sum_{i=1}^{n} \al_{i}$. Vectors of polynomials $\ve{g} \in
k[\ve{x}]^{m}$ can be referred to similarly, namely $\ve{g} =
\ve{g}(\ve{x}|\ma{B}) = \ma{B}\ve{x}^{\bm{\mc{A}}}$, where $\ma{B} \in \R^{m
\times b}$ is some real-valued matrix whose rows are indexed by the polynomials
$g_{1}, \ldots, g_{m}$, constituted of $b_{1}, \ldots, b_{m}$ terms
respectively, and whose $b$ columns are indexed by the elements of
$\bm{\mc{A}}$, whose cardinality is $|\bm{\mc{A}}| = b$. These are the unique
monomials of the pooled collection of $g_{i}$'s monomials.

In addition to expressions, polynomials can also be viewed as functions $g :
k^{n} \to k$ by substituting values of vectors $\ve{x} \in k^{n}$ into the
expression $g$ and evaluating. In this setting, the variety of $g$,
$\varty{V}(g)$, corresponds to the zero level set of the hypersurface defined by
$g$. Vectors of polynomials $\ve{g} \in k[\ve{x}]^{m}$ are similarly considered
from either perspective; functionally they are of the form $\ve{g}: k^{n} \to
k^{m}$, i.e.  vector fields. From this perspective, varieties can be seen to be
intersections of zero level sets of polynomials, the individual varieties
$\varty{V}(g_{i})$. By definition, points $\ve{x} \in \varty{V}(\ve{g})
\subseteq k^{n}$ are on the variety if and only if they satisfy the algebraic
system of equations $g_{1}(\ve{x}) = 0, \ldots, g_{m}(\ve{x}) = 0$. 

The field of algebraic geometry investigates the duality of varieties (geometric
objects) and ideals of $k[\ve{x}]$ (algebraic objects). For example, if $k$ is
an algebraically closed field, there is an elegant bijective correspondence
between certain ideals and varieties that enables geometric questions to be
reformulated as algebraic ones and vice versa \citep{iva2}. A common choice for
$k$ in algebraic geometry is thus $k = \C$, as $\C$ is an algebraically closed
field; however, in applications typically $k = \R$ or more likely a computer
representation of it. In this work we generally assume $k = \R$ so that
$\varty{V}(\ve{g}) \subseteq \R^{n}$ and the coefficients of the polynomials are
real. As a concrete example, suppose $g_{1} = x^{2} + y^{2} + z^{2} - 1$ and
$g_{2} = z - (x^{2} + y^{2})$ in $\R[x,y,z]$. A corresponding system of
equations may be
\begin{eqnarray*}
x^{2} + y^{2} + z^{2} &=& 1 \\ 
x^{2} + y^{2} &=& z,
\end{eqnarray*}
and $\varty{V}(\ve{g})$ is the intersection of $\varty{V}(g_{1})$, the unit
sphere $\mc{S}^{2}$ in $\R^{3}$, and $\varty{V}(g_{2})$, the upward opening
paraboloid in $\R^{3}$, which is a circle in $\R^{3}$ hovering above the
$xy$-plane. Such varieties can also be considered to be the real components of
varieties of polynomials with real coefficients over $\C^{n}$. Notice that when
$k = \R$, every variety can be generated by a single
polynomial $g = \ve{g}'\ve{g} = \sum_{i=1}^{m}g_{i}^{2}$, since $\ve{x} \in
\varty{V}(\ve{g}) \iff (\forall i = 1, \ldots, m)(g_{i}(\ve{x}) = 0) \iff
\sum_{i=1}^{m} g_{i}(\ve{x})^{2} = 0$.

Varieties generalize the geometry of linear algebra in interesting and complex
ways relevant to statistical science. In the linear case, polynomials are of
total degree at most one, and the varieties are linear varieties that are either
(1) empty, (2) consist of a single point, or (3) consist of a hyperplane of
points. In the general setting, $\varty{V}(\ve{g})$ may be (1) empty, (2)
consist of a finite number of points (roots) $\ve{x}_{1}, \ldots, \ve{x}_{r}$,
in which case it is said to be zero-dimensional, or (3) consist of
hypersurface(s) of points, in which case it is said to be positive dimensional.
If $g = 0$, then the entire ambient space is a variety. Positive dimensional
varieties are locally $C^{\infty}$ manifolds almost everywhere
\citep{bochnak1991real}, but otherwise can be quite complex: they can be
disconnected, of varying dimension, and exhibit cusp-like and self-intersecting
singularities.  Varieties also form the closed sets of a topology called the
Zariski topology: by taking products of the respective generating polynomials,
unions of varieties are varieties, and by pooling the polynomials, intersections
of varieties are varieties as well. In nice settings varieties can be
decomposed into unions of irreducible pieces, each varieties themselves, called
components.  A variety is irreducible if, when represented as a union of
distinct varieties, it is one of the varieties in the decomposition. An example
is $\varty{V}\big((x^{2} + y^{2})z\big)$, which is the union of the irreducible
varieties the $z$-axis $\varty{V}(x^{2} + y^{2})$ and the $xy$-plane
$\varty{V}(z)$.  Varieties arise in naturally in many domains of statistics,
typically as either subsets of parameter spaces defining models (e.g.
independence models in contingency tables and Gaussian graphical models) or as
response surfaces as in regression models, and these are a major topic of
interest in algebraic statistics.

Being able to computationally understand varieties is a fundamental problem of
applied mathematics: it corresponds to the ability to find solutions to systems
of algebraic equations, generalizing linear algebra.  Consequently, varieties
have been investigated, as solution sets of systems of polynomial equations, for
centuries. In algebraic geometry, in the 20th century two major lines of
innovation surfaced specifically tailored to these kinds of problems: symbolic
methods based on Gr\"{o}bner bases of ideals and numerical methods based on
homotopy continuation \citep{sturmfels2002solving}. The first approach is a
generalization of the Gaussian elimination/back substitution algorithm of linear
algebra and is able to provide closed form solutions to the system in the
zero-dimensional case, where closed form is understood in the sense that the
original system is reformulated as a hierarchy of univariate polynomial
root-finding problems, which can be efficiently and accurately computed
numerically \citep{iva4}. The challenge with this strategy is that the
reformulation of the problem, the computation of a Gr\"{o}bner basis, is
well-known to have worst-case behavior that is double exponential time, so for
practical purposes the methods in general must be assumed to be infeasible
\citep{mayr1982complexity}.\footnote{As an important exception, it should be
noted that many statistical models can be seen to be varieties whose
corresponding ideals have Gr\"{o}bner bases that can be determined
combinatorially. \citep{drton2008lectures}} The second numerical approach is a
clever iterative application of predictor-corrector methods: one forms a
homotopy from the solution set of a known system into that of the target system
and carefully tracks the solution set through the homotopy numerically.  The
algebraic structure of the problem can be exploited to guarantee that the
process is accurate and fast with certifiable solutions when properly
initialized \citep{sommese2005numerical}. While the method works well in a broad
array of practical applications and is implemented in software such as
\pl{Bertini(2)}, \pl{PHCpack}, \pl{Mathematica}, and others, the method, and
consequently the software, is not designed to \emph{explore} positive
dimensional varieties \citep{bates2013numerically, verschelde1999algorithm}.
Similarly, they are designed to work over $\C^{n}$, not $\R^{n}$. Still more
recent methods sample a variety by carefully moving a linear space through the
variety \citep{breiding2020random}.  By contrast, the methods described in this
article use novel probability distributions along with Monte Carlo methods to
generate a collection of points on or near a variety via an entirely different
mechanism.

The basic approach proposed in this article is composed of two parts: first,
constructing probability distributions that concentrate their mass near the
variety of interest; and second, sampling from those distributions. Depending on
the application of interest, if points actually on the variety are desired,
endgames from numerical algebraic geometry can be applied to move the points to
the variety reliably. For the probability distributions we propose new families
of multivariate distributions called variety distributions. We describe these in
Sections~\ref{sec:algfams} and \ref{sec:induced}, which also provide a framework
for semi-algebraic sets. The variety normal distributions introduced
Section~\ref{sec:algfams} form distinguished families of such distributions that
appear particularly useful in applications. To sample from the distributions,
we use Hamiltonian Monte Carlo (HMC), a Markov chain Monte Carlo (MCMC) variant
currently at the forefront of Bayesian computing and well suited for these kinds
of algebraic problems. This is presented in Section~\ref{sec:sampling} along
with a trick that allows the variety normal distribution to be expressed as a
posterior distribution of a Bayesian analysis. Among other things, this enables
the use of high quality software from the Bayesian statistics community to do
the sampling efficiently with very little effort. In
Section~\ref{sec:applications}, we showcase a few applications of the method.
We conclude in Section~\ref{sec:discussion} with a discussion on the proposed
method and future directions.

\section{Variety normal distributions}\label{sec:algfams}

In this section we construct a family of probability distributions that focus
their mass on a variety of interest based on the normal and multivariate normal
distributions.  We begin with an observation about the univariate normal
distribution that is later generalized greatly.

\subsection{The variety normal distribution of a single polynomial}\label{sec:univariety-normals}

The univariate normal distribution is characterized by a probability density
function (PDF) with respect to the Lebesgue measure on $\R$ of the form
$p(x|\mu,\sisq) =
\f{1}{\sqrt{2\pi}\si}\exp\{-\f{1}{2\sisq}\left(x-\mu\right)^{2}\}$, where $\mu
\in \R$ and $\sisq > 0$ are given quantities. Disregarding normalization
factors, the normal density is proportional to the exponential of a negative
quadratic form, written $p(x|\mu, \sisq) \propto
\exp\{-\f{1}{2\sisq}\left(x-\mu\right)^{2}\}$ or $\tilde{p}(x|\mu, \sisq) =
\exp\{-\f{1}{2\sisq}\left(x-\mu\right)^{2}\}$, where the tilde is used to
express an un-normalized density, which we sometimes refer to as a
pseudodensity. 

The normal distribution focuses its mass on $\mu$, not merely in the sense that
if $X \sim p(x|\mu,\sisq)$ the expected value $\E{X} = \mu$, but also and
perhaps more fundamentally in the sense that the mode of the distribution of $X$
is $\mu$ and the density exponentially and symmetrically decays as one moves
away from $\mu$.  The dispersion parameter $\sisq$, the variance of $X$,
describes the extent to which the distribution is concentrated on $\mu$: the
closer to $0$ $\sisq$ is, the more focused the distribution is on $\mu$.
Importantly, the normal distribution focuses its mass where the quantity in the
exponent vanishes. In the case of the univariate normal, this is the root of the
linear polynomial $g = x-\mu \in \R[x]$.  The variety is zero-dimensional:
$\varty{V}(g) = \set{\mu}$. 

We note in passing that the $\f{1}{2}$ can be absorbed into the polynomial
$p(x|\mu, \sisq) \propto
\exp\{-\f{1}{\sisq}\big(\f{1}{\sqrt{2}}x-\f{1}{\sqrt{2}}\mu\big)^{2}\}$ so that
the coefficient vector of the polynomial lies on the unit sphere (here circle).
This might have some benefit in providing a canonical representation of the
polynomial; however, as we are not here interested in the parameter space but
rather the individual distribution itself given $g$, and since the
representation using the standard normal density has benefits later on, we leave
the description as-is.

The observation that the normal distribution concentrates its mass near the
variety of the linear polynomial $g = x-\mu \in \R[x]$ suggests that the same is
true for an arbitrary multivariate polynomial $g(\ve{x}|\ve{\be}) \in
\R[\ve{x}]$: if one wishes to construct a distribution that places its mass near
$\varty{V}(g)$, simply swap $x - \mu$ for $g$.  This brings us to the
heteroskedastic variety normal distribution.

\begin{defn}\label{defn:hvn}
A random vector $\ve{X} \in \R^{n}$ is said to have the heteroskedastic variety
	normal (HVN) distribution, denoted $\ve{X} \sim \HMVpNorm{n}{g}{\sisq}$,
	if it admits a density
\begin{equation}
p(\ve{x}|g, \sisq) \propto 
\tilde{p}(\ve{x}|g, \sisq) :=
\exp\set{-\f{g(\ve{x}|\bm{\be})^{2}}{2\sisq}}
\label{eq:hvn}
\end{equation}
with respect to the Lebesgue measure on $\R^{n}$ for some $g(\ve{x}|\bm{\be})
	\in \R[\ve{x}]$ and $\sisq > 0$. $\ve{X}$ is said to
	have the truncated HVN distribution, denoted
	$\ve{X} \sim \HMVpNorm{n,\bm{\mc{X}}}{g}{\sisq}$, if
	\eqref{eq:hvn} only holds on some non-null $\bm{\mc{X}} \subset
	\R^{n}$, outside of which $p$ vanishes.
\end{defn}

\noindent The conditional notation $|g$, indicating $g$ is given, is
unconventional since it incorporates both the coefficients of the polynomial
$\ve{\be}$, which are parameters, and the variates $\ve{x}$ themselves.
Nevertheless, for the present purposes $|g$ comports a clarity that the vector
of coefficients does not, so we find it to be a useful device to communicate
intent. 

\begin{figure}[h!]
\begin{center}
\includegraphics[scale=.185]{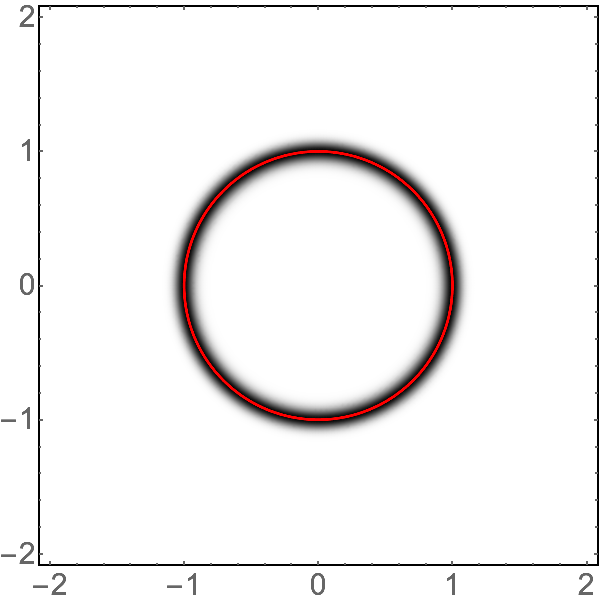} 
\includegraphics[scale=.185]{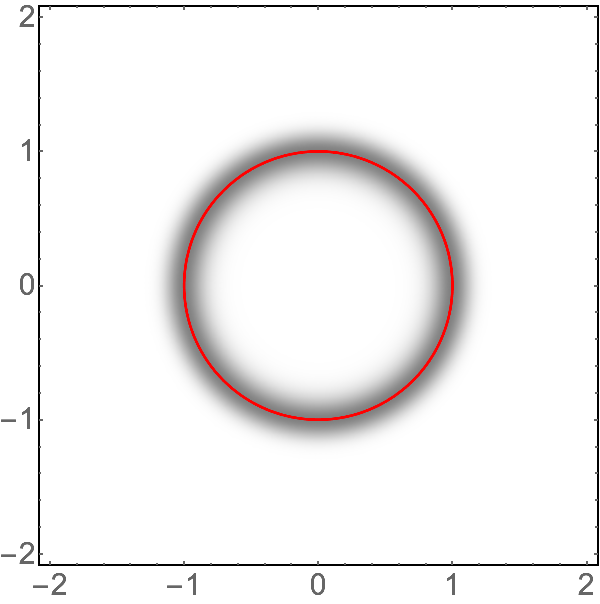} 
\includegraphics[scale=.185]{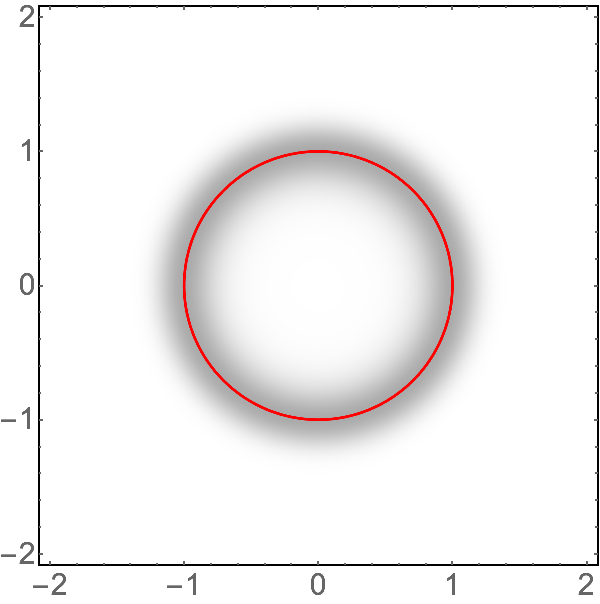} 
\includegraphics[scale=.185]{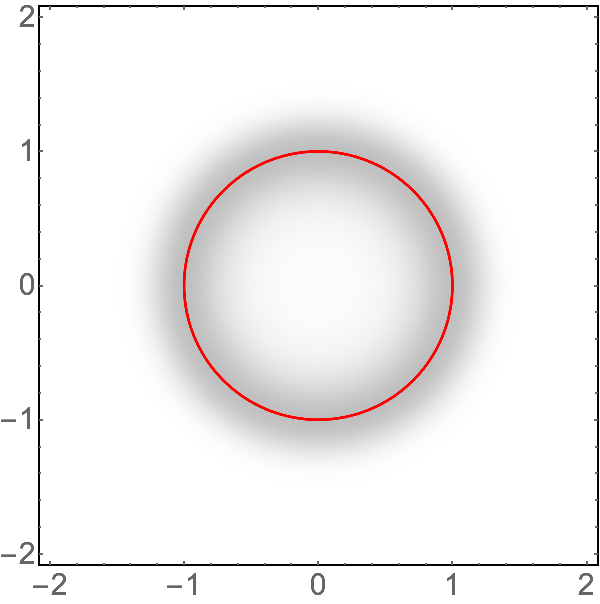} \\
\includegraphics[scale=.185]{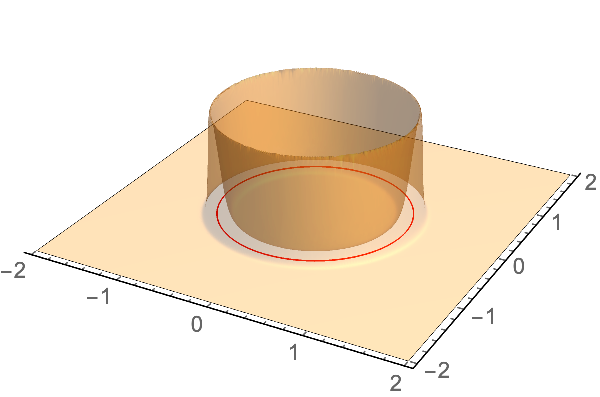} 
\includegraphics[scale=.185]{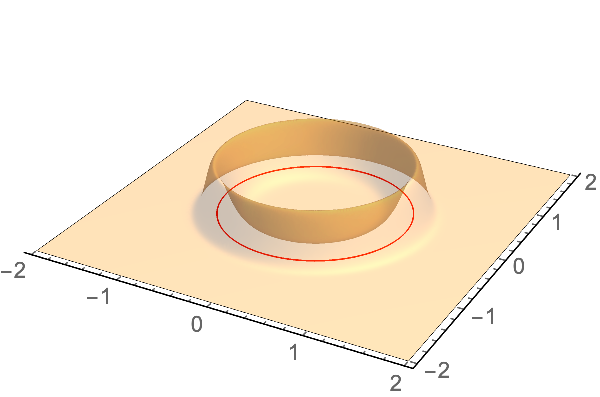} 
\includegraphics[scale=.185]{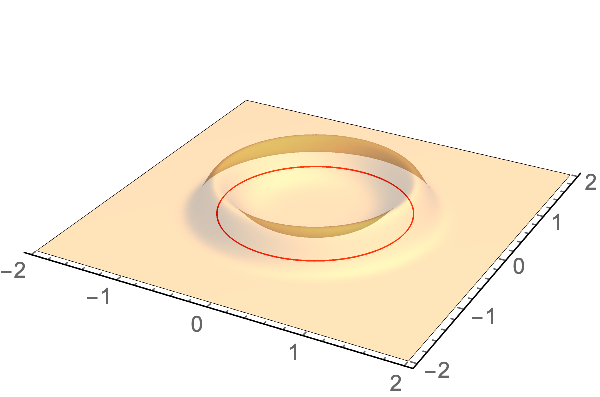} 
\includegraphics[scale=.185]{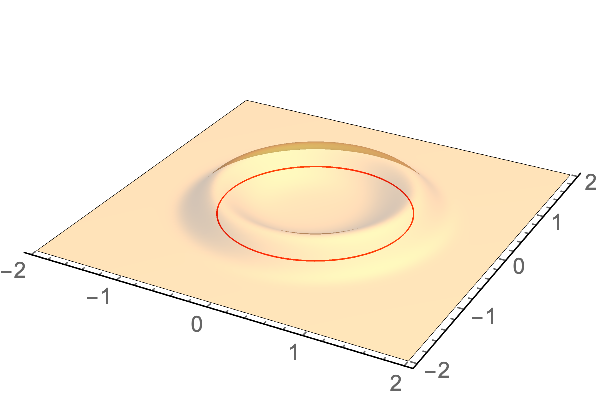} 
\end{center}
\caption{(Left to right.) $\HMVpNorm{2}{x^{2} + y^{2} - 1}{\sisq}$ from
	(\ref{defn:hvn}) with $\si = .1$, $.2$, $.3$,
	and $.4$; $\varty{V}(g)$ in red.}
\label{fig:donuts}
\end{figure}

Just as the univariate normal distribution focuses its mass on $\mu$, the HVN
focuses its mass on the variety $\varty{V}(g)$: the $-g^{2}$ in the exponent
demands that the pseudodensity is maximized at the variety (if nonempty) and
from there decreases (locally at least) so that the variety is the mode of the
distribution. Indeed, the HVN can be thought of as a blurred version of a
variety, where $\sisq$ governs the amount of blurring.  Probabilistically,
$\sisq$ is intended to gauge the likelihood of points falling a given distance
away from the variety. In the simple univariate normal case, the variety is
linear and zero-dimensional, and the empirical rule helps us interpret $\si =
\sqrt[+]{\sisq}$. The general case is more complex. An example of the bivariate
nonlinear case is illustrated in Figure~\ref{fig:donuts} with $g = x^{2} + y^{2}
- 1 \in \R[x,y]$.  Figure~\ref{fig:donuts} suggests the scheme works perfectly;
however, a few considerations meter this enthusiasm: normalizability and the
role of $\sisq$. We start with the role of $\sisq$ and defer the discussion
concerning normalizability to Section~\ref{sec:normalizability}.

\begin{figure}[h!]
\begin{center}
\includegraphics[scale=.20]{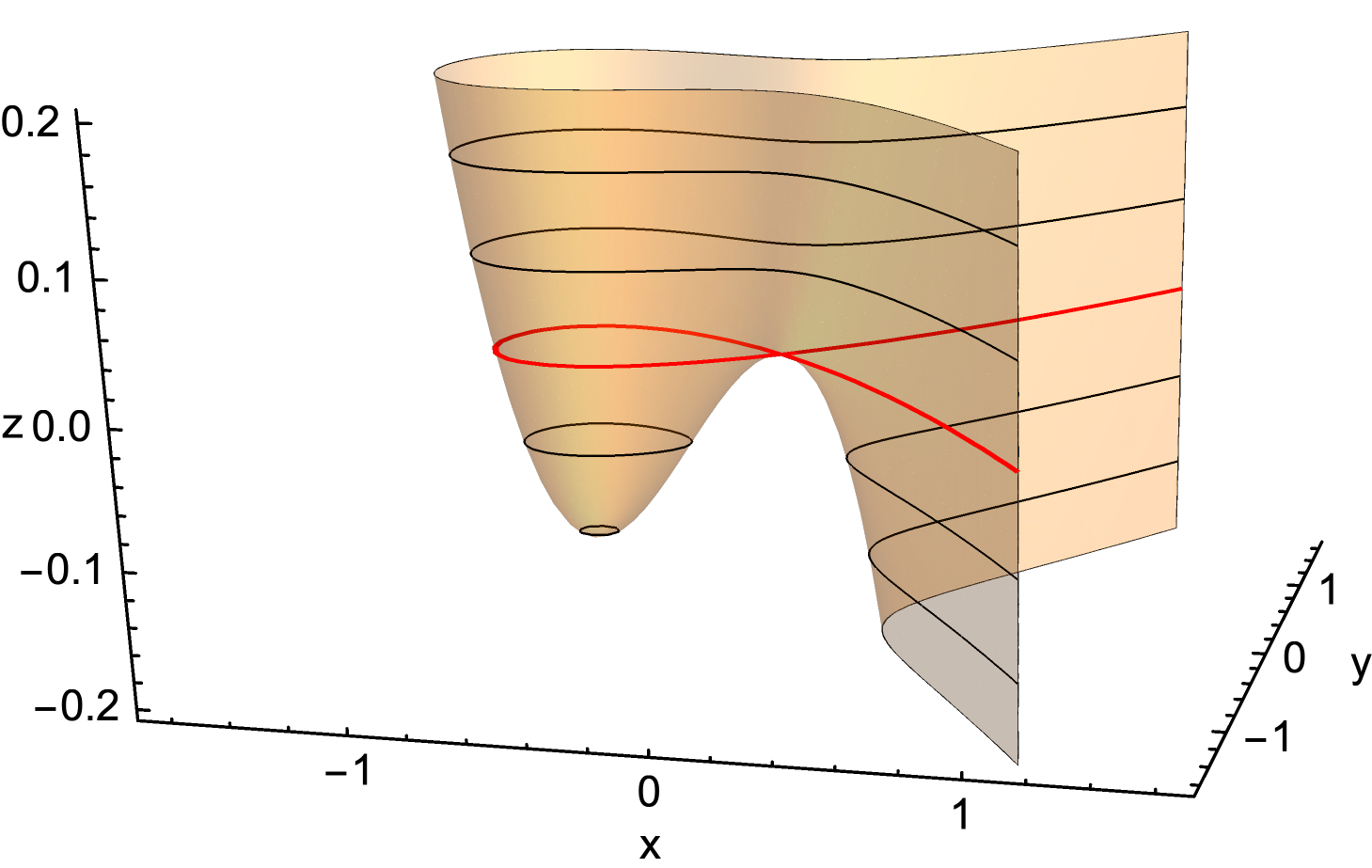}
\includegraphics[scale=.17]{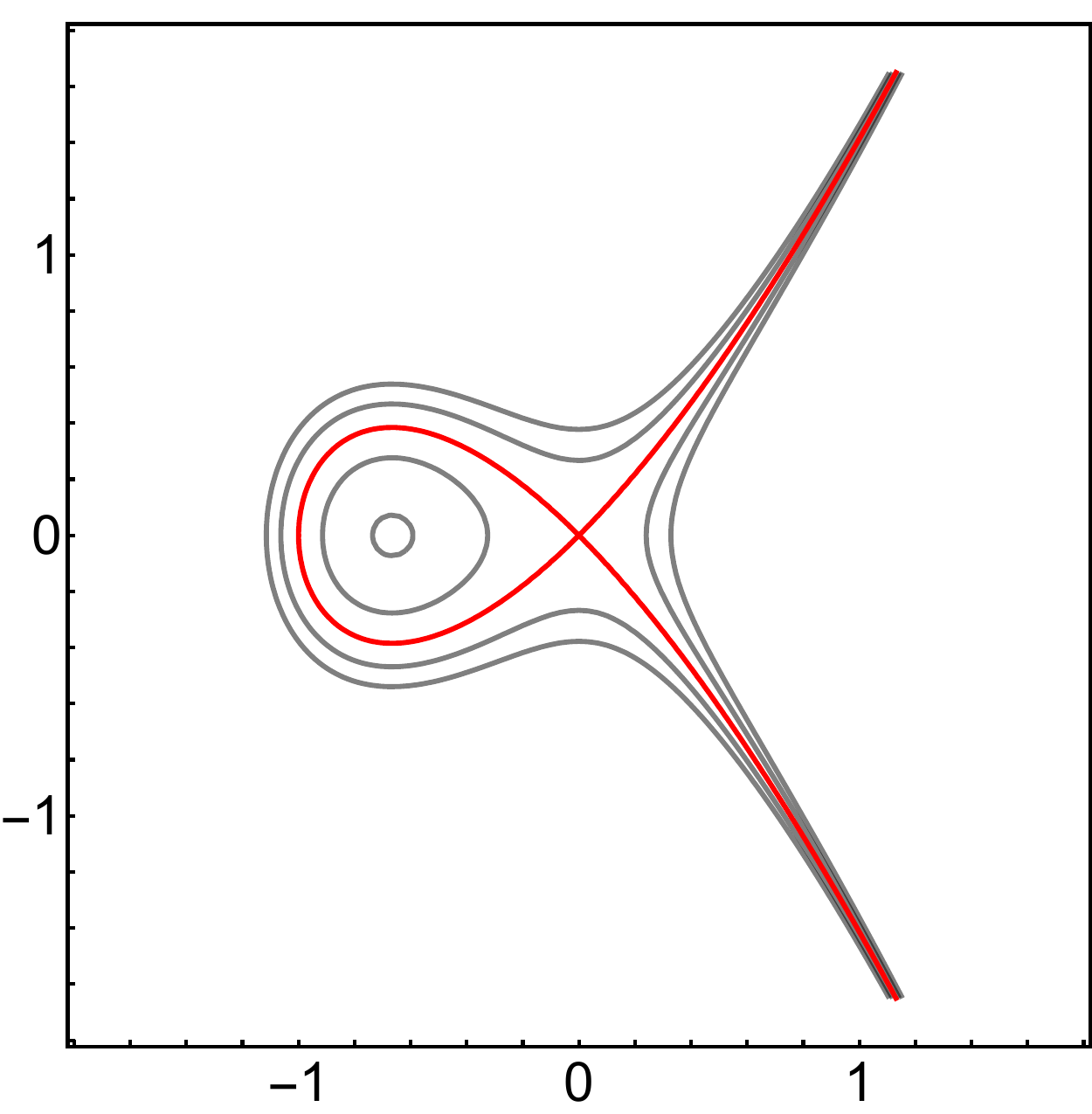} 
\includegraphics[scale=.17]{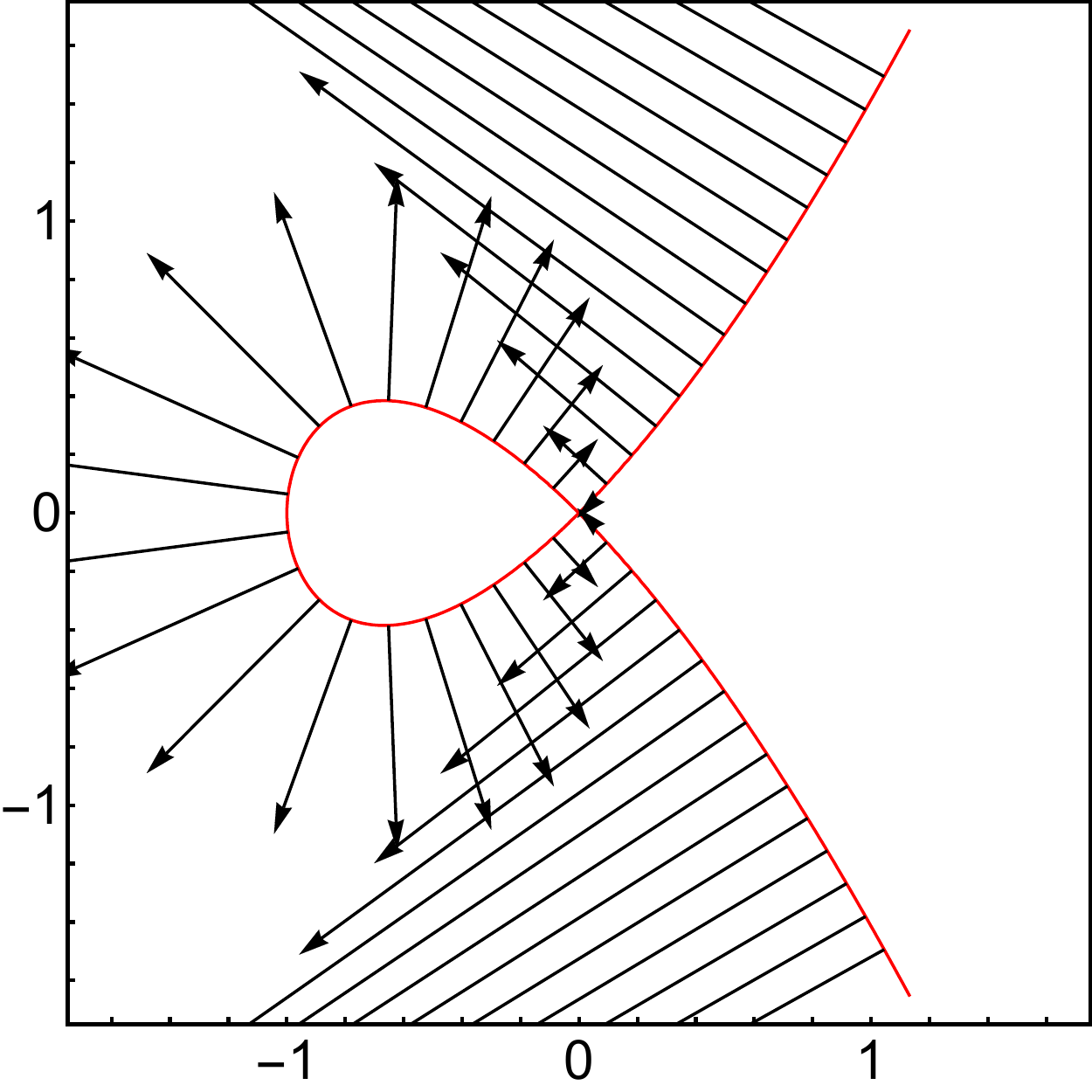}
\includegraphics[scale=.17]{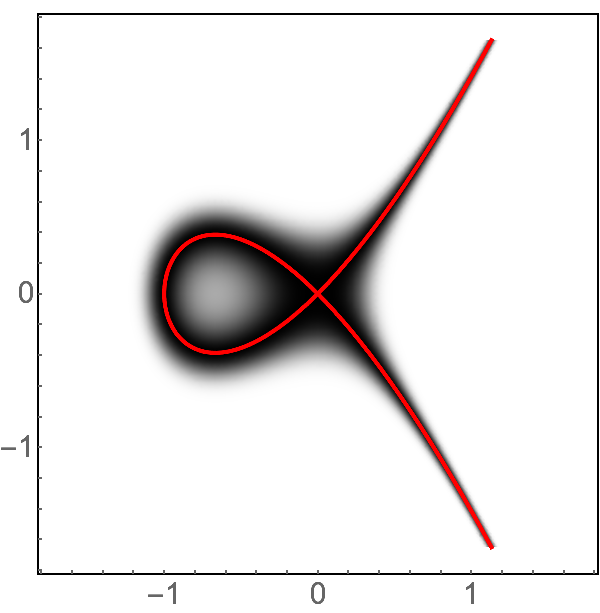}
\end{center}
\caption{Equidistant level sets of $g = y^{2} - (x^{3} + x^{2})$ do not result
	equidistant contours. From left to right: The graph of $g$ with equally
	spaced slices in $z$; those intersections projected down into the
	$xy$-plane; the gradient vector field on the variety; and the density
	plot of the HVN with $\si = .1$.}
\label{fig:naive}
\end{figure}

Coming from the normal distribution, it feels natural to want to interpret
$\sisq$ as gauging the amount of variability uniformly across the variety in the
sense any two points in $\R^{n}$ the same distance away from the variety should
have the same likelihood of occurring, at least approximately.  By distance here
we simply mean the Euclidean distance from the points to the nearest point on
the variety. (These closest points may not be unique, of course.) Unfortunately,
however, the HVN distribution does not obey this basic intuition, as is easily
seen with the polynomial $g = y^{2} - (x^{3} + x^{2}) \in \R[x,y]$, whose
variety is the alpha curve displayed in red in Figure~\ref{fig:naive}. This
explains the adjective heteroskedastic, which does not here refer to variability
changing on account of another variable per se but across the variety itself.
It is the phenomenon where two points, similarly situated with respect to the
variety, can have quite different likelihoods of occurring.

Why does this happen? Recalling that varieties are zero level sets, the HVN
distribution can be thought of as shifting the graph of $g$, a surface in
$\R^{3}$, up or down by a random amount following a $\Norm{0}{\sisq}$
distribution and selecting a point on the resulting zero level set. However, the
shifting has varying effects across the variety: equal amounts of shifting do
not result in equal amounts of movement of the zero level set across the
intersection with the $\ve{x}$-plane.  Figure~\ref{fig:naive} illustrates this
with five equally spaced level sets of $g$. 

\begin{figure}[h!]
\begin{center}
\includegraphics[scale=.4]{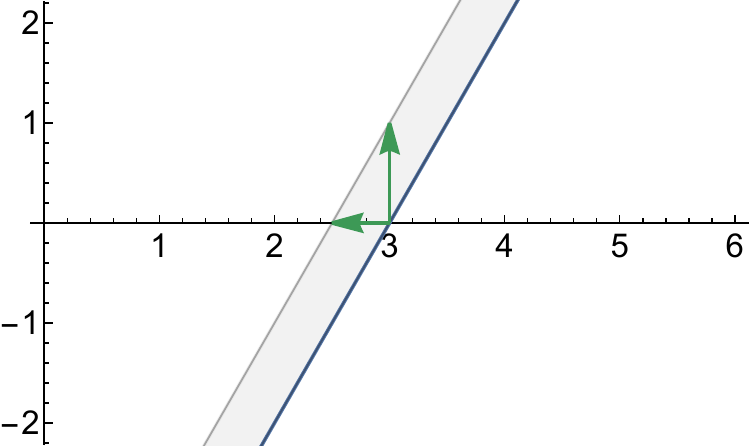}
\includegraphics[scale=.4]{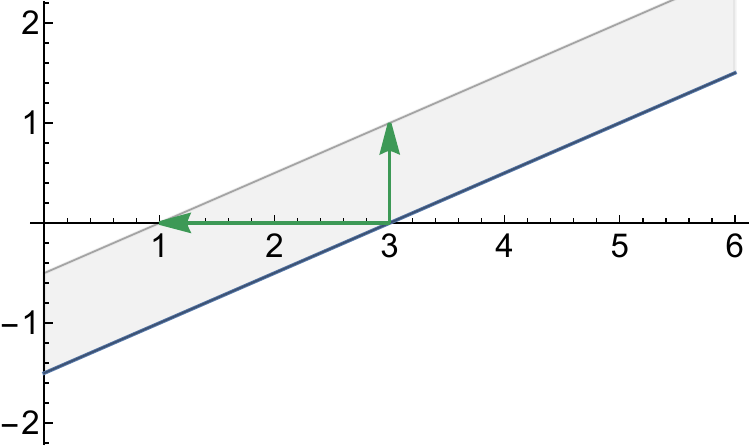}
\includegraphics[scale=.4]{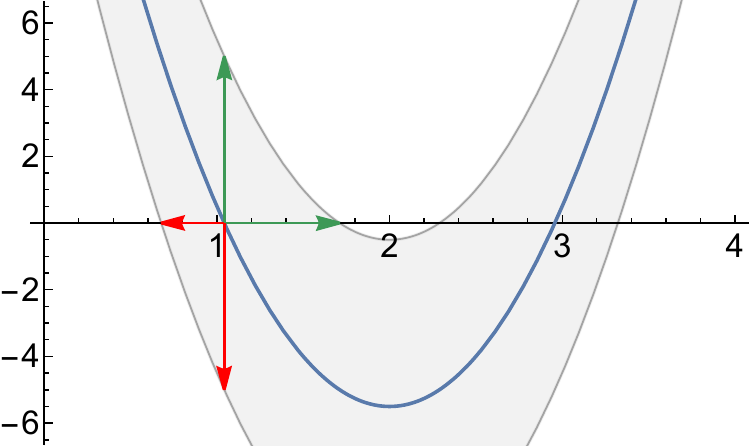}
\end{center}
\caption{For a fixed amount of shifting up/down, varieties of rapidly changing
	functions exhibit less change than those of slowly changing
	functions. Curvature modulates this change in different
	directions.}
\label{fig:shift}
\end{figure}

This is in fact a general feature of level sets related to the inverse function
theorem: parts of the variety where the function changes rapidly exhibit small
changes, and places where it changes gradually exhibit large changes.  This
basic phenomenon is easily seen in univariate linear and quadratic polynomials;
see Figure~\ref{fig:shift}.  In one dimension, for a nonconstant univariate
linear polynomial $g = \be_{0} + \be_{1}x \in \R[x]$, $\varty{V}(g) =
\{-\be_{0}/\be_{1}\}$. Shifting the graph of $g$ up by $\ep$ moves the root from
$-\be_{0}/\be_{1}$ to $-\be_{0}/\be_{1} - \ep/\be_{1}$, i.e. an amount
proportional to $\be_{1}^{-1}$ and in the direction of the opposite of the sign
of $\be_{1}$. Shifting it down by $\ep$ moves it the same amount, but in the
other direction--the direction of the sign of $\be_{1}$, the derivative/slope of
$g$. 

In general, root mobility is inversely related to the magnitude of the gradient
and in the opposite direction if the shift is up ($+\ep$) or the same direction
if the shift is down ($-\ep$). This is because the gradient $\nabla_{\ve{x}}g$
points in the direction of steepest ascent, so near the variety (where $g$ is
$0$) shifting the graph up requires a negative offset (a decrease from 0) to
balance to 0, and this occurs in the direction of the negative gradient,
assuming simple roots of course. Similarly, shifting the graph down moves the
root in the direction of the gradient, the direction that would be required to
move the value of the function $g$ from 0 to a sufficiently positive value to
offset the shift down. Curvature of the graph, a departure from linearity,
affects an asymmetry in the magnitude of root mobility in same-sized upward and
downward shifts, but this asymmetry is negligible for sufficiently small shifts
up and down.  In the current setting, this phenomenon can be observed at
different points on the same variety, e.g. the images of Figure~\ref{fig:naive}.
The varying amount of change in the gradient's magnitude results in a
disproportionate amount of the distribution's mass being placed on areas where
the surface near the variety is relatively flat.

The chief problem with the HVN distribution is that because $g(\ve{x}|\bm{\be})$
changes at different rates across the variety, the volume traversed in the
ambient space $\R^{n}$ by the surface as it is shifted up and down is uneven
across the variety, and so the HVN distribution does not faithfully represent
the notion that $\sisq$ gauges proximity to the variety in a constant way across
the whole of the variety.  Depending on the context and use case, this may be
considered a feature of the HVN; however, for our purposes we consider it a
defect.

\begin{figure}[h!]
\begin{center}
\includegraphics[scale=.20]{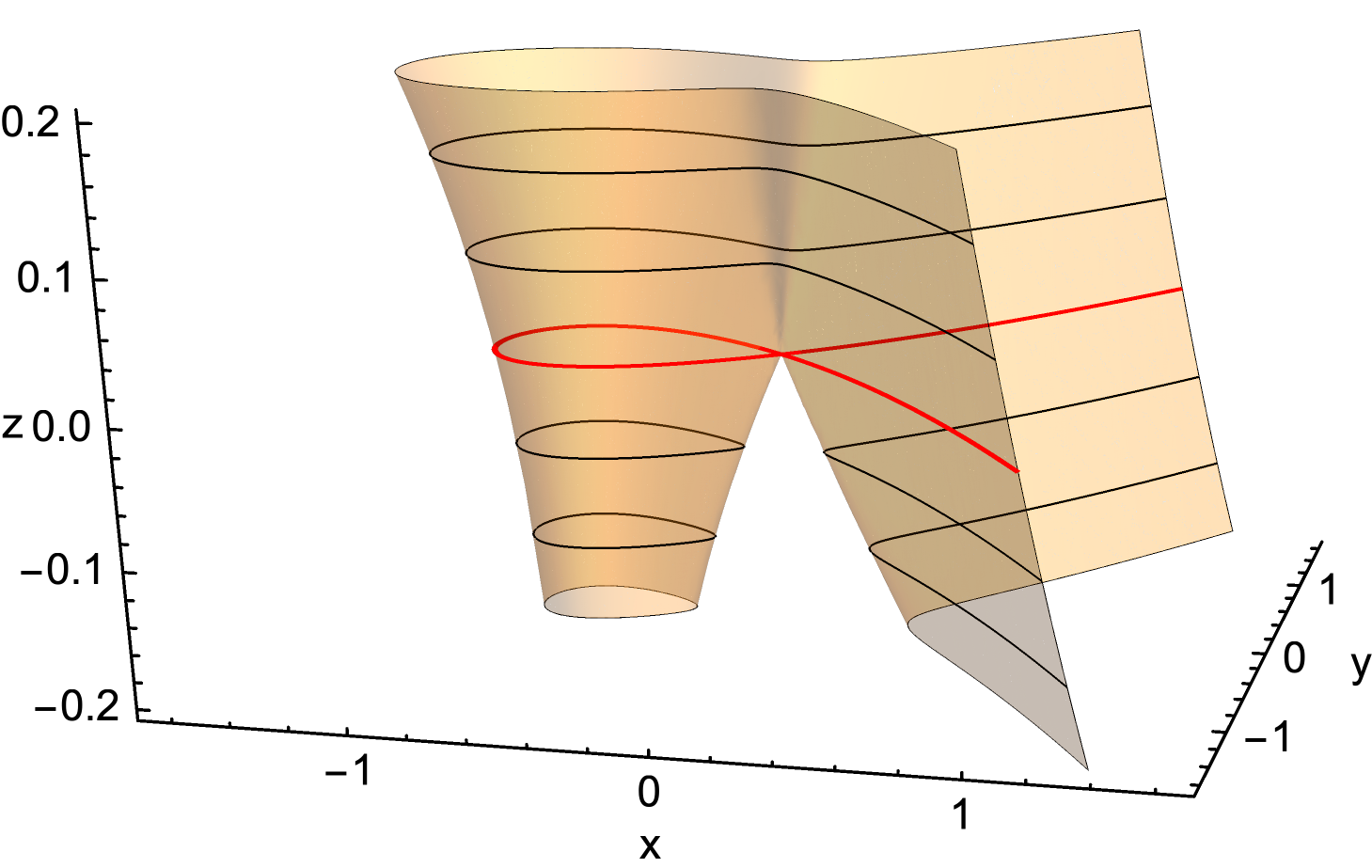}
\includegraphics[scale=.17]{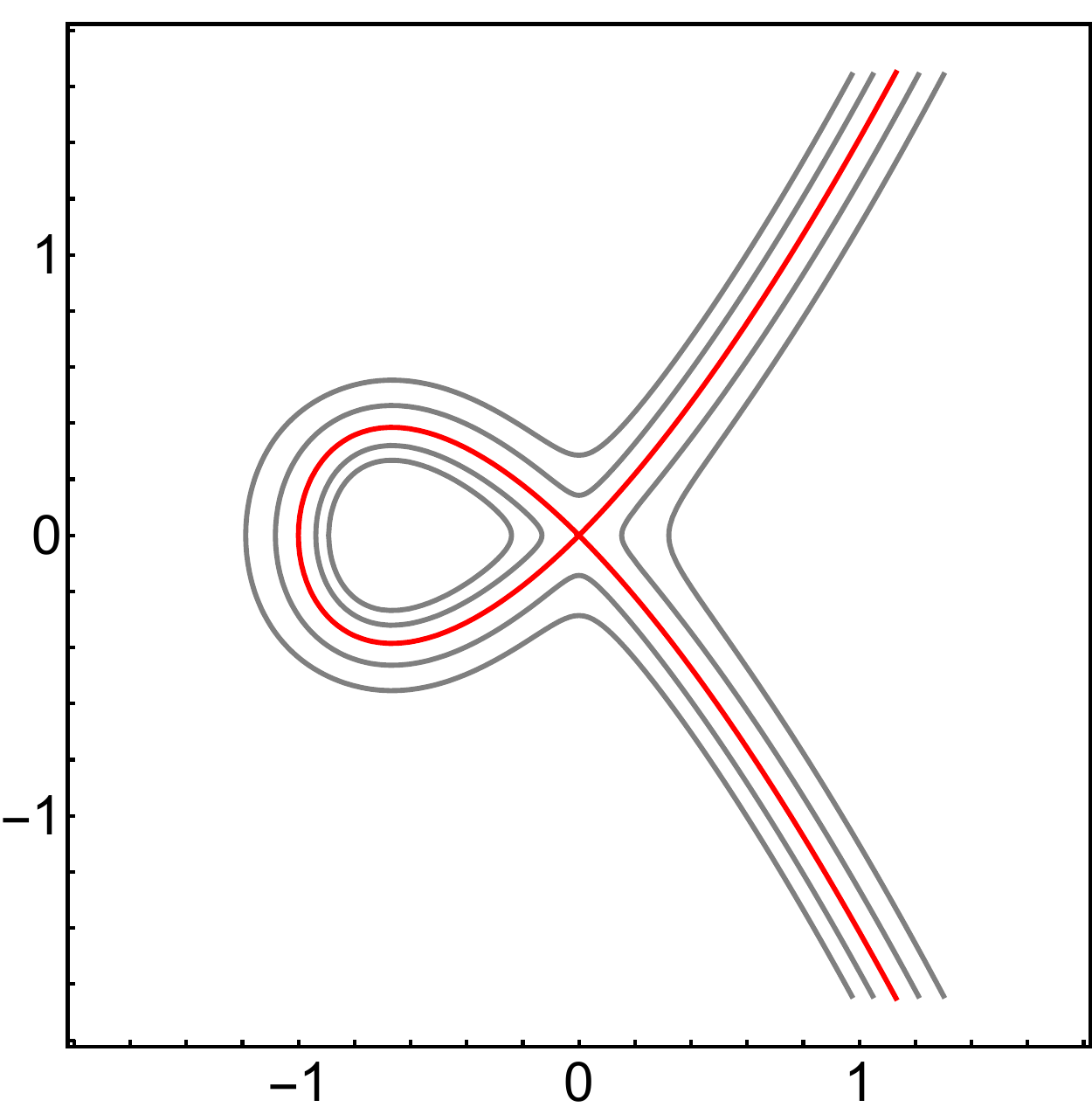} 
\includegraphics[scale=.17]{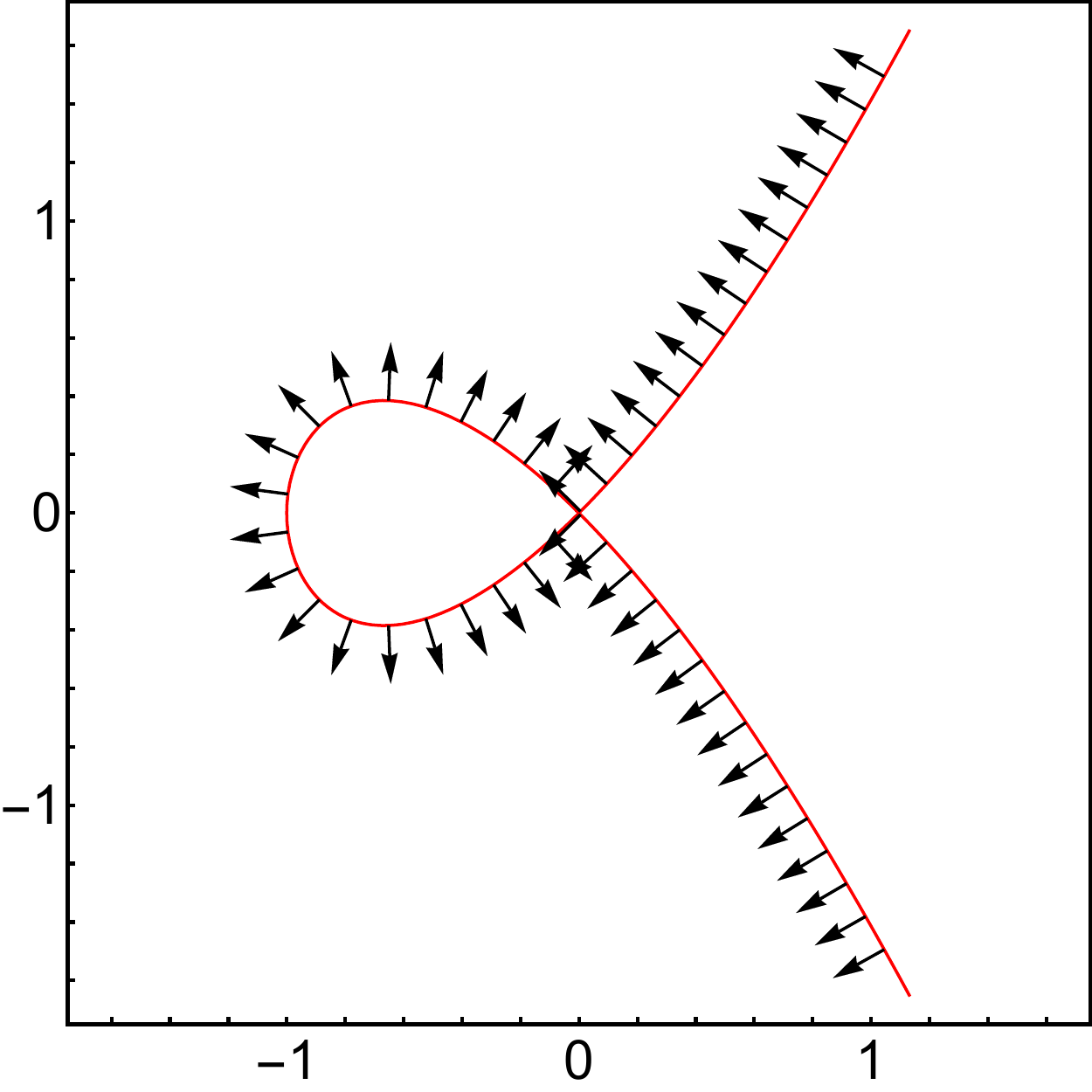}
\includegraphics[scale=.17]{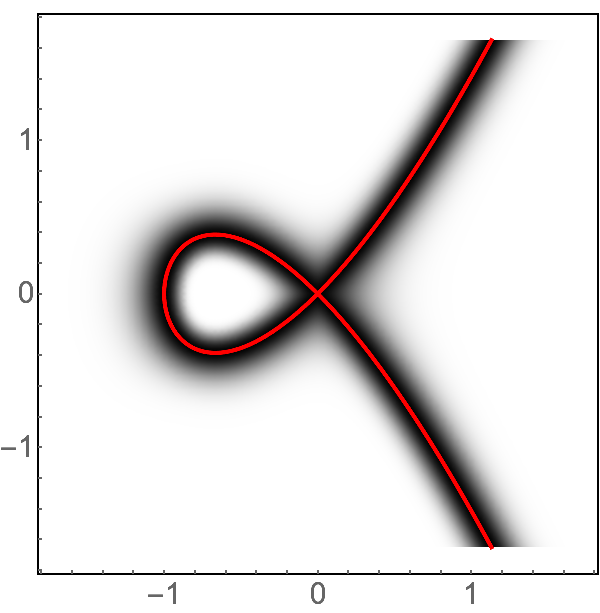}
\end{center}
\caption{Equidistant level sets of $\bar{g} =
	g\big/\lpnorm{}{\nabla_{\ve{x}} g}$ result in approximately
	equidistant contours  ($\si = .1$). Vectors drawn with $2\si$
	magnitude.}
\label{fig:correct}
\end{figure}

One solution to this problem is to locally linearize the surface to a constant
scale by normalizing $g$ by the size of its gradient; that is, to use $\bar{g} =
\bar{g}(\ve{x}|\bm{\be}) = \f{g}{\lpnorm{}{\nabla_{\ve{x}}g}}$ instead of $g$ in
place of the linear form $x - \mu$ in the normal density. (Norms this article
are assumed to be the 2-norm.) This can be justified by expanding $g$ in a
Taylor series about a point $\ve{x} \in \varty{V}(g) \subset \R^{n}$. Since
$g(\ve{x}) = 0$,
$$g(\ve{x}+\ve{h}) 
= \nabla g(\ve{x})'\ve{h} + o(\lpnorm{}{\ve{h}})
= \lpnorm{}{\nabla g(\ve{x})}\lpnorm{}{\ve{h}}\cos \th + o(\lpnorm{}{\ve{h}}),$$
where $\th$ is the angle between the gradient $\nabla g(\ve{x})$ and $\ve{h}$.
Dividing by the norm of the gradient provides
$$\f{g(\ve{x}+\ve{h})}{\lpnorm{}{\nabla g(\ve{x})}} = \lpnorm{}{\ve{h}}\cos \th
+ o(\lpnorm{}{\ve{h}}),$$ where the $o$ term is unaffected since
$\lpnorm{}{\nabla g(\ve{x})}$ is a constant with respect to $\ve{h}$. This of
course assumes the gradient is nonzero.  Note two facts: 1) where
$\lpnorm{}{\nabla g} \neq 0$, the zero locus of $\bar{g}$ is precisely that of
$g$, and 2) $\lpnorm{}{\nabla g} = 0$ on a set of Lebesgue measure 0, since
$\lpnorm{}{\nabla_{\ve{x}} g} = 0 \iff \lpnorm{}{\nabla_{\ve{x}} g}^{2} = 0$,
and the latter only occurs on a set of measure 0 because
$\lpnorm{}{\nabla_{\ve{x}} g}^{2}$ is polynomial.\footnote{And not the zero
polynomial. In this article $g$ is always nonconstant.} As PDFs are only defined
up to equivalence classes that allow for modifications on sets of measure zero,
the un-normalized density that uses the normalized expression $\bar{g}$ is, up
to normalizability at least, properly defined.  Analogues to
Figure~\ref{fig:naive} using the normalized version of $g$ are presented in
Figure~\ref{fig:correct}.

\begin{figure}[h!]
\begin{center}
\includegraphics[scale=.18]{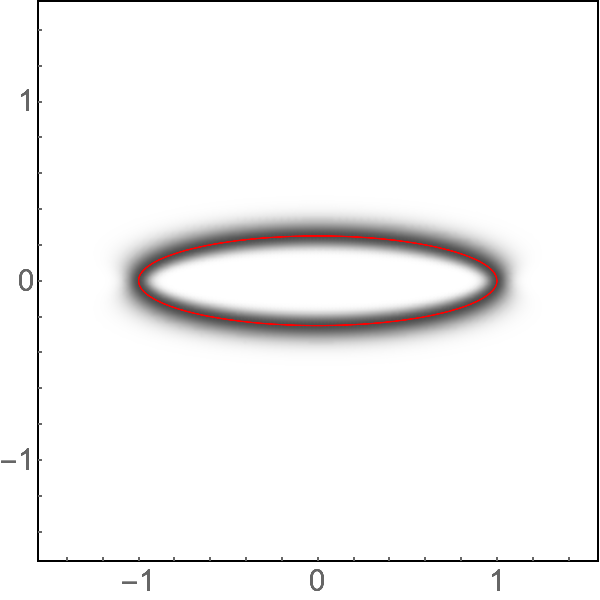}
\includegraphics[scale=.18]{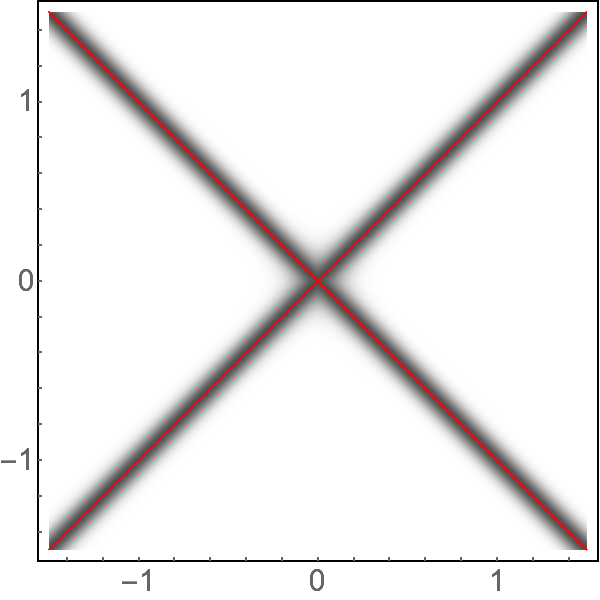}
\includegraphics[scale=.18]{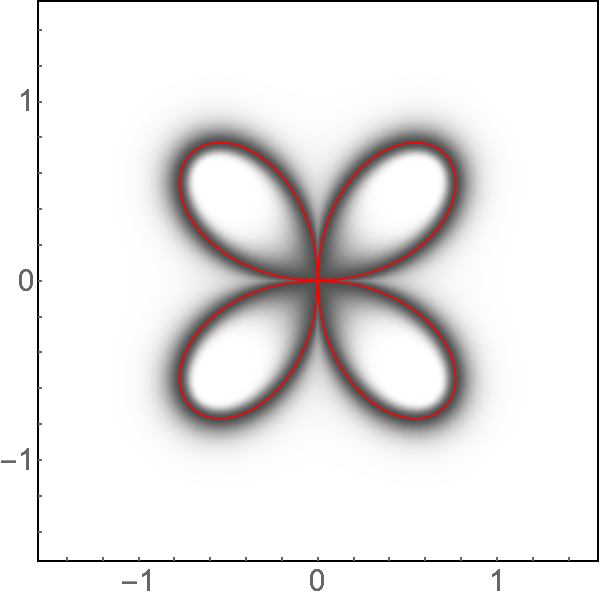}
\includegraphics[scale=.18]{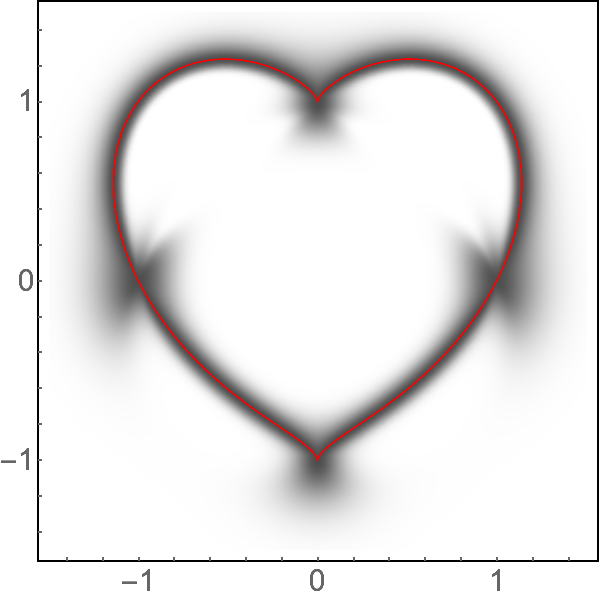}
\end{center}
\caption{From left to right, density plots of
	$\MVpNorm{2,\bm{\mc{X}}}{\bar{g}}{\sisq}$, truncated to the window,
	with $g = x^{2}+(4y)^{2}-1$, $(y-x)(y+x)$,
	$(x^{2}+y^{2})^{3}-4x^{2}y^{2}$, and $(x^{2}+y^{2}-1)^{3}-x^{2}y^{3}$
	and $\si = .05$. Aesthetic scales are consistent across the images. Note
	abnormalities at singularities in the last image.}
\label{fig:2dexs}
\end{figure}

With these things in mind, we are now able to present the definition we adopt in
this work as the variety normal distribution.
\begin{defn}\label{defn:vn}
A random vector $\ve{X} \in \R^{n}$ is said to have the (homoskedastic) variety
	normal (VN) distribution, denoted $\ve{X} \sim \MVpNorm{n}{g}{\sisq}$,
	if it admits a density
\begin{equation}
p(\ve{x}|g, \sisq) \propto 
\tilde{p}(\ve{x}|g, \sisq) :=
\exp\set{-\f{\bar{g}(\ve{x}|\bm{\be})^{2}}{2\sisq}} =
\exp\set{-\f{1}{2\sisq}\left(\f{g(\ve{x}|\bm{\be})}{\lpnorm{}{\nabla_{\ve{x}} g(\ve{x}|\bm{\be})}}\right)^{2}}
\label{eq:vn}
\end{equation}
with respect to the Lebesgue measure on $\R^{n}$ for some $g(\ve{x}|\bm{\be})
	\in \R[\ve{x}]$ such that $\bar{g}(\ve{x}|\bm{\be}) =
	\f{g(\ve{x}|\bm{\be})}{\lpnorm{}{\nabla_{\ve{x}} g(\ve{x}|\bm{\be})}}$
	and $\sisq > 0$.  $\ve{X}$ is said to have the truncated (homoskedastic)
	variety normal distribution, denoted $\ve{X} \sim
	\MVpNorm{n,\bm{\mc{X}}}{g}{\sisq}$, if \eqref{eq:vn} only holds on some
	non-null $\bm{\mc{X}} \subset \R^{n}$, outside of which $p$ vanishes.
\end{defn}
\noindent We present density plots for several bivariate variety normal
distributions in Figure~\ref{fig:2dexs}.

\begin{figure}[h!]
\begin{center}
\frame{\includegraphics[scale=.31]{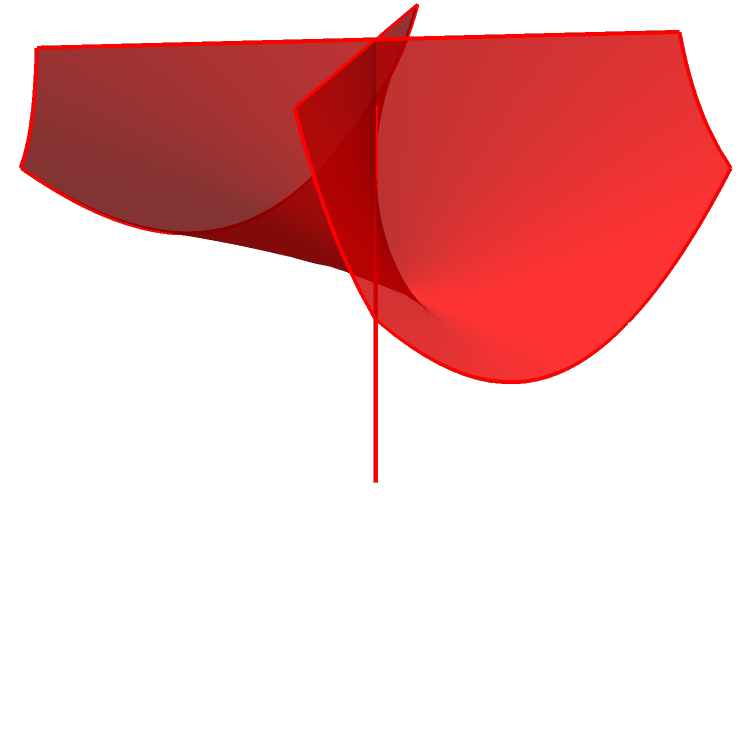}}
\frame{\includegraphics[scale=.31]{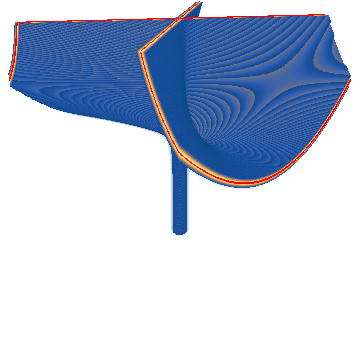}} 
\frame{\includegraphics[scale=.31]{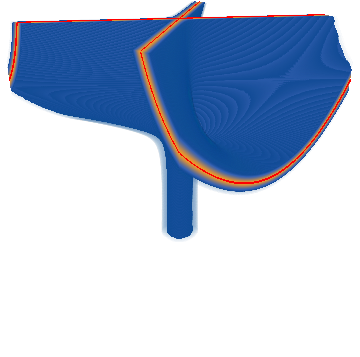}} 
\frame{\includegraphics[scale=.31]{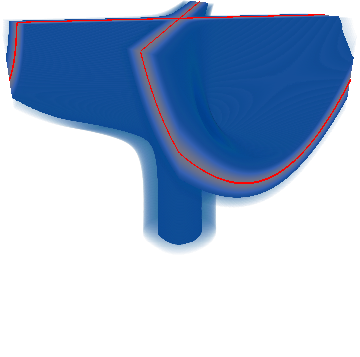}}
\end{center}
\caption{(Left) The Whitney umbrella defined by $g(x,y,z) = x^2 - y^2 z$ is the
	union of a one-dimensional handle and a two-dimensional canopy in
	$\R^{3}$. (Right) Three 3D density plots of the truncated VN
	distribution defined by the Whitney umbrella with $\si = .025$, $.05$,
	and $.10$.}
\label{fig:whitney}
\end{figure}

Of course, the VN distribution is not only defined for distributions on
$\R^{2}$, but any dimensional real space; the distributions just become more
difficult to view.  In $\R^{3}$ the Whitney umbrella given by $g(x,y,z) = x^2 -
y^2 z$ is illustrative of the general character of these distributions.
Figure~\ref{fig:whitney} provides 3D density visuals of this distribution with
different amounts of variability; as the Whitney umbrella is not compact, a
truncation is used.

In light of this example, a helpful way to think of these distributions is as
probabilistic thickenings of the variety into the ambient space $\R^{n}$.  As
the Whitney umbrella demonstrates, the thickening expands all components of the
variety, regardless of dimension, into the full dimension of the ambient space.
Moreover, the thickening is locally uniform across the variety in the sense
that, away from singularities, two points equidistant from the variety have
arbitrarily close probability density provided $\sisq$ is selected sufficiently
small. By construction, probability density decreases exponentially as one moves
orthogonally away from the variety--in the direction parallel to the gradient.
It is also approximately symmetric, with asymmetries arising from the curvature
in $g(\ve{x}|\bm{\be})$ relative to the size of $\sisq$, the same effect as seen
in Figure~\ref{fig:shift} (right). In Section~\ref{sec:induced} we describe how
different kinds of thickenings can be performed.

\subsection{The variety normal distribution as shifting surfaces}\label{sec:shifting-surfaces}

\begin{figure}[h!]
\includegraphics[scale=.2]{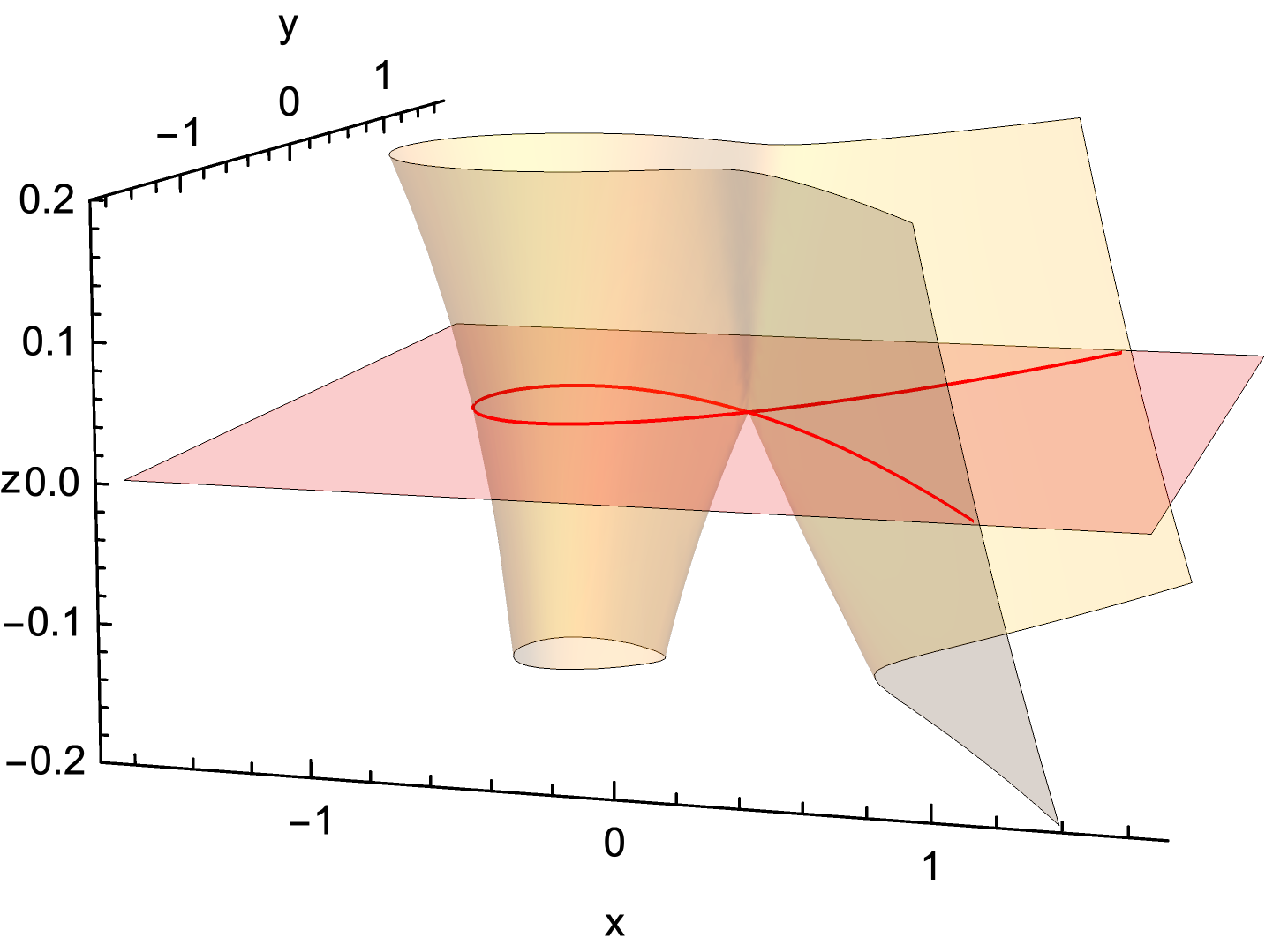}
\includegraphics[scale=.2]{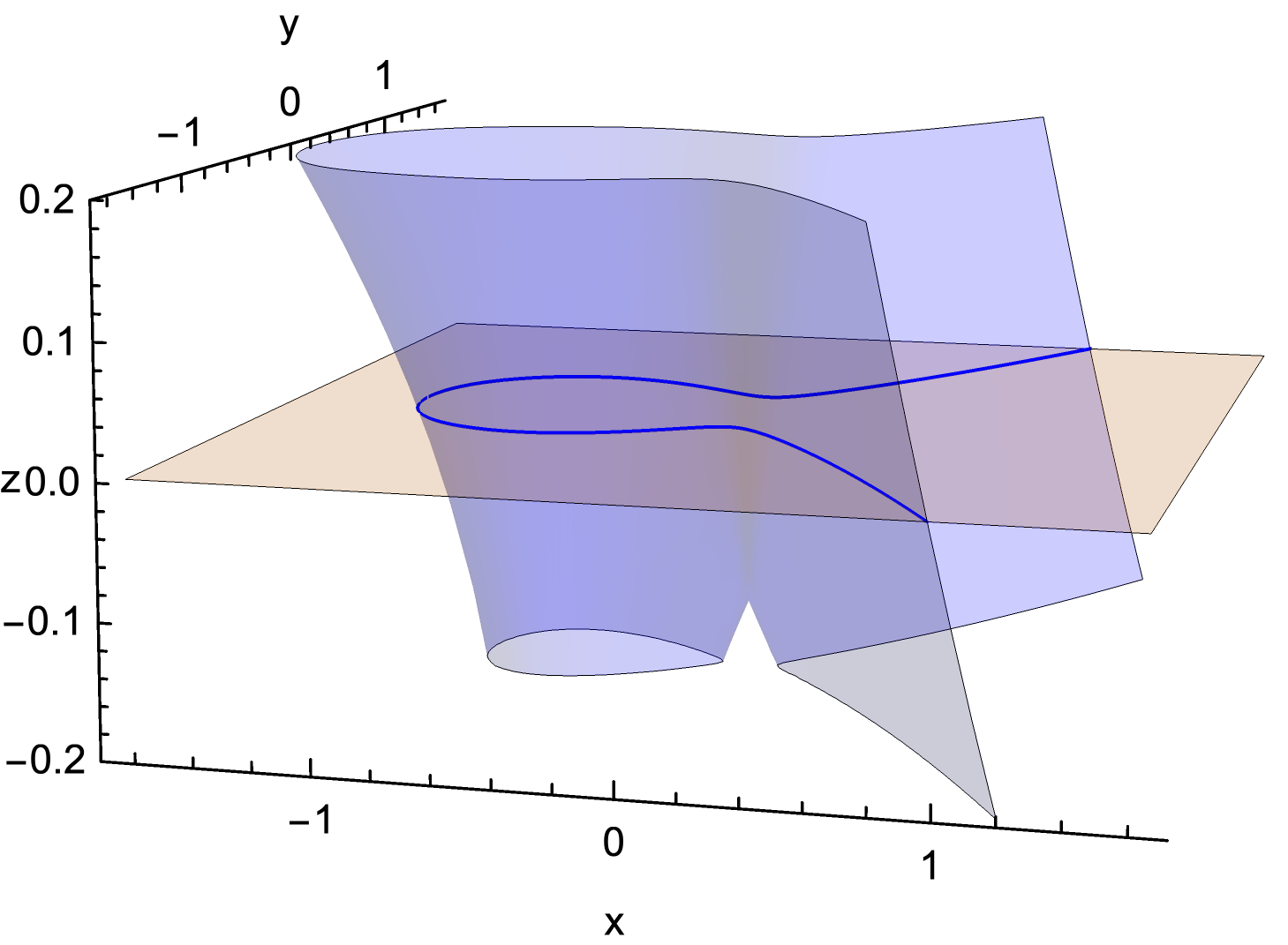}
\includegraphics[scale=.2]{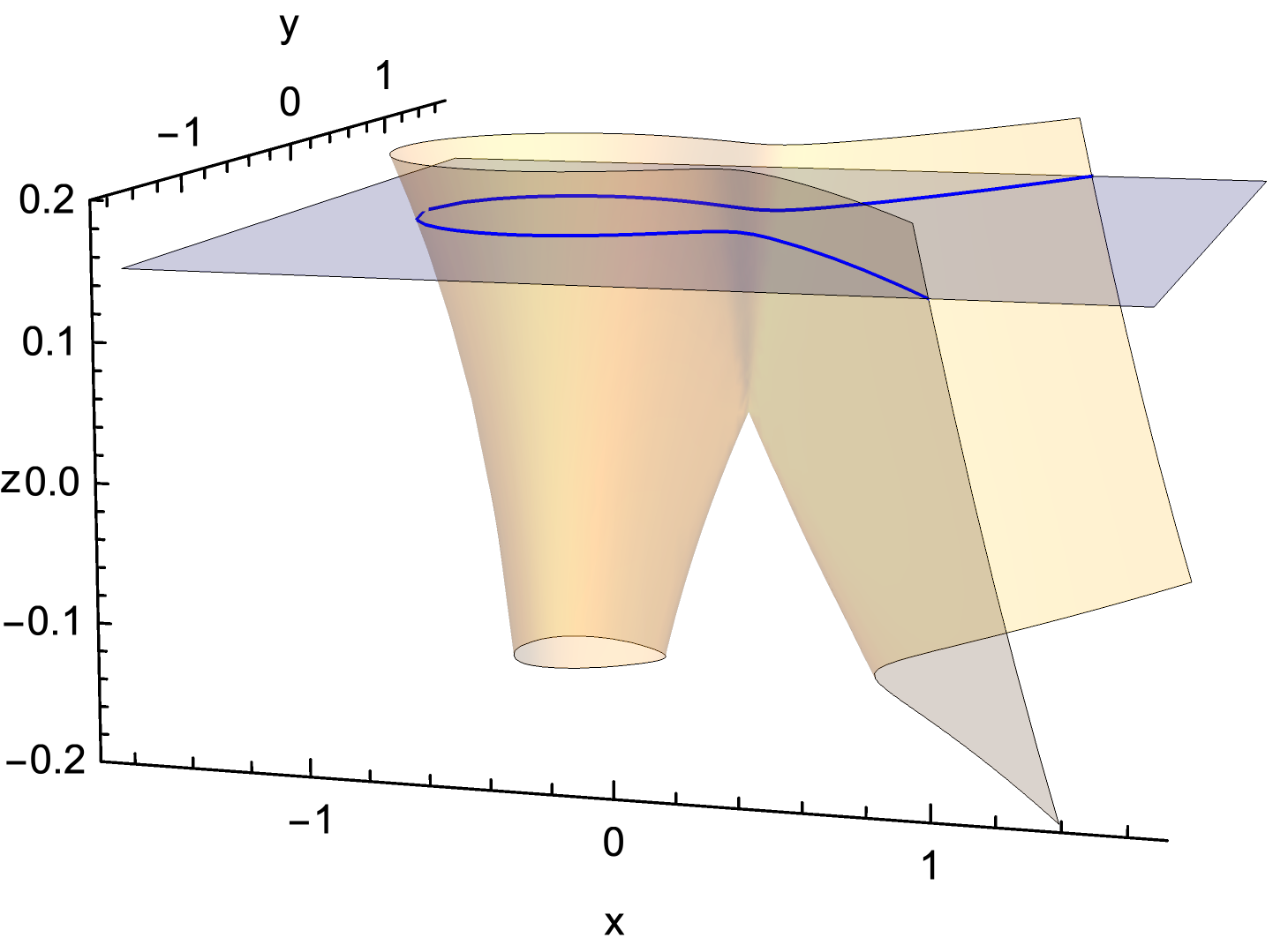} \\
{\color{white}.} \hskip 2.5in 
$\mc{G}_{\bar{g}-X} \cap \mc{G}_{0}$ \hskip 1.45in 
$\mc{G}_{\bar{g}} \cap \mc{G}_{X}$  \\
\includegraphics[scale=.2]{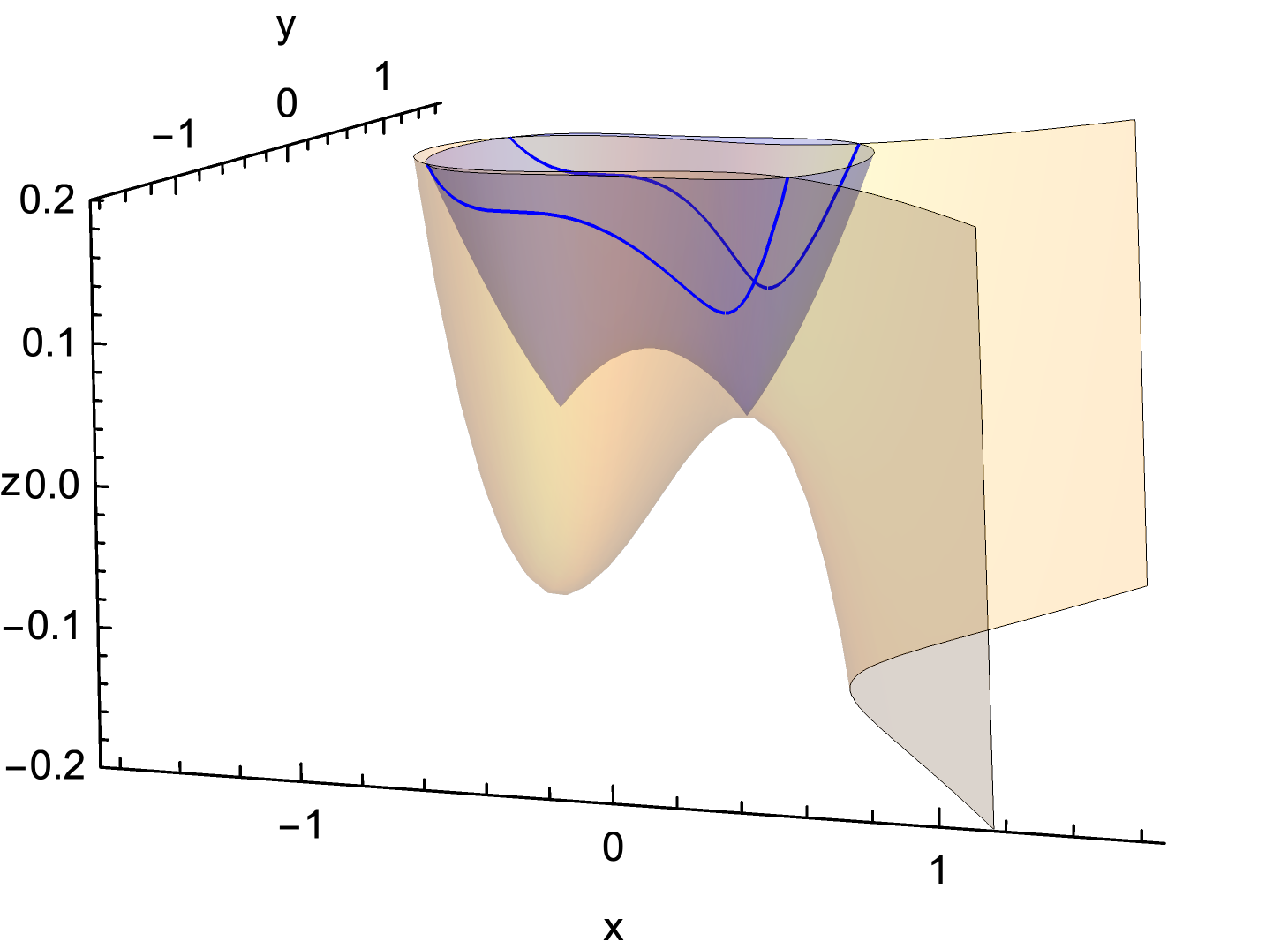}
\includegraphics[scale=.2]{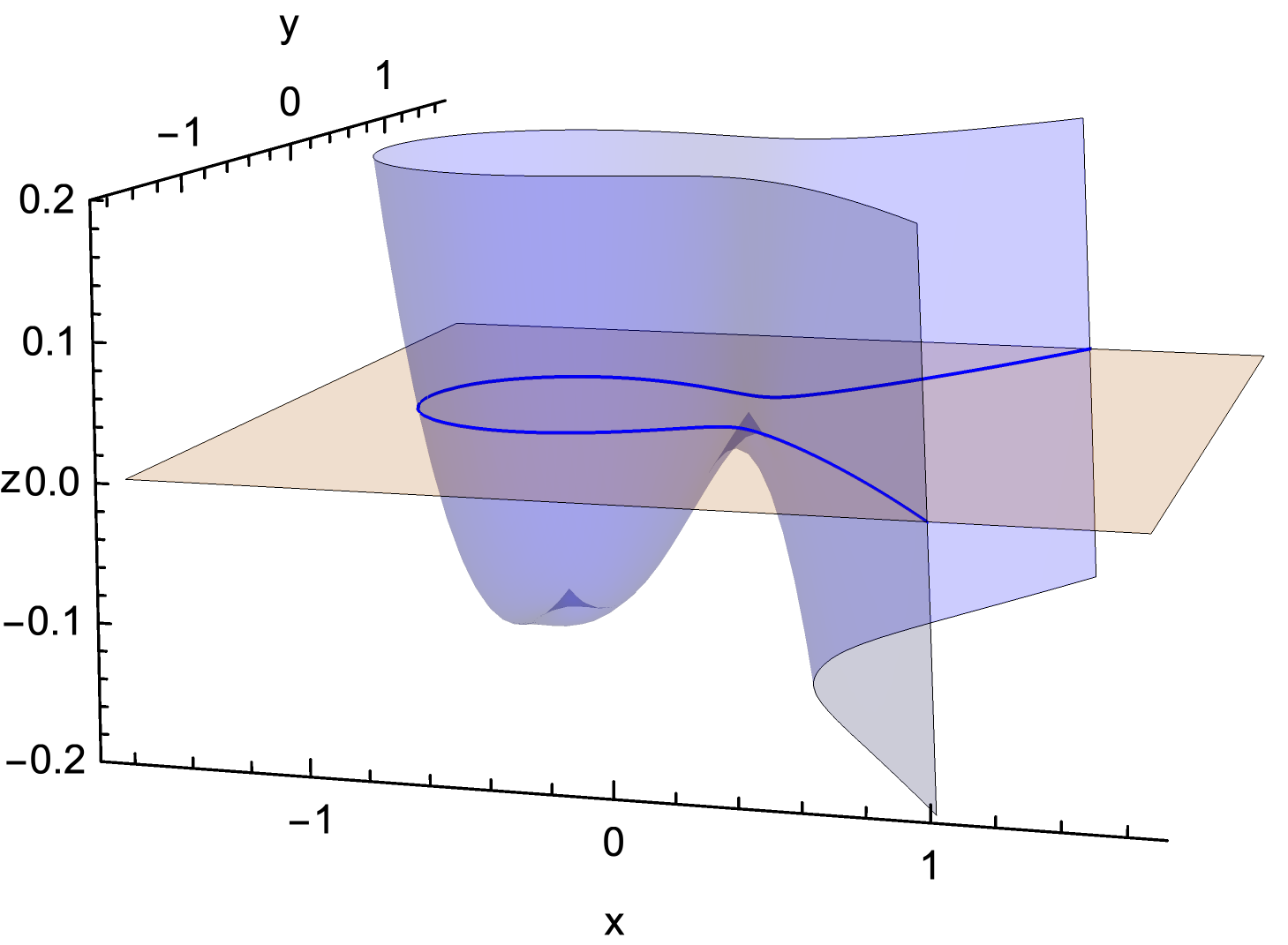}
\includegraphics[scale=.2]{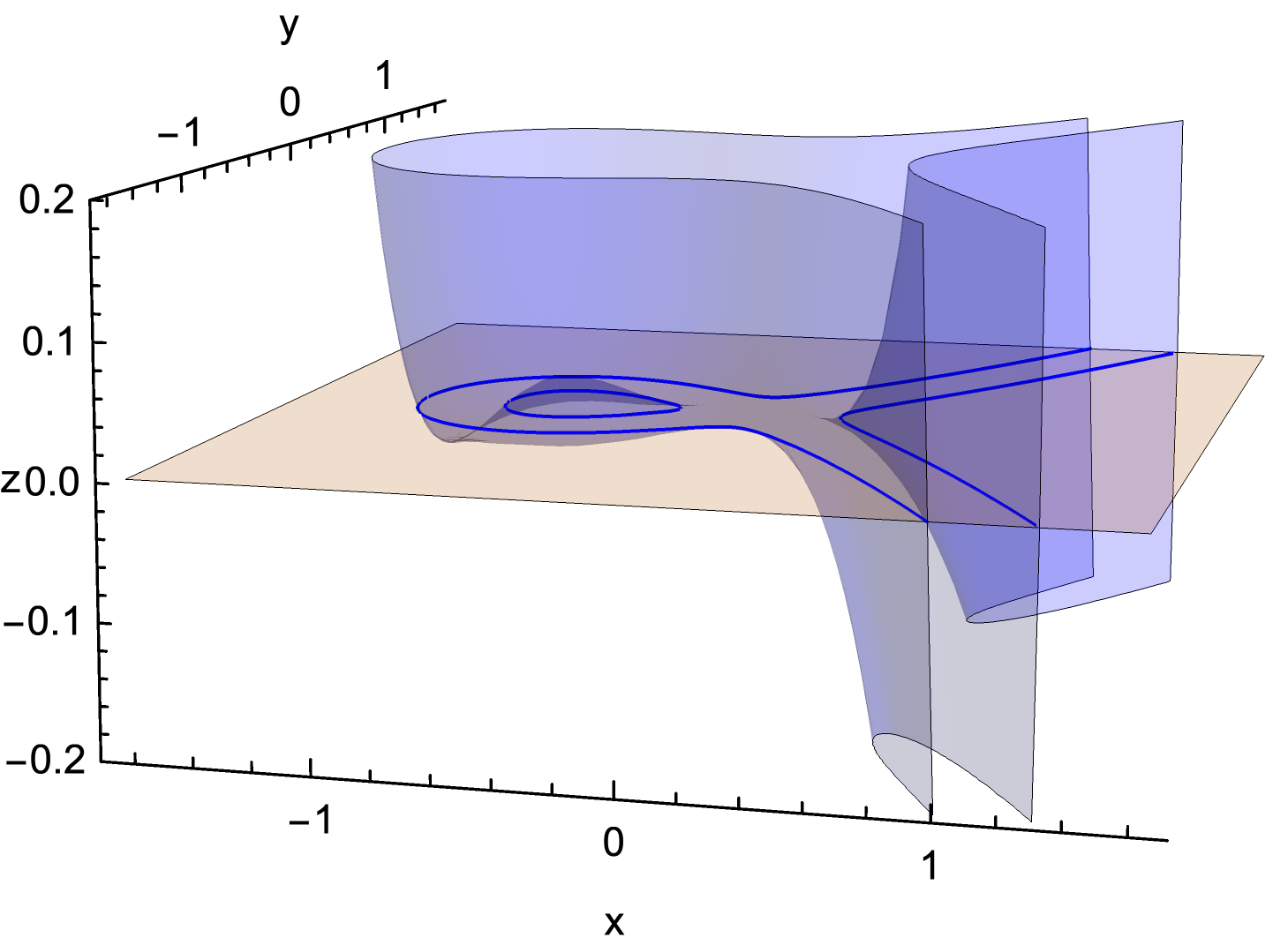} \\
{\color{white}.} \hskip .5in 
$\mc{G}_{g} \cap \mc{G}_{X\lpnorm{}{\nabla_{\ve{x}}g}}$ \hskip 1in 
$\mc{G}_{g-X\lpnorm{}{\nabla_{\ve{x}} g}} \cap \mc{G}_{0}$ \hskip .75in 
$\mc{G}_{g^{2}-X^{2}\lpnorm{}{\nabla_{\ve{x}} g}^{2}} \cap \mc{G}_{0}$
\caption{Geometrically, the variety normal distribution can be seen from various
	perspectives as selecting a point at the intersection of a fixed and a
	random surface (projected into $\R^{n}$). Perspectives I.--V. are
	illustrated, with the random surface in blue. In each case, the random
	surface illustrated was generated with $X = .15$, i.e. at 1.5 ``standard
	deviations'' from the variety with $\si = .1$.}
\label{fig:perspectives}
\end{figure}

Before discussing normalizability and the multivariety normal distribution, we
share a descriptive insight helpful in understanding the geometry of the VN
distribution. The VN distribution admits an interesting interpretation as a four
step procedure of selecting a random surface, intersecting it with a fixed
surface, projecting the intersection, and randomly selecting a point in the
resulting set. Let $\pi : \R^{n+1} \to \R^{n}$ denote the coordinate projection
$(x_{1}, \ldots, x_{n}, x_{n+1}) \mapsto (x_{1}, \ldots, x_{n})$ and $\mc{G}_{f}
= \set{\big(\ve{x},f(\ve{x})\big): \ve{x} \in \R^{n}} \subset \R^{n+1}$ the
graph of the function $f: \R^{n} \to \R$, a $n$-dimensional hypersurface in
$\R^{n+1}$. For $c \in \R$, let $\mc{G}_{c}$ denote the flat $n$-dimensional
hyperplane in $\R^{n+1}$ at height $c$ parallel to the $\ve{x}$-axis. Sampling
the VN distribution then admits the following five perspectives, among others,
each corresponding to different selections of hypersurfaces derived from
manipulations of the equation $\f{g}{\lpnorm{}{\nabla_{\ve{x}} g}} - X = 0$ with
$X \sim \Norm{0}{\sisq}$. The first is the one that motivated our previous
discussion.
\begin{enumerate}
\item[I.] $\mc{G}_{\bar{g}-X} \cap \mc{G}_{0}$: randomly shift
	$\mc{G}_{\bar{g}}$ up/down in $\R^{n+1}$ by a $\Norm{0}{\sisq}$ draw,
		intersect it with the $\ve{x}$-axis, project the intersection
		onto the first $n$ components, and sample uniformly from the
		resulting set. 

\item[II.] $\mc{G}_{\bar{g}} \cap \mc{G}_{X}$, randomly shift $\mc{G}_{0}$
	up/down in $\R^{n+1}$ by a $\Norm{0}{\sisq}$ draw, intersect it with
		$\mc{G}_{\bar{g}}$, project the intersection onto the first $n$
		components, and sample uniformly from the resulting set. 

\item[III.] $\mc{G}_{g} \cap \mc{G}_{X\lpnorm{}{\nabla_{\ve{x}}g}}$, which has
	two interpretations:
\begin{enumerate}
\item[(1)] differentially scale the random flat plane $\mc{G}_{X}$ by
	$\lpnorm{}{\nabla_{\ve{x}}g}$, intersect it with $\mc{G}_{g}$, project
		the intersection onto the first $n$ components, and sample
		uniformly from the resulting set, or
\item[(2)] randomly scale $\mc{G}_{\lpnorm{}{\nabla_{\ve{x}}g}}$ with
	a $\Norm{0}{\sisq}$ draw, intersect it with $\mc{G}_{g}$, project the
		intersection onto the first $n$ components, and sample uniformly
		from the resulting set.
\end{enumerate}

\item[IV.] $\mc{G}_{g-X\lpnorm{}{\nabla_{\ve{x}} g}} \cap \mc{G}_{0}$:
	intersect a carefully selected random hypersurface in $\R^{n+1}$ with
		the $n$-dimensional $\ve{x}$-axis, project the intersection onto
		the first $n$ components, and sample uniformly from the
		resulting set.

\item[V.] $\mc{G}_{g^{2}-X^{2}\lpnorm{}{\nabla_{\ve{x}} g}^{2}} \cap
	\mc{G}_{0}$, similar to the previous perspective but where the
		hypersurface is a variety,  which is of interest as it is more
		algebraic. This approximates the offset variety
		\citep{horobet2018offset}.
\end{enumerate}
These perspectives are illustrated in Figure~\ref{fig:perspectives}.

\subsection{Normalizability, emptiness, and near roots}\label{sec:normalizability}


\subsection{The multivariety normal distribution}\label{sec:multivariety-normal}

It is natural at this point to consider the relationship between the VN
distribution and the multivariate normal distribution: both are multivariate
distributions and both generalize the univariate normal distribution.  One key
difference is that, like the univariate normal distribution, the multivariate
normal distribution always concentrates its mass around a single point, the
zero-dimensional variety $\varty{V}(\ve{x}-\ve{\mu}) = \{\bm{\mu}\}$, whereas
the VN distribution can support a positive-dimensional variety.  A second
difference can be found in the variability about the variety: in the VN
distribution, $\sisq$ gauges deviations from the variety globally and
isotropically, whereas the multivariate normal distribution can support
differing amounts of variability anisotropically. In fact, the VN distribution
can be generalized to support this kind of variability as well. This leads us to
a more general formulation of the VN distribution.

\begin{defn}\label{defn:mvn}
A random vector $\ve{X} \in \R^{n}$ is said to have the multivariety normal
	(MVN) distribution, or simply the variety normal distribution, denoted
	$\ve{X} \sim \MVpNorm{n}{\ve{g}}{\bm{\Si}}$, if it admits a density
\begin{eqnarray}
p(\ve{x}|\ve{g}, \bm{\Si}) &\propto& 
\exp\set{-\f{1}{2}\bar{\ve{g}}(\ve{x}|\ma{B})'\bm{\Si}^{-1}\bar{\ve{g}}(\ve{x}|\ma{B})} \label{eq:mvn} \\
&=& \exp\set{-\f{1}{2}\big(\ma{J}_{\ve{x}}^{+}\ve{g}(\ve{x}|\ma{B})\big)'\bm{\Si}^{-1}\big(\ma{J}_{\ve{x}}^{+}\ve{g}(\ve{x}|\ma{B})\big)} 
\end{eqnarray}
with respect to the Lebesgue measure on $\R^{n}$ for some
	$\ve{g}(\ve{x}|\ma{B}) \in \R[\ve{x}]^{m}$ such that
	$\bar{\ve{g}}(\ve{x}|\ma{B}) =
	\ma{J}_{\ve{x}}^{+}\ve{g}(\ve{x}|\ma{B})$, where
	$\ma{J}_{\ve{x}}^{+}$ is the $n \times m$ pseudoinverse of
	the $m \times n$ $\ve{x}$-Jacobian matrix of
	$\ve{g}(\ve{x}|\ma{B})$, and $\bm{\Si} \in \S^{n}_{+}$, the
	cone of symmetric positive definite $n \times n$ matrices.
\end{defn}

\begin{figure}[h!]
\hskip .65in
$\varty{V}(x,y) \hskip 1in
\varty{V}(y - x, y + x) \hskip .65in
\varty{V}(y - x^{2}, y + x^{2}) \hskip .2in
\varty{V}(x y^{3} - x^{3} y, x^{2} + y^{2} - 1)$
\begin{center}
\includegraphics[scale=.185]{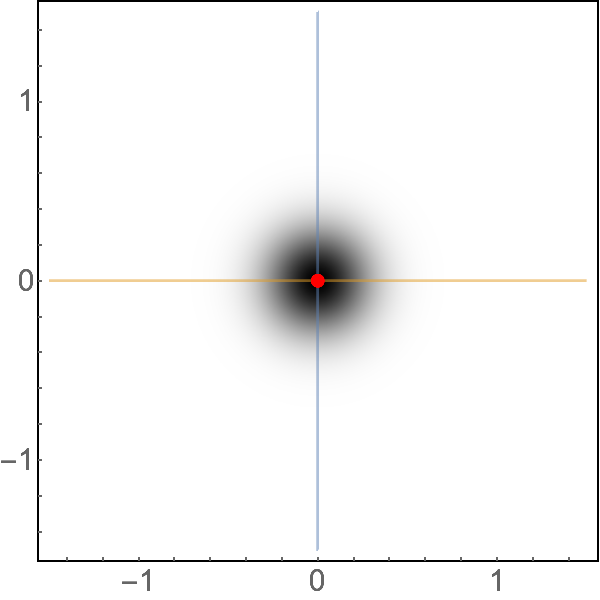}
\includegraphics[scale=.185]{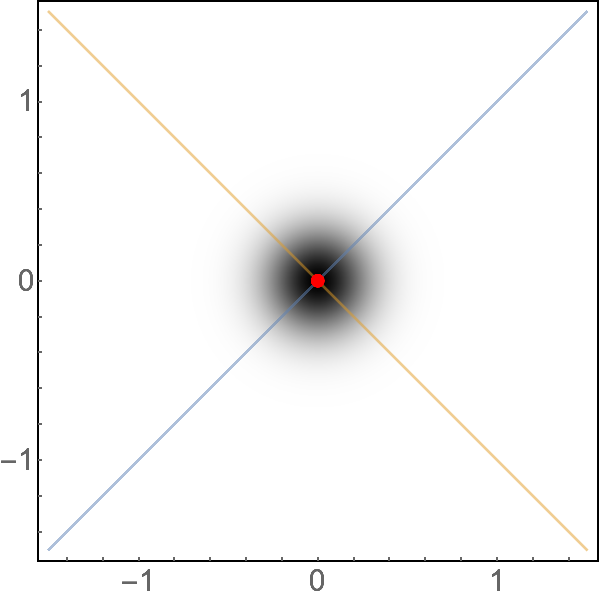}
\includegraphics[scale=.185]{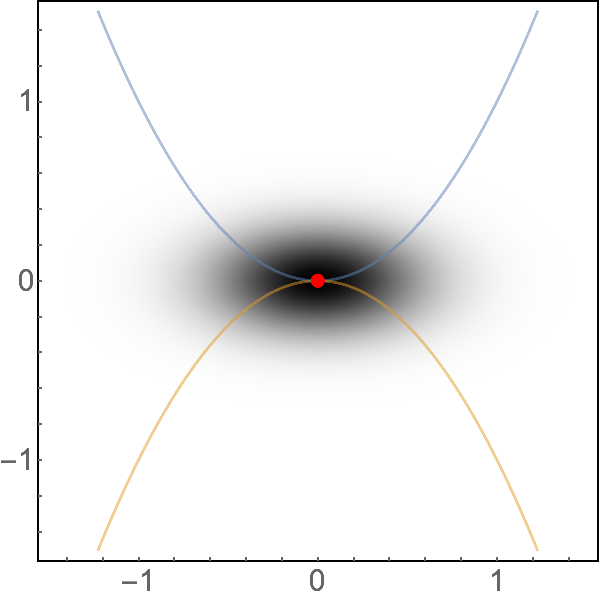}
\includegraphics[scale=.185]{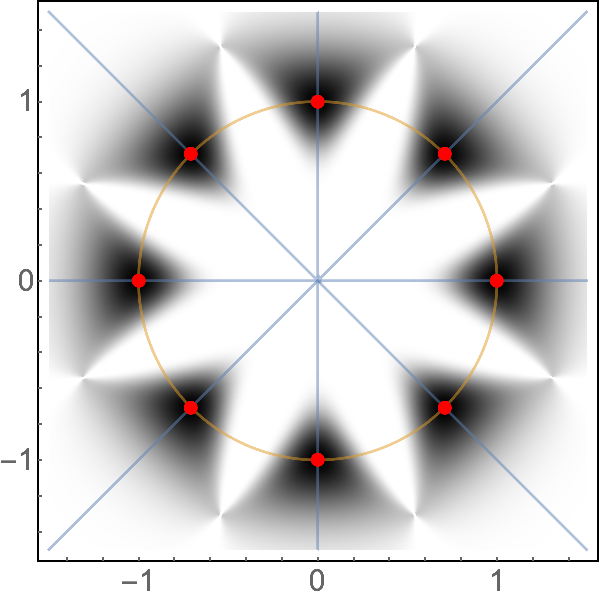}
\end{center}
\begin{center}
\includegraphics[scale=.185]{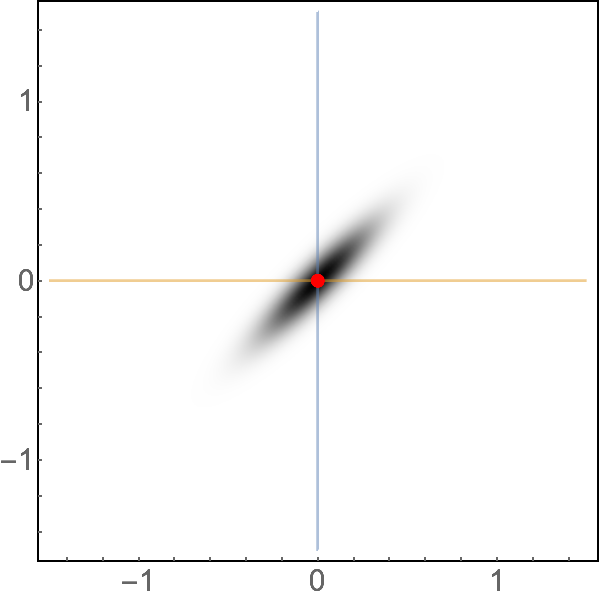}
\includegraphics[scale=.185]{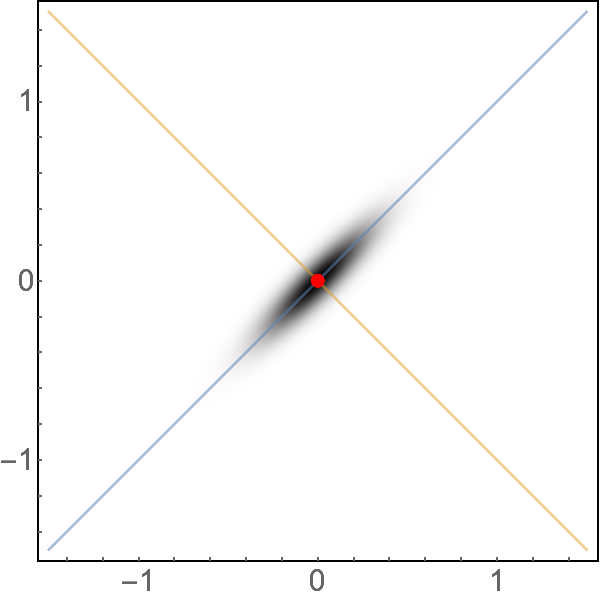}
\includegraphics[scale=.185]{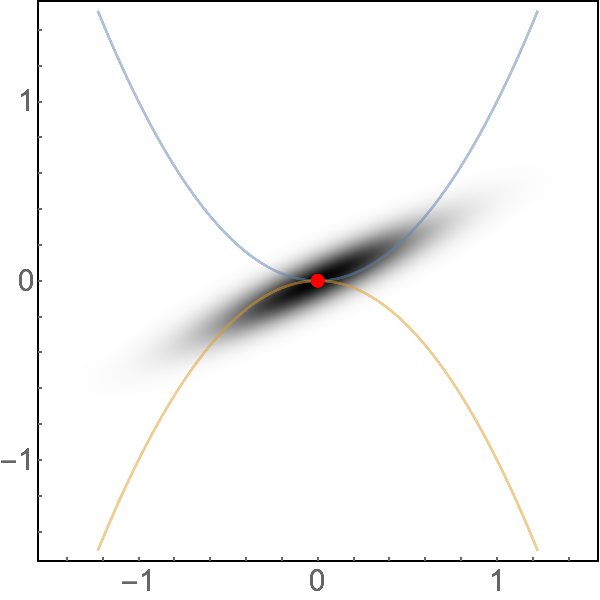}
\includegraphics[scale=.185]{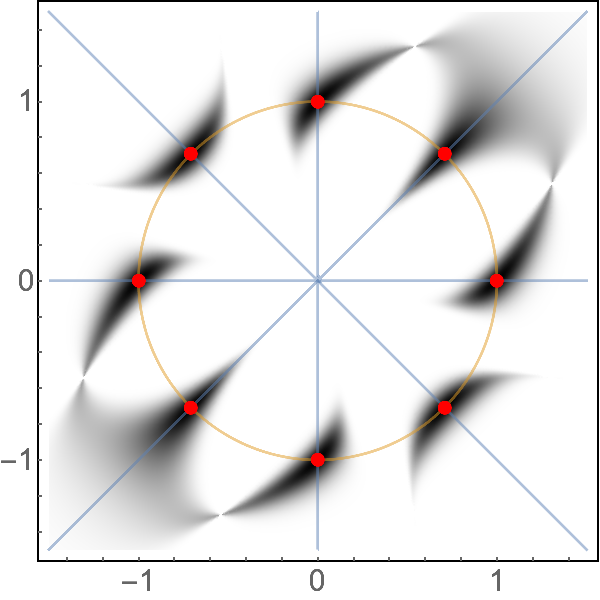} 
\end{center}
\caption{Multivariety normal distributions over zero-dimensional varieties and
	their associated systems with no ``correlation'' (top) and positive
	``correlation'' ($\rho = .9$, bottom). In each case $\bm{\Si} = \sisq
	\ma{I}$. While $\sisq = .2^{2}$ in each case, the density color scales
	and $z$-axes ranges vary across graphics.}
\label{fig:multivariety-1}
\end{figure}

\noindent $\bar{\ve{g}}(\ve{x}|\ma{B}) =
\ma{J}_{\ve{x}}^{+}\ve{g}(\ve{x}|\ma{B})$ generalizes the notion of normalizing
by the gradient.  Considering $\ve{g}$ as the vector field $\ve{g}: \R^{n} \to
\R^{m}$, the local linear approximation to points on the variety (properly
translated) is given by $\ma{J}_{\ve{x}} \in \R^{m \times n}$, which is
``inverted'' by $\ma{J}_{\ve{x}}^{+} \in \R^{n \times m}$. In the square case,
at smooth points the inverse function theorem guarantees that
$\bar{\ve{g}}(\ve{x}|\ma{B})$ will be properly normalized so that $\ma{\Si}$
gauges ``covariance'' globally across $\varty{V}(\ve{g})$. Several such square
MVN distributions over zero-dimensional varieties are illustrated in
Figure~\ref{fig:multivariety-1}. If there are more variables than equations ($n
> m$), the Jacobian represents an underdetermined system corresponding to a
positive dimensional variety; $\ma{J}_{\ve{x}}$ is full row rank for almost all
$\ve{x} \in \R^{n}$, and $\ma{J}_{\ve{x}}^{+} =
\ma{J}_{\ve{x}}'(\ma{J}_{\ve{x}}\ma{J}_{\ve{x}}')^{-1}$.  Previous examples of
the variety normal, a single equation in many variables, are in fact examples of
this situation. If there are more equations than variables ($m > n$) but the
variety is nonempty, the Jacobian represents an overdetermined system;
$\ma{J}_{\ve{x}}$ is full rank for almost every $\ve{x} \in \R^{n}$; and
$\ma{J}_{\ve{x}}^{+} = (\ma{J}_{\ve{x}}'\ma{J}_{\ve{x}})^{-1}\ma{J}_{\ve{x}}'$.
Either way, $\ma{J}_{\ve{x}}^{+}$ transforms the original $m$-dimensional
$\ve{g}$ into an $n$-dimensional vector $\bar{\ve{g}}$ compatible with
$\bm{\Si}$.

As one might expect, the multivariety normal family of distributions subsumes
the variety normal and multivariate normal families.  And, since the
multivariety normal subsumes the multivariate normal, it subsumes the
matrix-variate normal and tensor-variate normal (also called the multilinear
normal) families as well \citep{ohlson2013multilinear}.  We now explain how this
works.

If $m = 1$ and $\bm{\Si} = \sisq \ma{I}_{n}$, the multivariety normal
distribution reduces to the variety normal distribution. To see this, define
$\ve{d} = \nabla_{\ve{x}}g(\ve{x}|\bm{\be})$ and note that in the $m = 1$ case
$\ve{g}(\ve{x}|\ma{B}) = g(\ve{x}|\bm{\be}) = g$, $\ma{J}_{\ve{x}} = \ve{d}'$,
and $\ma{J}_{\ve{x}}^{+} = \ve{d}(\ve{d}'\ve{d})^{-1} =
\f{1}{\lpnorm{}{\ve{d}}^{2}}\ve{d}$.  So if $\bm{\Si} = \sisq \ma{I}_{n}$,
\begin{eqnarray*}
\exp\set{-\f{1}{2}\big(\ma{J}_{\ve{x}}^{+}\ve{g}\big)'\bm{\Si}^{-1}\big(\ma{J}_{\ve{x}}^{+}\ve{g}\big)} 
&=&
\exp\set{-\f{1}{2}\left(\Big(\f{1}{\lpnorm{}{\ve{d}}^{2}}\ve{d}\Big)g\right)'(\sisq \ma{I}_{n})^{-1}\left(\Big(\f{1}{\lpnorm{}{\ve{d}}^{2}}\ve{d}\Big)g\right)} \\
&=&
\exp\set{-\f{1}{2\sisq}\left(\f{g}{\lpnorm{}{\ve{d}}}\Big(\f{\ve{d}}{\lpnorm{}{\ve{d}}}\Big)'\Big(\f{\ve{d}}{\lpnorm{}{\ve{d}}}\Big)\f{g}{\lpnorm{}{\ve{d}}}\right)} \\
&=&
\exp\set{-\f{1}{2\sisq}\left(\f{g}{\lpnorm{}{\nabla_{\ve{x}}g}}\right)^{2}} \ \ 
= \ \ \exp\set{-\f{\bar{g}^{2}}{2\sisq}},
\end{eqnarray*}
which is \eqref{eq:vn}, the PDF of the variety normal distribution.  More
generally, if the system is not overdetermined, $\ma{J}_{\ve{x}}$ typically has
$m$ independent rows and so if $\bm{\Si} = \sisq \ma{I}_{n}$,
\begin{eqnarray}
p(\ve{x}|\ve{g}, \bm{\Si}) &\propto& \exp\set{-\f{1}{2\sisq}\ve{g}'(\ma{J}_{\ve{x}}\ma{J}_{\ve{x}}')^{-1}\ve{g}}.  \label{eq:varty-normal-simple-not-over}
\end{eqnarray}
If $\ve{g} = \ve{x} - \ve{\mu}$, $m = n$ and $\ma{J}_{\ve{x}} = \ma{I}_{n}$, and
the distribution reduces to the multivariate normal.  The individual varieties
of the $g_{i}$'s are the $n$ hyperplanes of dimension $n-1$ that run parallel to
the coordinate axes and intersect at the point $\ve{\mu}$.

\begin{figure}[h!]
\begin{center}
\includegraphics[scale=.185]{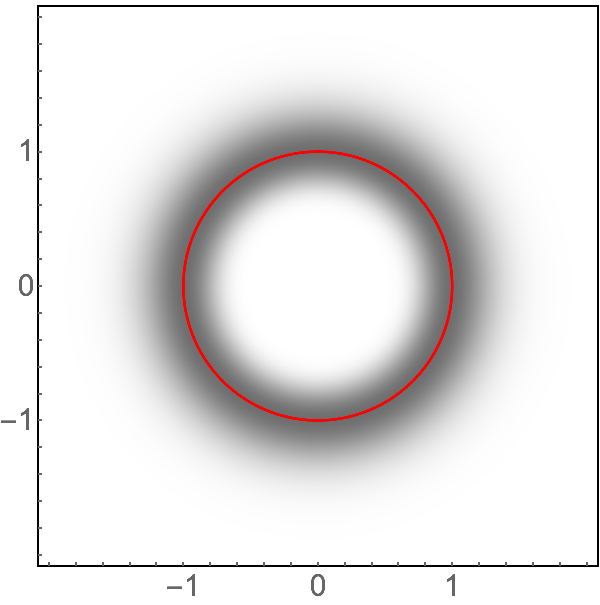}
\includegraphics[scale=.185]{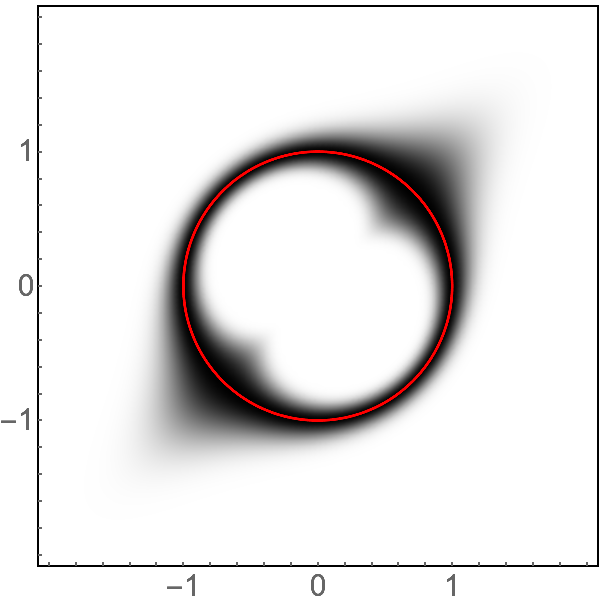}
\includegraphics[scale=.185]{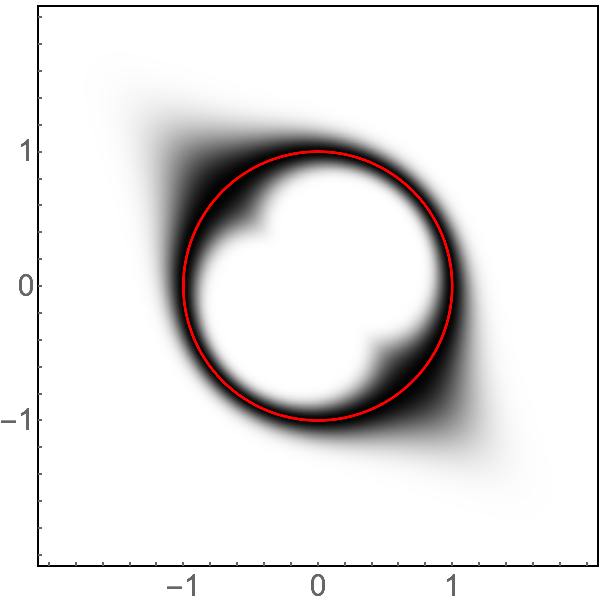}
\end{center}
\caption{Multivariety normal densities with $g = x^{2} + y^{2} - 1$,
	$\si_{1} = \si_{2} = .2$, and $\rho = 0, .9, -.9$.}
\label{fig:multivariety-circle}
\end{figure}

\begin{figure}[h!]
\begin{center}
\includegraphics[scale=.18]{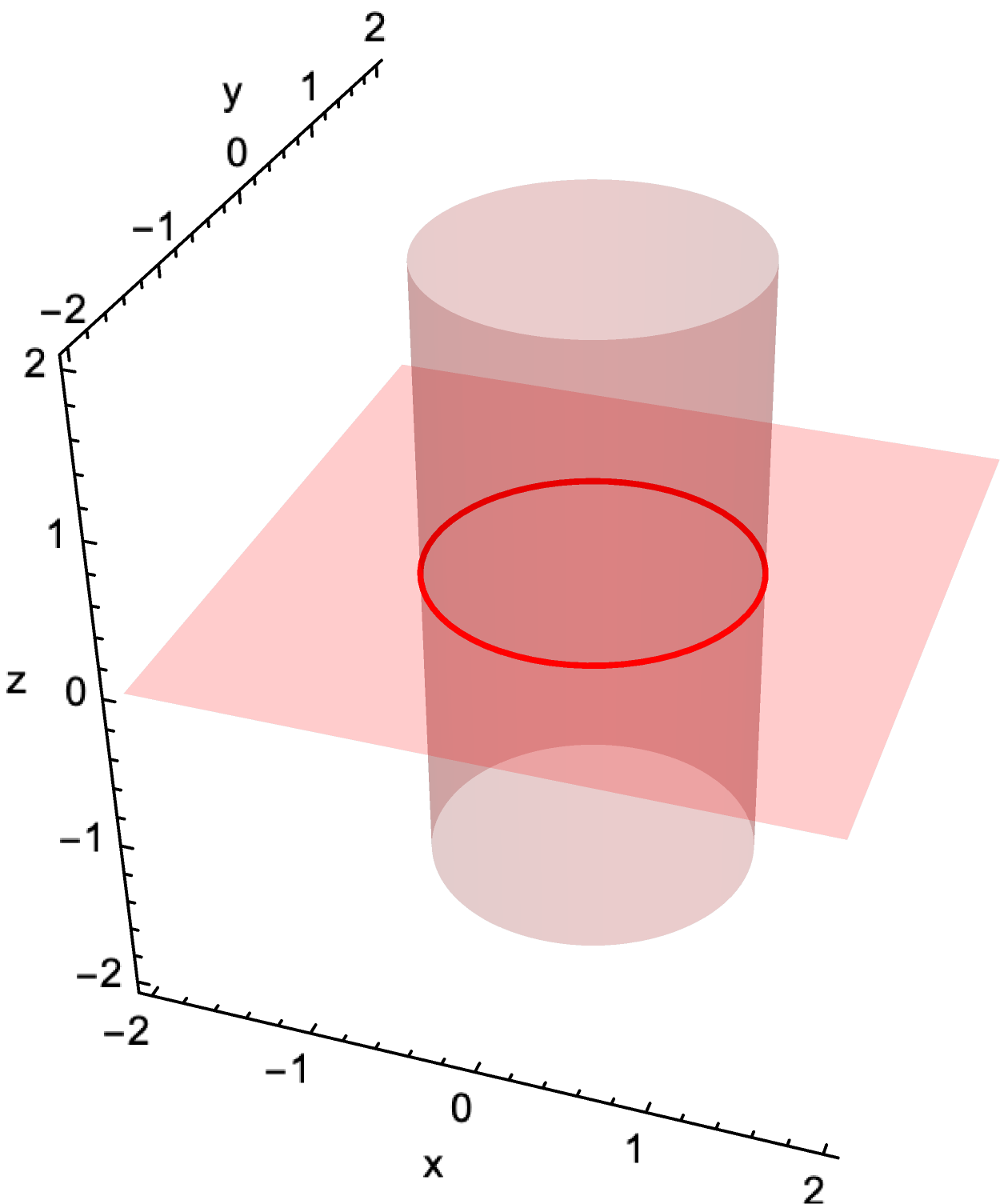}
\includegraphics[scale=.18]{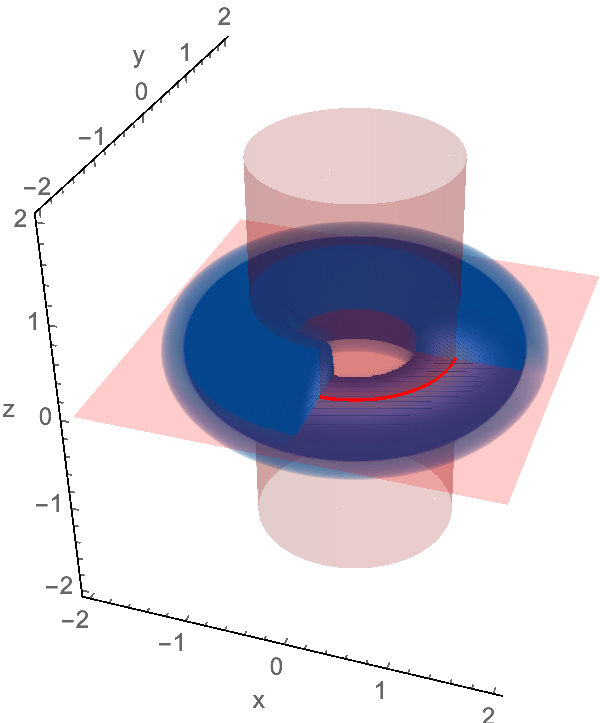}
\includegraphics[scale=.18]{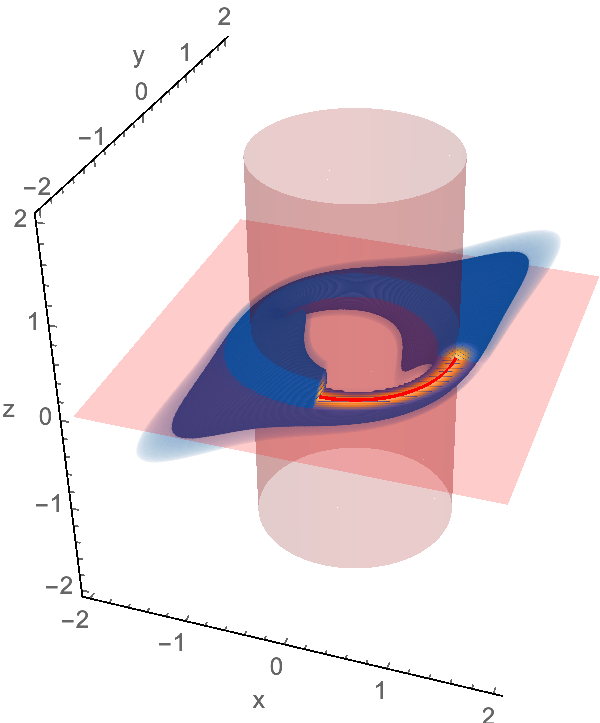}
\includegraphics[scale=.18]{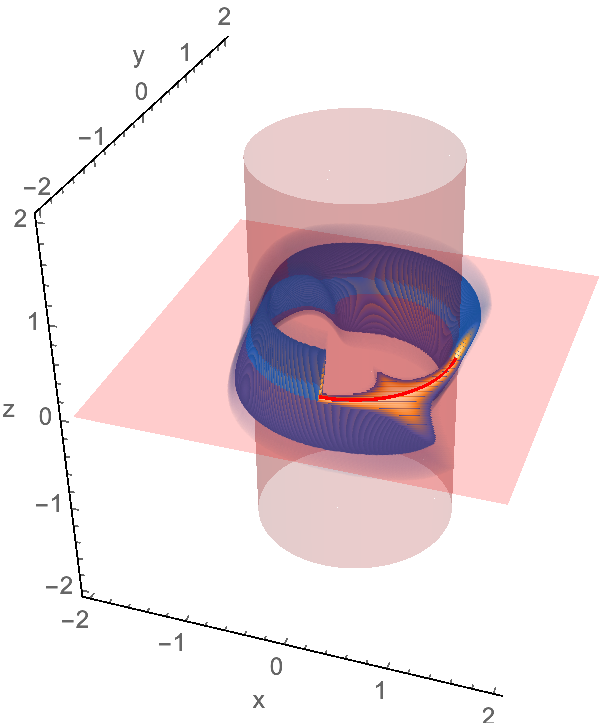}
\end{center}
\caption{From left to right, the varieties of $g_{1} = x^{2} + y^{2} - 1$ and
	$g_{2} = z$ and their intersection $\varty{V}(g_{1}, g_{2})$, the unit
	circle in the $xy$ plane; and three 3D density plots of the MVN
	distribution with $\si_{1} = \si_{2} = \si_{3} = .2$ and $\rho_{12} =
	\rho_{13} = \rho_{23} = 0$, $\rho_{12} = \rho_{13} = \rho_{23} = .9$,
	and $\rho_{12} = -.9$, $\rho_{13} = .5$, and $\rho_{23} = -.1$.
	Aesthetic scales are consistent across graphics, and the $x,z \geq 0$,
	$y \leq 0$ orthant has been removed to see inside the regions.}
\label{fig:plane-cylinder}
\end{figure}

An added nuance of the MVN distribution not present in the VN distribution is
that of covariance, or rather what would be covariance in a multivariate normal
distribution.  The covariance for MVN distributions is far more complex than the
parameters themselves and not of primary interest, like other moments for
variety normal distributions. Figure~\ref{fig:multivariety-circle} considers the
variety corresponding to the polynomial $g = x^{2} + y^{2} - 1$ using $\bm{\Si}
=
\twomat{\si_{1}}{0}{0}{\si_{2}}\twomat{1}{\rho}{\rho}{1}\twomat{\si_{1}}{0}{0}{\si_{2}}$
with $\si_{1} = \si_{2} = .2$ and $\rho = 0$, $.9$, and $-.9$. When $\rho = 0$,
the variability about $\varty{V}(\ve{g})$ is isotropic; this is the
``independence'' case. When it is positive, variability is oriented towards
regions of same-signed coordinates, and the opposite when it is negative.
Figure~\ref{fig:plane-cylinder}, the one dimensional variety generated by the
two polynomials $g_{1} = x^{2} + y^{2} - 1$ and $g_{2} = z$, illustrates the
effect of ``correlation'' among the three variables, which can be very complex.

Understanding a variety tends to be easiest when correlations are zero, so that
$\bm{\Si}$ is diagonal, and the magnitude of the thickening (variability) in
each coordinate axis is uniform, so that $\bm{\Si} = \sisq \ma{I}_{n}$. In
principle, the smaller the $\sisq$, the better; however, if $\sisq$ is chosen to
be too small, the sampling schemes described in Section~\ref{sec:sampling} tend
to not perform as well. We revisit this topic there and in
Section~\ref{sec:Kuramoto}.

\section{Induced variety distributions}\label{sec:induced}

The construction of the variety normal distributions admits a different
interpretation that leads to a nice generalization whereby one can put
essentially any distribution around a variety in the same way that those
distributions put a normal around it.  In the VN case, instead of thinking of
$\bar{g}(\ve{x}|\bm{\be})$ as replacing $x - \mu$ in the normal kernel, one can
think of taking any distribution in the normal family, applying a location
transformation so that the mode is zero, and then substituting
$\bar{g}(\ve{x}|\bm{\be})$ for $x$; it just so happens that in the normal case
this location transformation resets $\mu$ to 0. The same is true for the MVN
case by replacing $\ve{x}-\bm{\mu}$ with $\bar{\ve{g}}(\ve{x}|\ma{B})$. This
same process applies generally:
\begin{enumerate}
\item Select a univariate or multivariate distribution with features of
	interest, e.g. unimodality,
\item Shift the distribution's location to the origin, rescaling as desired, and
\item Substitute $\bar{\ve{g}}(\ve{x}|\ma{B})$ for $\ve{x}$ in the resulting
	expression.
\end{enumerate}
\noindent We refer to these distributions as induced variety distributions.

\begin{figure}[h!]
\begin{center}
\includegraphics[scale=.21]{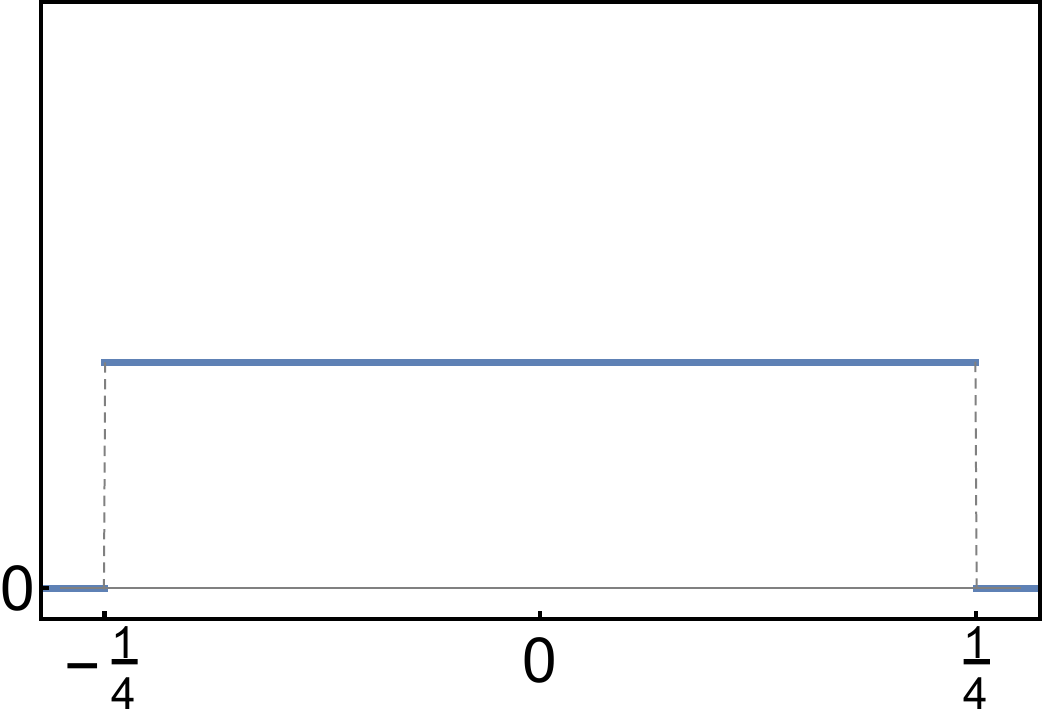} \hskip .05in
\includegraphics[scale=.21]{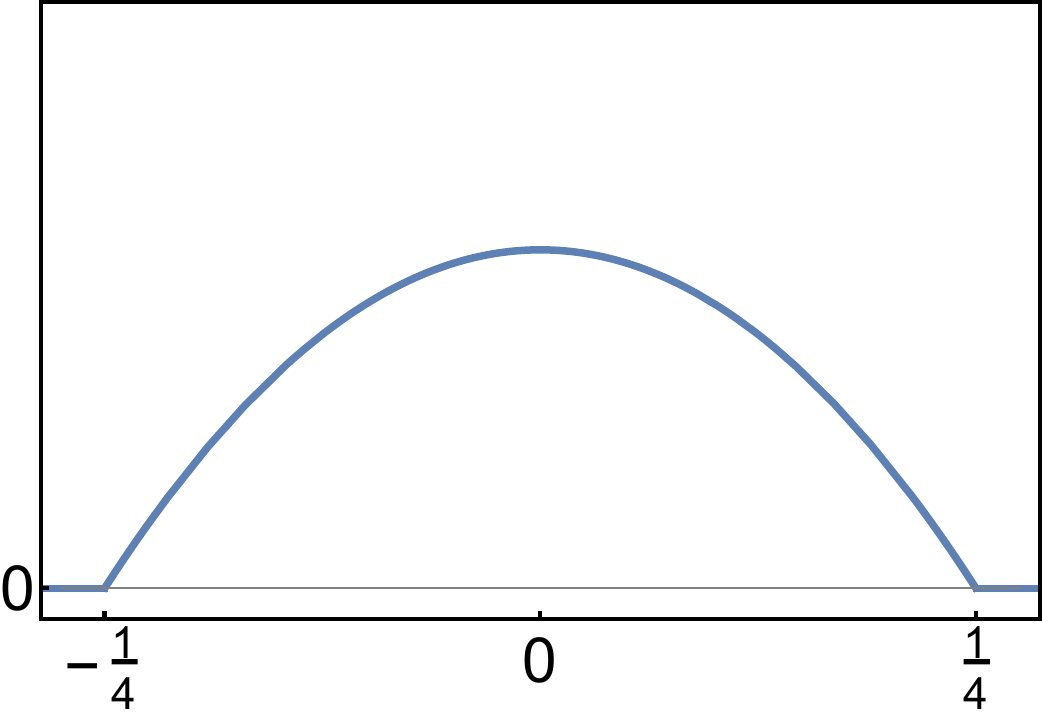} \hskip .05in
\includegraphics[scale=.21]{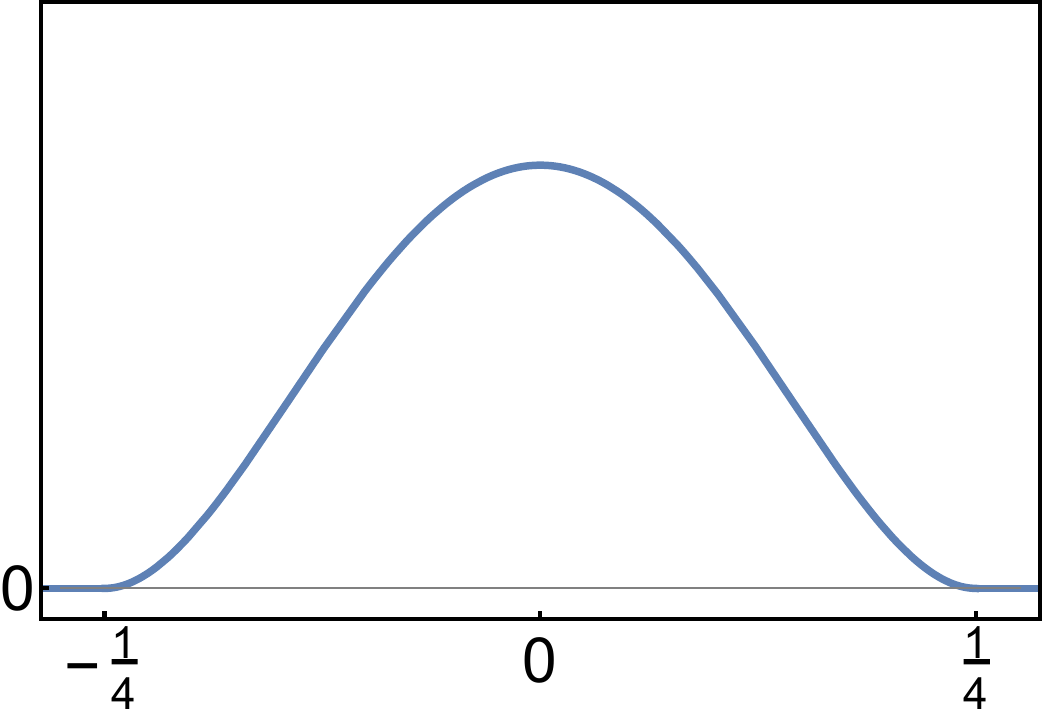} \hskip .05in
\includegraphics[scale=.21]{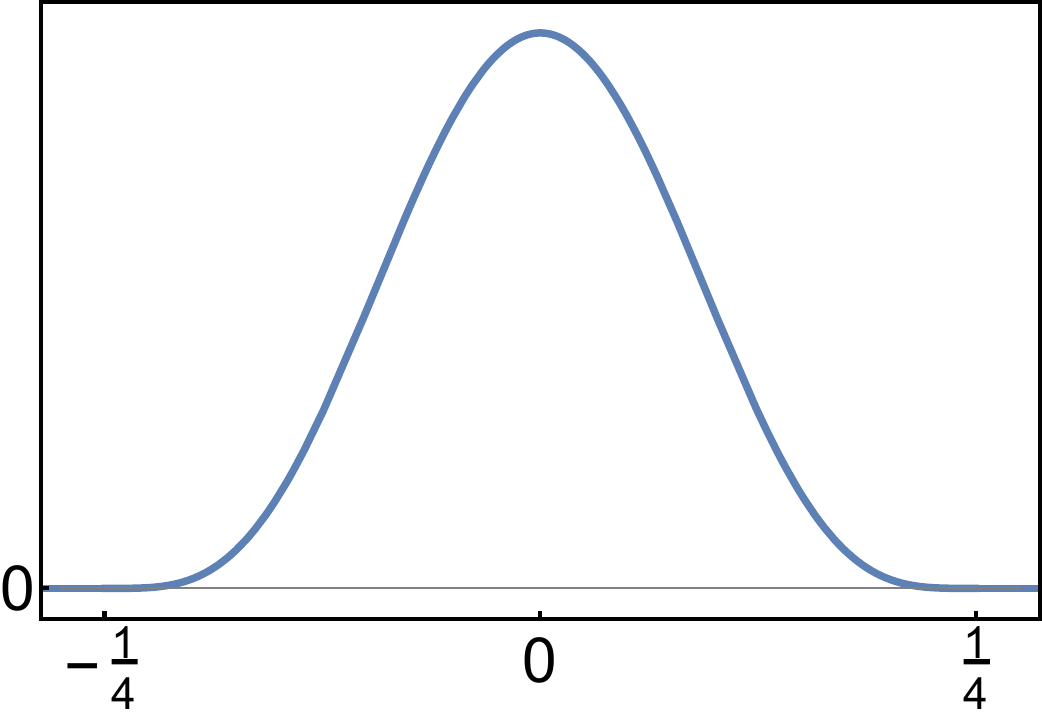} \\
\includegraphics[scale=.18]{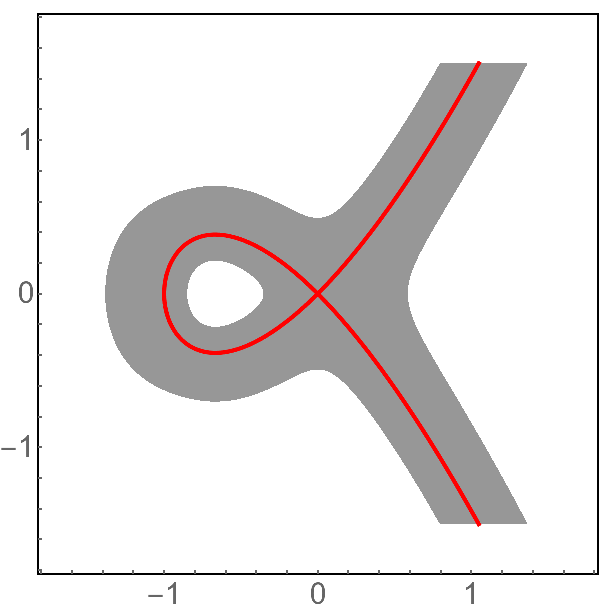}
\includegraphics[scale=.18]{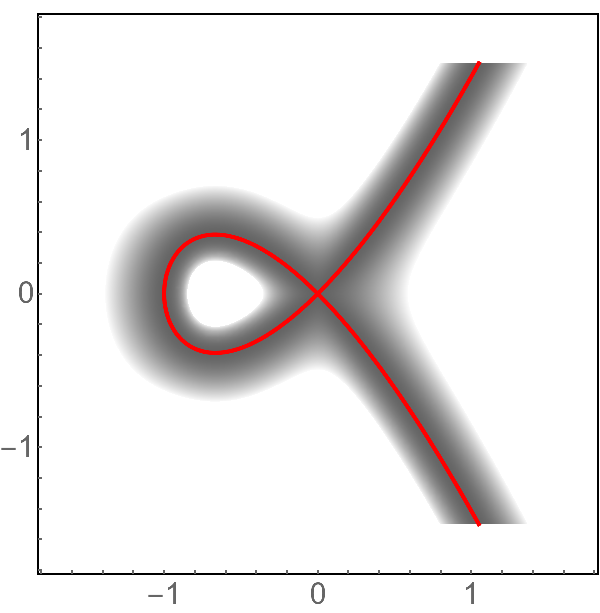}
\includegraphics[scale=.18]{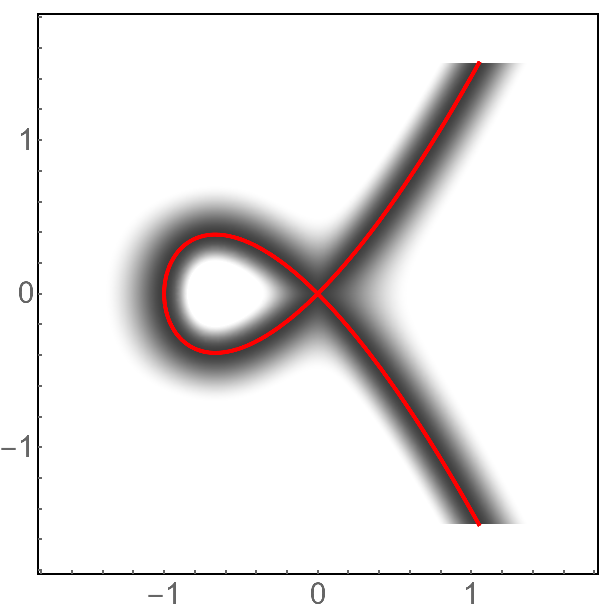}
\includegraphics[scale=.18]{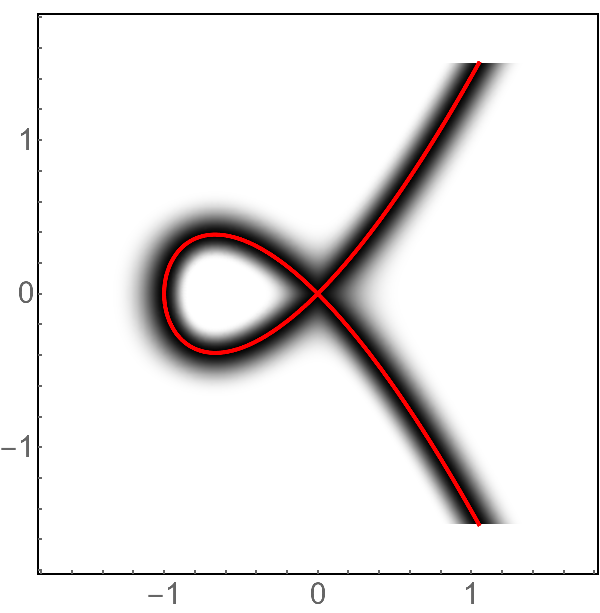}
\end{center}
\caption{The induced variety beta distribution on the alpha curve with $\si =
	1/2$, $\mu = -\si/2$, and $(\al,\be) = (1,1), (2,2), (3,3)$, and
	$(5,5)$, from left to right.}
\label{fig:alpha-beta}
\end{figure}

It is helpful to illustrate the process on a univariate family. Suppose again we
are interested in the alpha curve, the variety of $g = y^{2} - (x^{3}-x^{2})$,
but want to use a uniform distribution, a member of the beta family, instead of
a normal.  The PDF of the beta distribution is $p(x|\al,\be) \propto
x^{\al-1}(1-x)^{\be-1}$ on $0 < x < 1$ for $\al, \be > 0$; the uniform occurs
when $\al = 1$ and $\be = 1$.  If $X \sim p(x|\al,\be)$, it is a basic fact that
if $\si > 0$ and $\mu \in \R$ the distribution of $\si X + \mu$ is proportional
to $p(\f{x-\mu}{\si}|\al,\be)$ for $x \in (\mu,\mu+\si)$, so choosing $\mu =
-\f{\si}{2}$ is reasonable to center the distribution on zero. The resulting
bivariate distribution described by the method is therefore
$p(x,y|\al,\be,\mu,\si) \propto p(\f{\bar{g}-\mu}{\si} \Big| \al,\be)$ with
$\bar{g} = (y^{2}-(x^{3}+x^{2})) / \sqrt{(3x^{2}+2x)^{2} + (2y)^{2}}$. As
$\varty{V}(g)$ is not compact, we truncate the distribution to $\bm{\mc{X}} =
[-1.5,1.5]^{2}$. This is illustrated in Figure~\ref{fig:alpha-beta} with $\si =
1/2$ and various settings of $\al$ and $\be$. The asymmetry observed is an
artifact of selecting $\si$ too large relative to the linear approximation
produced by the gradient normalization.

While this strategy works, it doesn't seem to have many advantages over the
variety normal distributions. One limitation is that if the selected
distribution has compact support, the resulting distribution will not have the
full support of $\R^{n}$, and consequently Hamiltonian samplers may be prone to
more problems, e.g. initialization and feasible proposals. However, one
observation about these distributions enables the creation of distributions
about semi-algebraic sets.

\subsection{Semi-algebraic distributions}\label{sec:semialg}

Although the specific definition varies slightly, varieties are commonly
referred to as algebraic sets in the literature.  Semi-algebraic sets are
generalizations of algebraic sets that allow for inequalities.
\begin{defn}\label{defn:semialg}
A semi-algebraic set is a set of the form $\set{\ve{x} \in \R^{n}:
	\ve{g}(\ve{x}) = \ve{0}_{m} \ \mbox{and} \ \ve{h}(\ve{x}) > \ve{0}_{l}}$
	for $\ve{g}(\ve{x}) \in \R[\ve{x}]^{m}$ and $\ve{h}(\ve{x}) \in
	\R[\ve{x}]^{l}$.
\end{defn}
\noindent A few minor modifications to the strategies described above allow us
to create distributions that focus their mass on semi-algebraic sets in the same
way that variety distributions focus their mass on varieties. We note a quick
trick at the outset, however, which is both common and useful in practice: if a
single variable needs to be non-negative, it suffices to swap the variable with
the square of another variable. As the square is always non-negative (over the
reals), one simply retains the draws of the squared variable.

There are two basic approaches one can take to creating semi-algebraic
distributions.  
\begin{enumerate}
\item The first strategy is similar to the induced variety approach, but instead
	of picking a distribution that centers its mass on 0, pick a
		distribution that borders 0 by piling all its mass up against
		one side. We call the resulting distribution a border (induced
		variety) distribution.  While this strategy seems like a good
		fit, it suffers from two drawbacks. First, while it works for
		varieties generated by a single polynomial, it is not clear how
		it could be conveniently applied to semi-algebraic sets in
		general, i.e. sets specified with several polynomials and a mix
		of equalities and inequalities.  Second, the usefulness of the
		resulting distribution for understanding the semialgebraic set
		is very sensitive to the selection of the base distribution,
		often placing mass in the immediate viscinity of the boundary of
		the semi-algebraic set and almost entirely missing parts from
		from it.  
\item The second strategy is to eliminate inequalities by converting them to
	equalities by introducing slack variables, placing a distribution about
		the corresponding variety in the higher dimensional space, and
		marginalizing out the slack variables. Geometrically, this
		corresponds to creating a variety in higher dimensional space
		whose projection is the semi-algebraic sets of interest. This
		strategy appears to largely alleviate both of the challenges
		faced by the first strategy. 
\end{enumerate}
As a simple illustration of these concepts, consider sampling from the unit disc
$\set{(x,y) \in \R^{2} : x^{2} + y^{2} < 1}$, which in canonical form is
$\set{(x,y) \in \R^{2}: h(x,y) > 0}$ with $h(x,y) = 1  - (x^{2} + y^{2})$:
\begin{enumerate}
\item For a border distribution, we may select the uniform distribution as the
	base distribution: $p(x|0,\be) = 1[0 \leq x \leq \be]$.  Following the
		approach of the previous section, no shifting of the
		distribution is required, since it comes up against 0 from
		above, and we can simply set the distribution of interest to be
		$p(x,y|0,\be) \propto 1[0 \leq \bar{h}(x,y) \leq 1]$, where
		$\bar{h}(x,y) = \f{1  - (x^{2} + y^{2})}{2\sqrt{x^2 + y^2}}$.
		This is illustrated in Figure~\ref{fig:semialg-variety-induced}. 

\begin{figure}[h!]
\begin{center}
\includegraphics[scale=.21]{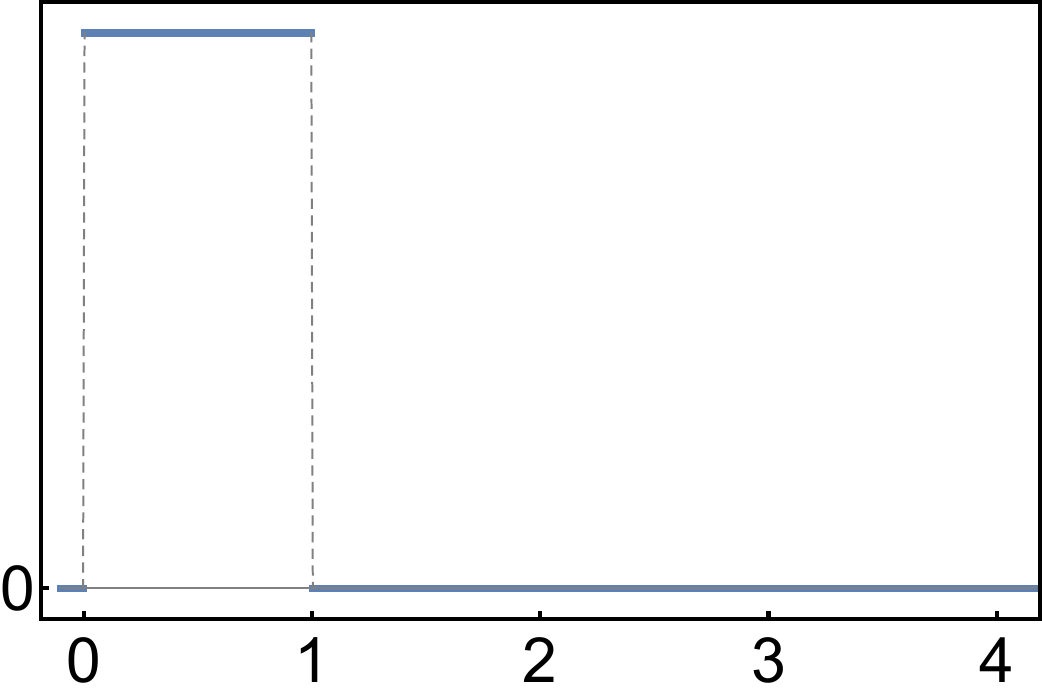} \hskip .04in 
\includegraphics[scale=.21]{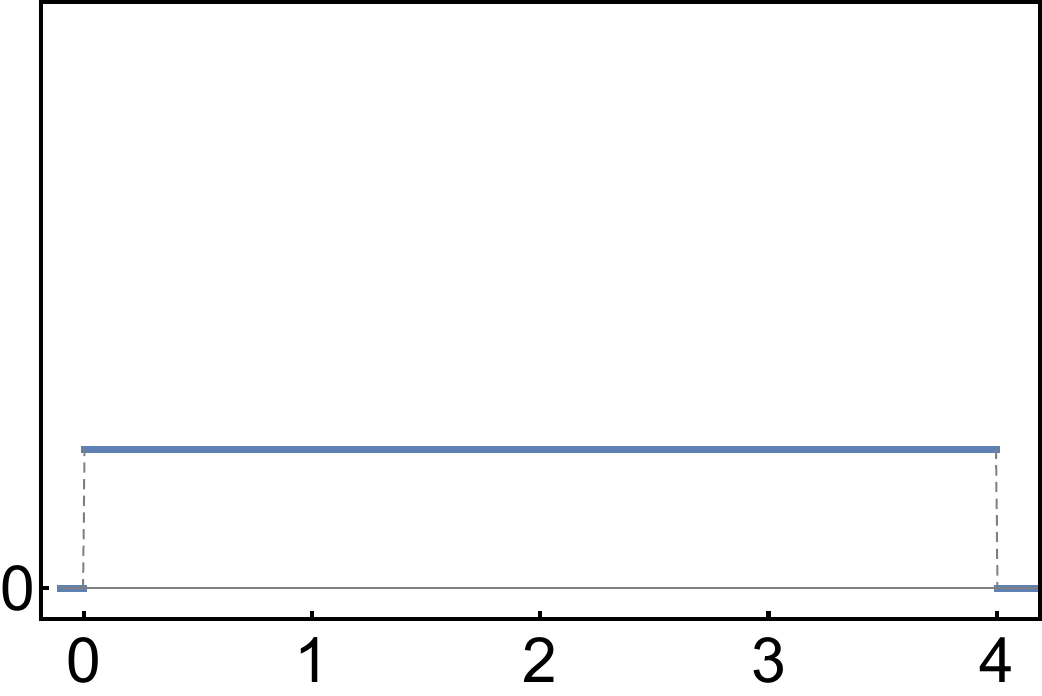} 
\includegraphics[scale=.21]{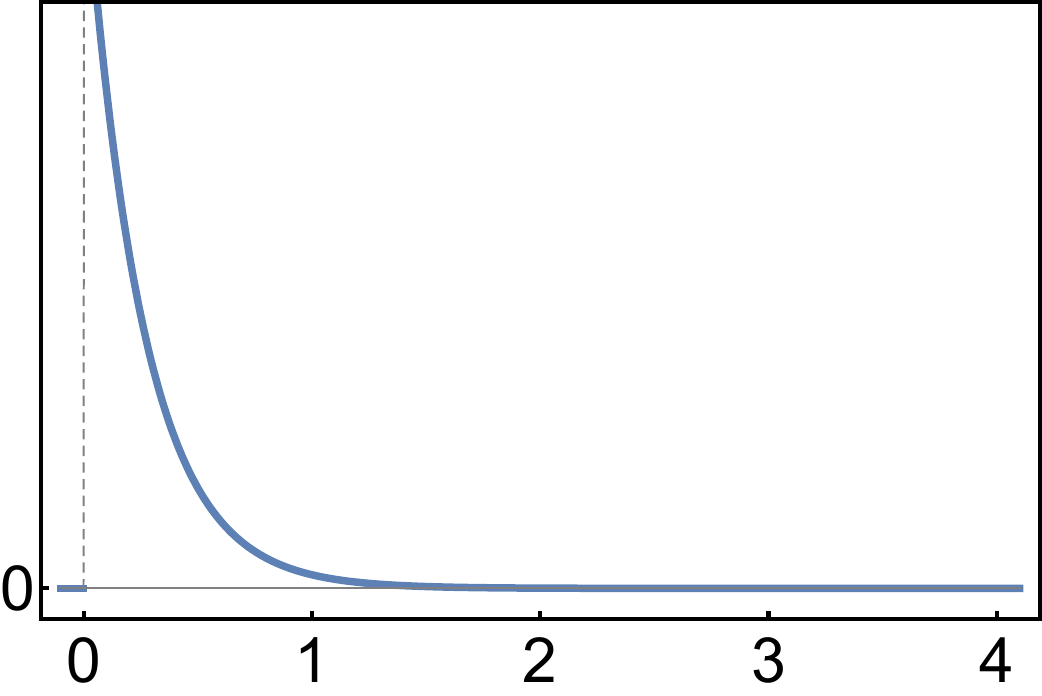} \hskip .04in 
\includegraphics[scale=.21]{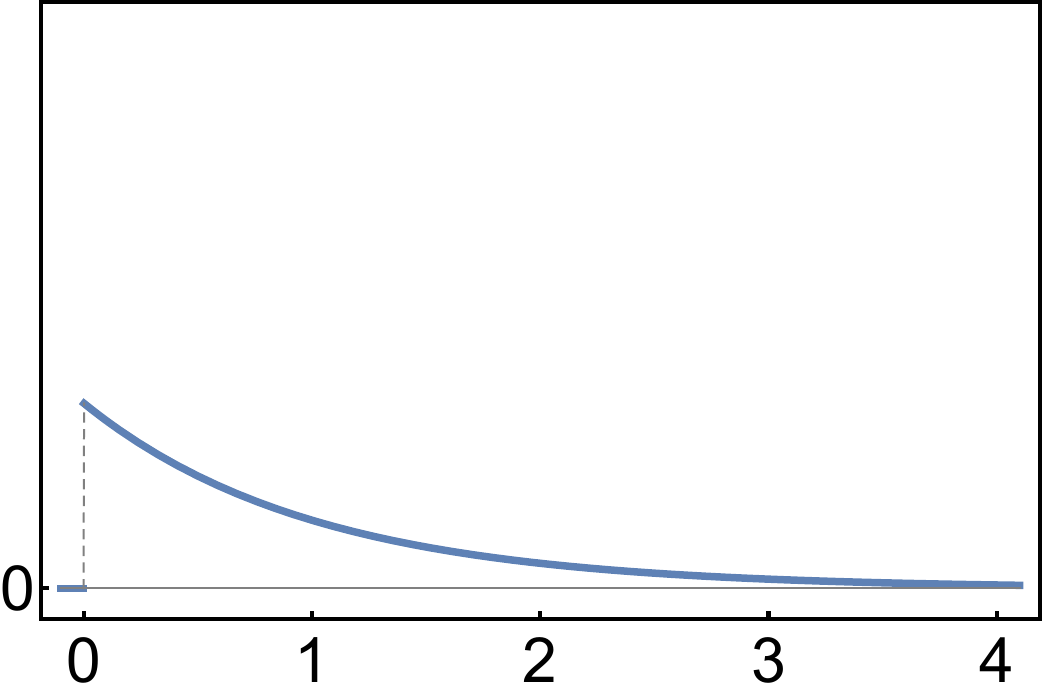}  \\
\includegraphics[scale=.18]{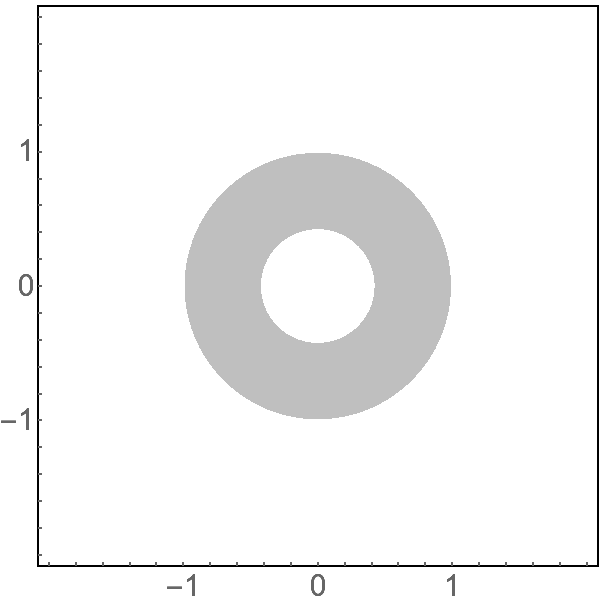} 
\includegraphics[scale=.18]{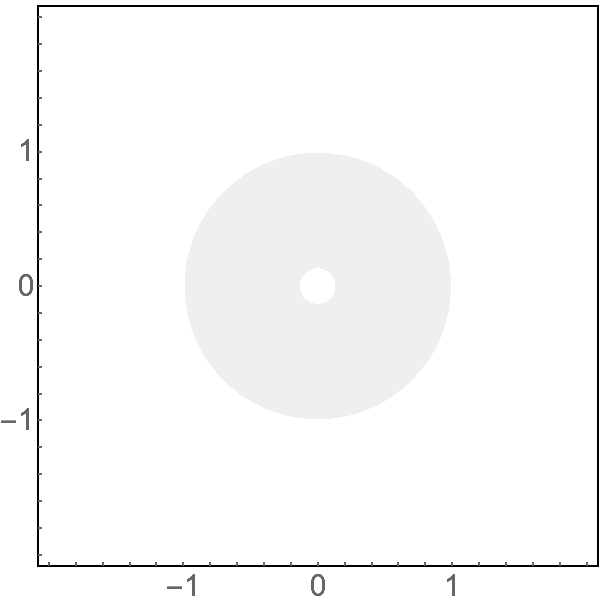} 
\includegraphics[scale=.18]{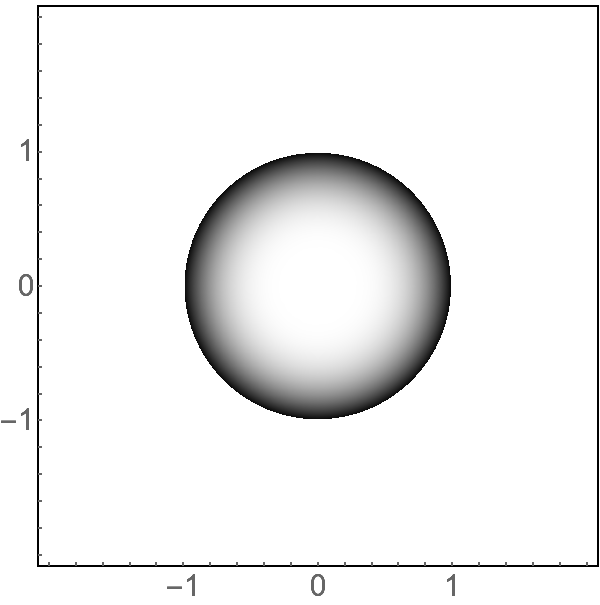} 
\includegraphics[scale=.18]{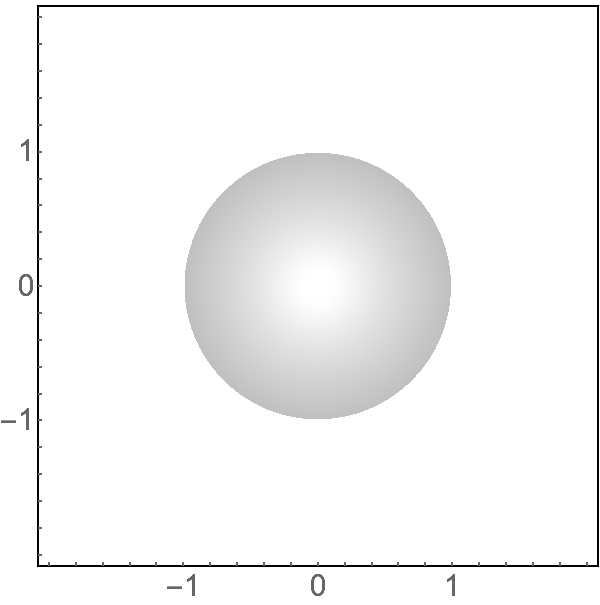} 
\end{center}
\caption{Distributions about semi-algebraic sets can be constructed by bordering
	distributions on one side of 0. From left to right, this is illustrated
	using the $\Unif{0}{1}$, $\Unif{0}{4}$, $\ExpD{4}$, and $\ExpD{1}$
	distributions.}
\label{fig:semialg-variety-induced}
\end{figure}

Notice that the resulting distribution always exhibits a hole in the middle,
		resulting in a distribution on an annulus instead of the disk.
		While this is alleviated by forcing $\be$ to be large, it always
		induces the artificial hole.  As an alternative, one may choose
		to use a base distribution without an upper bound such as the
		exponential distribution, whose density is $p(x|\la) = \la
		e^{-\la x}$ for $x \geq 0$. This is illustrated in
		Figure~\ref{fig:semialg-variety-induced}.  Unfortunately this,
		too, results in a noticeably non-uniform distribution. As a
		uniform distribution cannot be placed on all of $\R_{+}$, this
		strategy generally appears unsatisfying.

\item By contrast, the second strategy proceeds by recognizing that the
	condition $h(x,y) > 0$ implies the existence of a positive number
		$s^{2}$ such that $g(x,y,s) = h(x,y) - s^{2} = 0$. Since this
		describes a variety in $\R^{3}$, we can use any of the methods
		of the previous sections to place a distribution on it. We can
		then marginalize out the lifting/slack/auxiliary variable $s$ to
		obtain the distribution of interest. In this case, since
		$g(x,y,s) = 1 - x^{2} - y^{2} - s^{2}$, this corresponds to
		creating a distribution about the sphere in $\R^{3}$ and
		marginalizing. This is illustrated in Figure~\ref{fig:semialg}.

\begin{figure}[h!]
\begin{center}
\includegraphics[scale=.2]{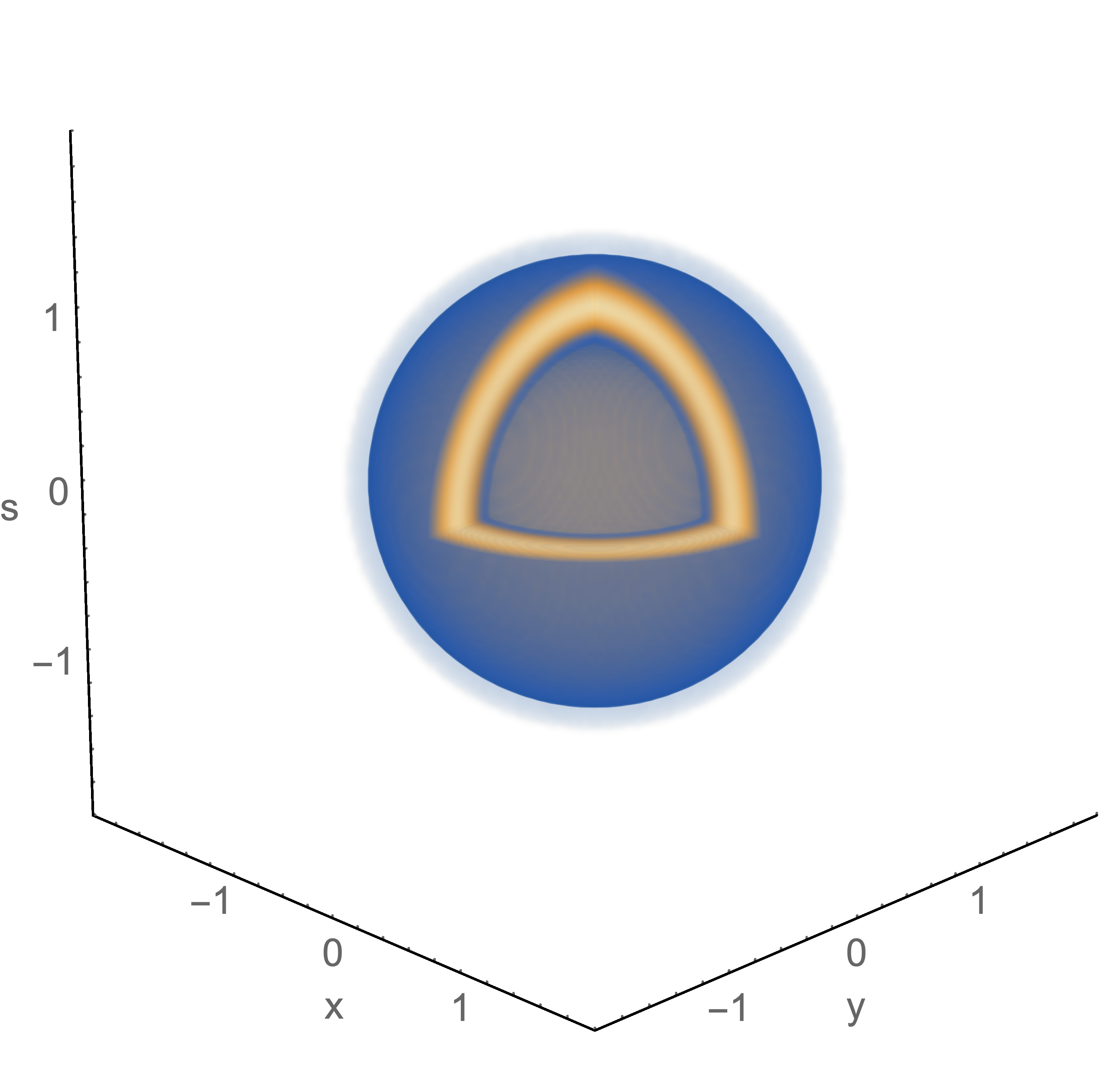} \hskip .1in
\includegraphics[scale=.2]{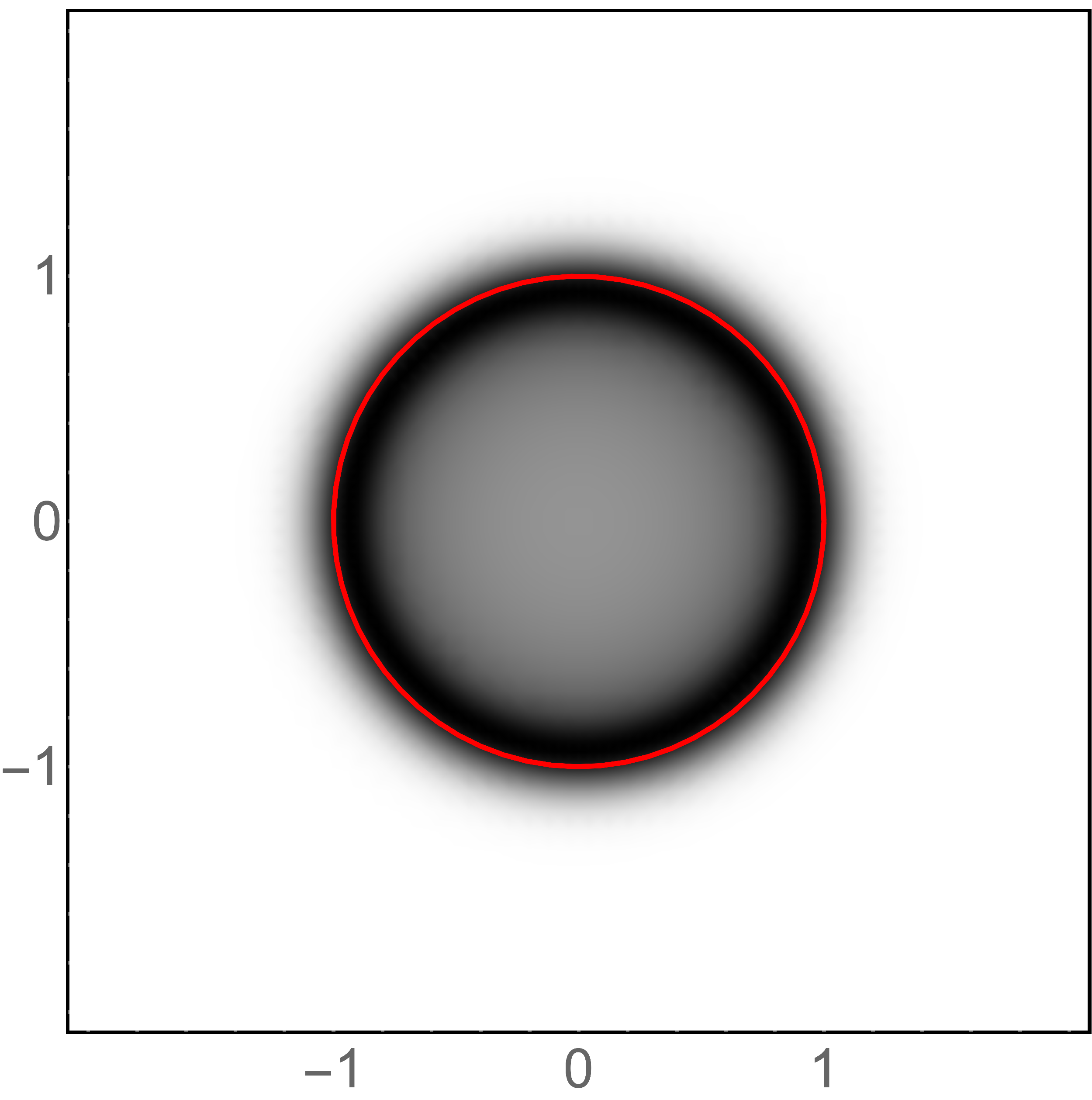} \hskip .1in
\includegraphics[scale=.3]{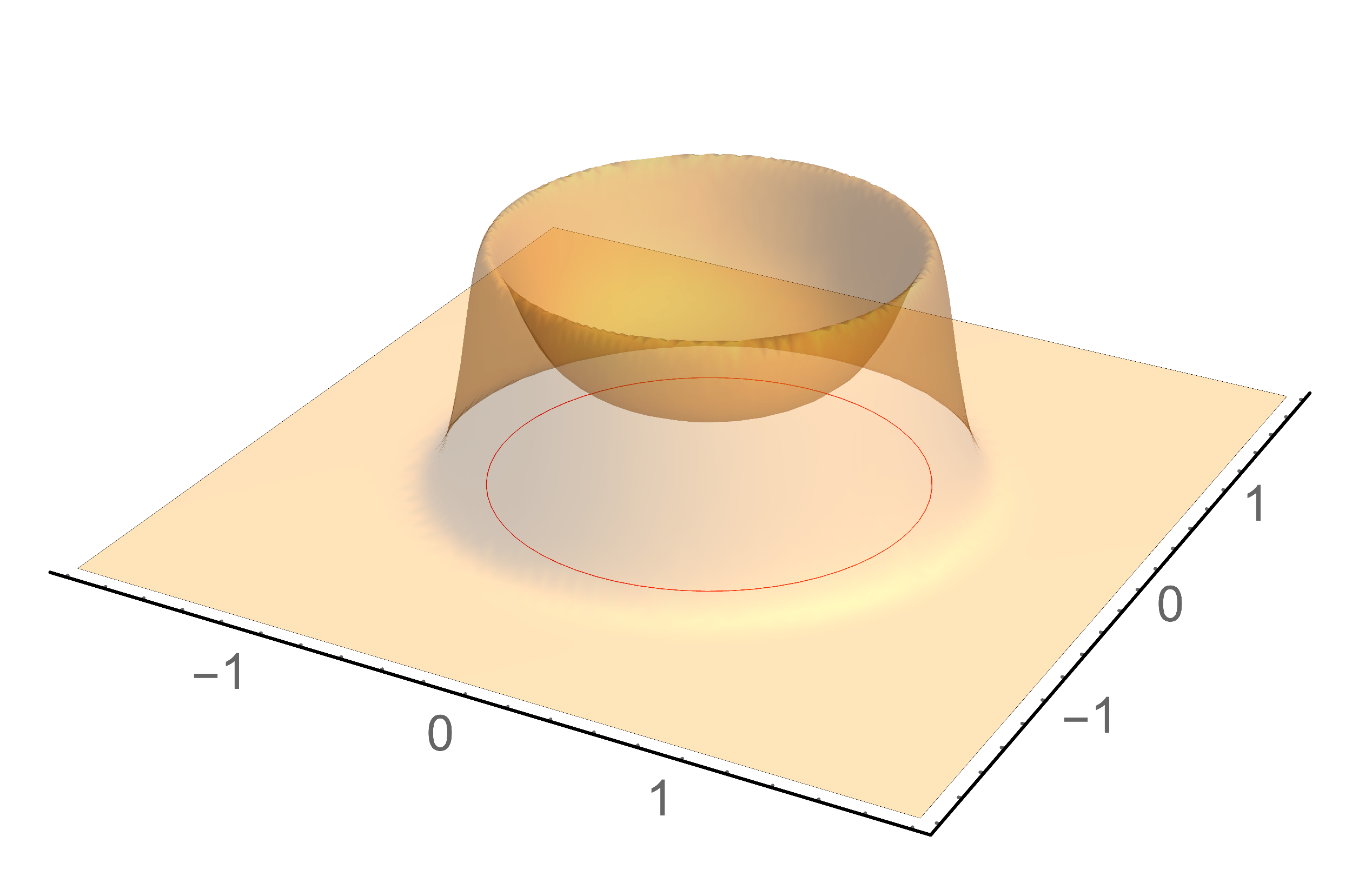}
\end{center}
\caption{Distributions about semi-algebraic sets can be constructed as marginal
	distributions of variety distributions. (Left) A variety normal density
	on $\mc{S}^{2}$ along with a cutaway to visualize variability. (Middle)
	A density plot of the marginalied distribution over the semi-algebraic
	set of the open unit disc.  (Right) The graph of the joint PDF of the
	distribution of $(X,Y)$.}
\label{fig:semialg}
\end{figure}

\end{enumerate}

For the reasons described above, the second strategy generally seems preferable
to the first, and so we recommend it when presented with semi-algebraic sets.
Nevertheless it, too, is imperfect, and more work is needed to develop a more
satisfactory solution.  Experimentally, the strategy presents some boundary
effects where the distribution is slightly more concentrated near the boundary
of the projected region; however, these appear to be relatively small in
practice.  It should also be noted that alternative algebraic formulations can
be used to represent the semi-algebraic set of interest.  For example, instead
of moving from $h(x,y) > 0$ to $g(x,y,s) = h(x,y) - s^{2} = 0$ in the above, we
might have opted for $g(x,y,s) = s^{2} h(x,y) - 1 = 0$.  Unfortunately, the
representation matters, and a bad representation may not work at all from a
sampling perspective because the resulting varieties may be unbounded.

In general, if a semi-algebraic set is presented as above, it can be represented
as the projection of the variety defined by $g_{1}(\ve{x}), \ldots,
g_{m}(\ve{x}), h_{1}(\ve{x}) - s_{1}^{2}, \ldots, h_{l}(\ve{x}) - s_{l}^{2} \in
\R[\ve{x},\ve{s}]$ in $\R^{p+l}$ \citep[p.27]{bochnak1991real}. In fact, Motzkin
has shown that this can always be done by introducing only one slack variable
\citep{motzkin1970real, bochnak1991real}; however, as the current strategy is
simple and effective, we saw no need to pursue it further.
\cite{pecker1990elimination} has found a similar yet simpler strategy.

It should be noted that semi-algebraic sets represent a very diverse collection
of subsets of $\R^{p}$.  In particular, they can allow puncturing of sets and,
more generally, the removal from larger dimensional sets smaller dimensional
geometries.  These cannot be faithfully represented by the present construction
because variety distributions always thicken the variety into the fullness of
the ambient space.

\section{Sampling and the stochastic exploration of real varieties}\label{sec:sampling}
In this section we describe practical strategies to sample from the
distributions presented in the previous sections to generate points near the
variety as well as methods to move those points to the variety. To ease the
exposition and accentuate how nice the situation can be, we focus our
presentation mainly on the variety normal distribution and note relevant
differences for other induced variety distributions where key features stick
out.

\subsection{Sampling}\label{sec:hmc}

There are many strategies one can use to sample from a probability distribution
\citep{gentle2006random, robert2013monte}. For a given situation, the best
sampler depends heavily on features of the distribution of interest, the target
distribution. Perhaps most importantly from a sampling perspective, all the
distributions described in this work are continuous on $\R^{n}$ and only known
up to an unknown constant of proportionality; this limits the strategies that
can be used.  Fortunately, there is already a wealth of knowledge on sampling
from un-normalized distributions since these distributions form the foundation
of Bayesian statistics. As such, Bayesian statisticians and others have
developed very powerful general purpose tools to perform such sampling, both
theoretically and in the form of freely accessible computer implementations.
Before progressing to these more complex techniques, however, it is worth noting
a simple one that works in many low-dimensional settings: rejection sampling.

\subsubsection{Rejection sampling}\label{sec:rejection}

Rejection sampling relies on the following basic fact referred to by
\citet{robert2013monte} as the Fundamental Theorem of Simulation (FTS): Given a
(potentially un-normalized) PDF $\tilde{p}(\ve{x})$, sampling from
$\tilde{p}(\ve{x})$ is equivalent to sampling from the uniform distribution on
the region below its graph, $$\set{(\ve{x}, u) \in \R^{n+1}: 0 \leq u \leq
\tilde{p}(\ve{x})}.$$  Rejection sampling refers to any algorithm that obtains
such uniform draws by sampling uniformly from the region below a proposal
distribution $q(\ve{x})$ for which $M q(\ve{x}) \geq \tilde{p}(\ve{x})$ for some
$M \in \R$ and then checking if the draws are also under the graph of
$\tilde{p}(\ve{x})$. Specifically: 1) sample $\ve{x}' \sim q(\ve{x})$, 2) sample
$u \sim \Unif{0}{M q(\ve{x}')}$, and 3) if $u \leq \tilde{p}(\ve{x}')$, $\ve{x}'
\sim p(\ve{x})$ otherwise repeat the procedure. The resulting retained/accepted
draws consitute an independent and identically distributed sample from
$p(\ve{x})$, the normalized distribution corresponding to $\tilde{p}(\ve{x})$.
The efficiency of the sampler depends on how similar $q(\ve{x})$ is to
$p(\ve{x})$ and how small $M$ is.

By construction, the un-normalized (multi)variety normal distributions' PDFs
\eqref{eq:vn} and \eqref{eq:mvn} are bounded above by 1, which is tight on the
variety.  If a spatial extent (range of the $\ve{x}$ variables) is known a
priori, one can therefore simply apply a rejection sampling algorithm over that
extent using the uniform distribution. 

\begin{figure}
\begin{center}
    \includegraphics[scale=.27]{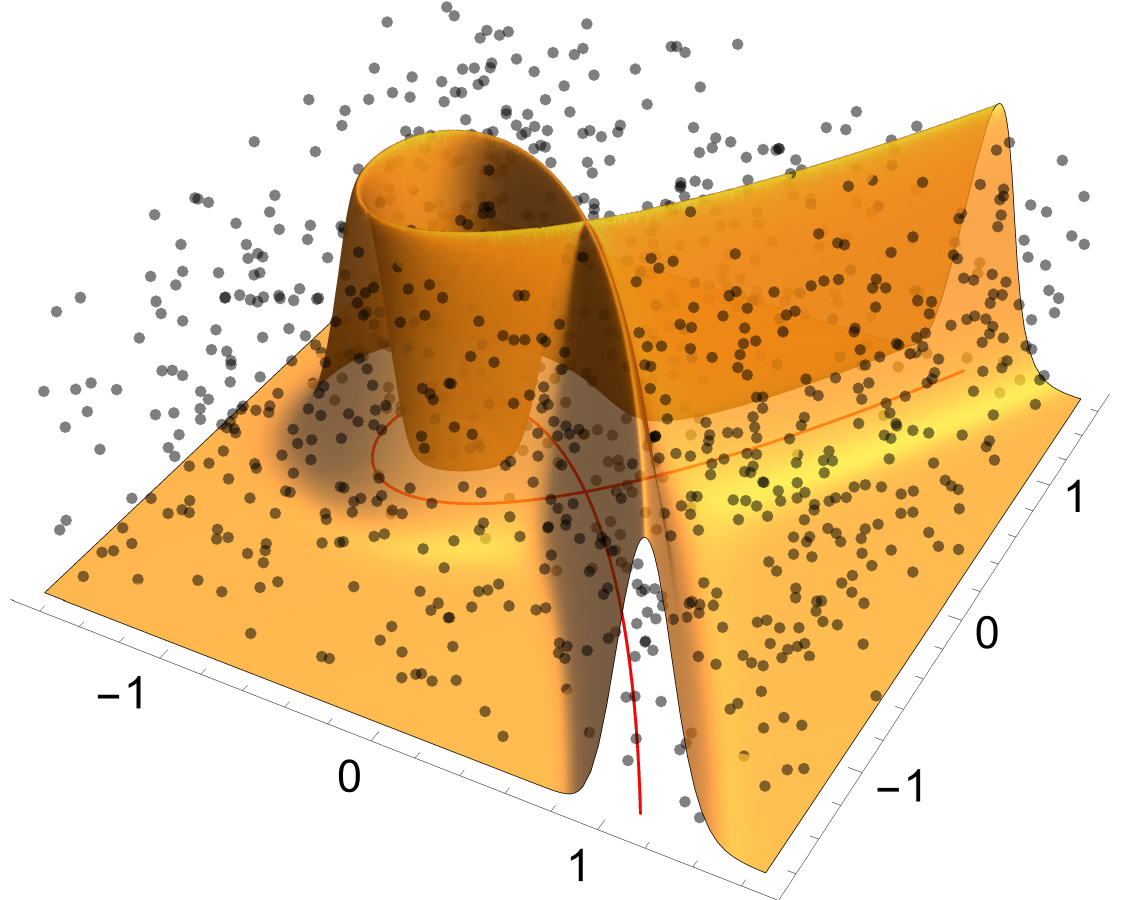}
    \includegraphics[scale=.24]{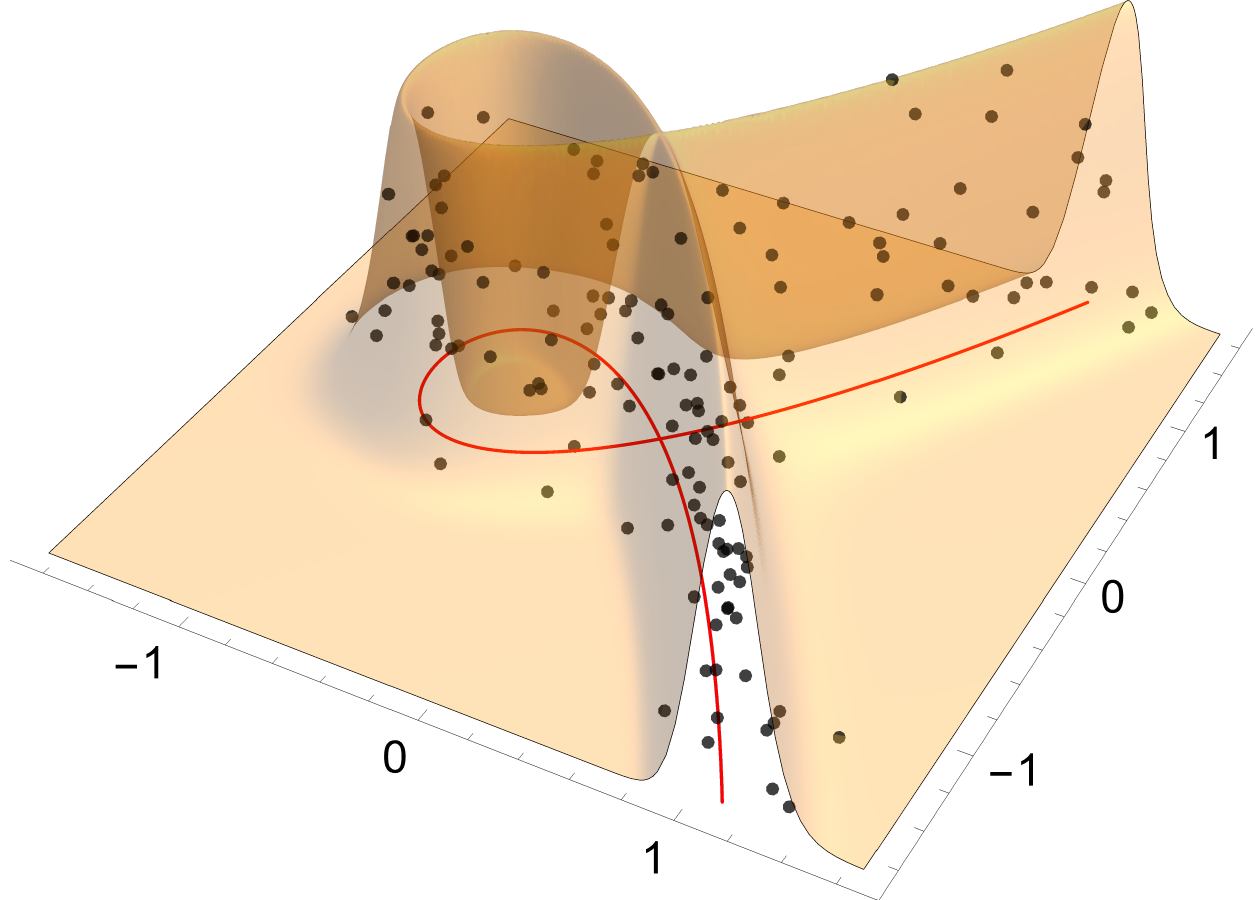}
    \includegraphics[scale=.26]{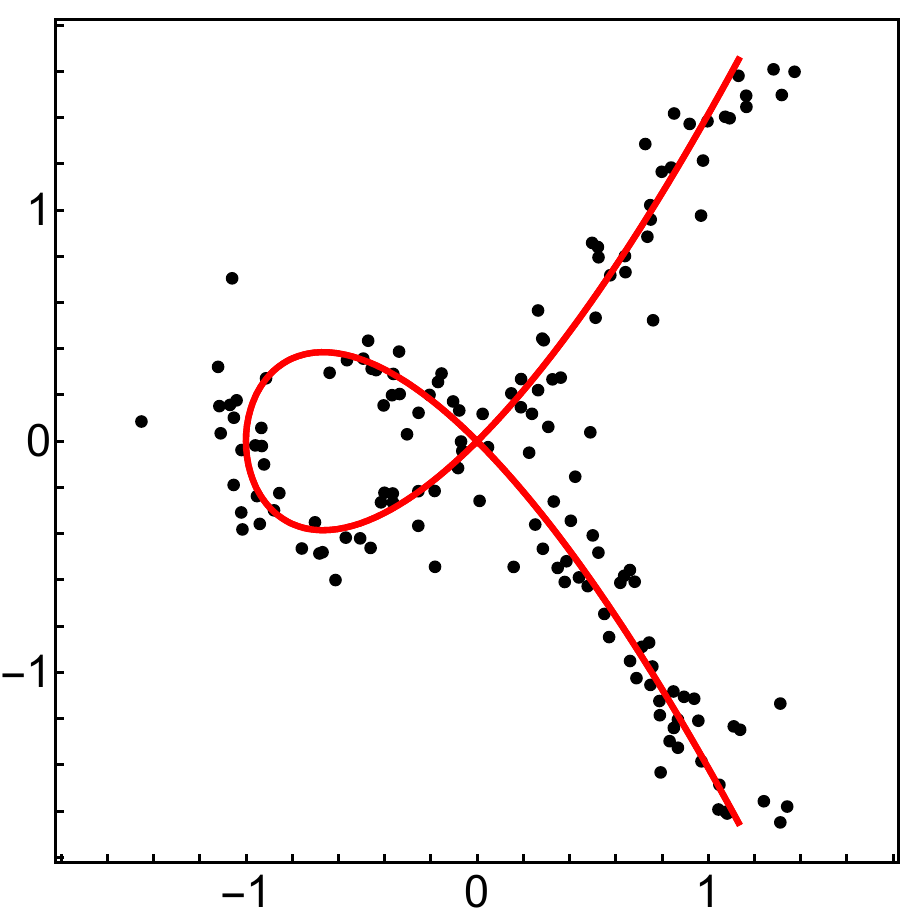}
\end{center}
\caption{Rejection sampling is a simple algorithm that can be used to sample
	from low-dimensional variety distributions. Here we sample from the
	$\MVpNorm{2,\bm{\mc{X}}}{y^{2} - (x^{3}-x^{2})}{\f{1}{10^{2}}}$
	distribution using a uniform proposal distribution over $\bm{\mc{X}} =
	[-1.5,1.5]^{2}$. We begin with 10,000 points (left), of which only 1,948
	were found to be under $\tilde{p}(x,y|g,\sisq)$, or about 1 in 5 draws
	(middle). The projected values, the $(x,y)$ coordinates of the draws,
	are then retained (right). Probabilistically this corresponds to
	marginizaling the height variable $u$.}
\label{fig:rejection}
\end{figure}

As an example, consider sampling from the variety normal on the alpha curve,
$\MVpNorm{2,\bm{\mc{X}}}{y^{2} - (x^{3}-x^{2})}{\f{1}{10^{2}}}$ with
$\bm{\mc{X}} = [-1.5,1.5]^{2}$, whose normalized polynomial was seen in the
previous section to be $\bar{g} = (y^{2}-(x^{3}+x^{2})) / \sqrt{(3x^{2}+2x)^{2}
+ (2y)^{2}}$. From \eqref{eq:vn}, its un-normalized PDF is $\tilde{p}(\ve{x}'|g,
\sisq) = \exp\left\{-50\bar{g}^{2}\right\}.$ Thus, rejection sampling from the
distribution would take two draws $x, y \sim \Unif{-1.5}{1.5}$ and one $u \sim
\Unif{0}{1}$ and check if $u \leq \tilde{p}(\ve{x}'|g, \sisq)$. If so, we retain
$(x, y)$ as one draw from $p(x,y|g, \sisq)$, otherwise we repeat the procedure.
This is illustrated in Figure~\ref{fig:rejection}.

\begin{figure}
\begin{center}
    \includegraphics[scale=.35]{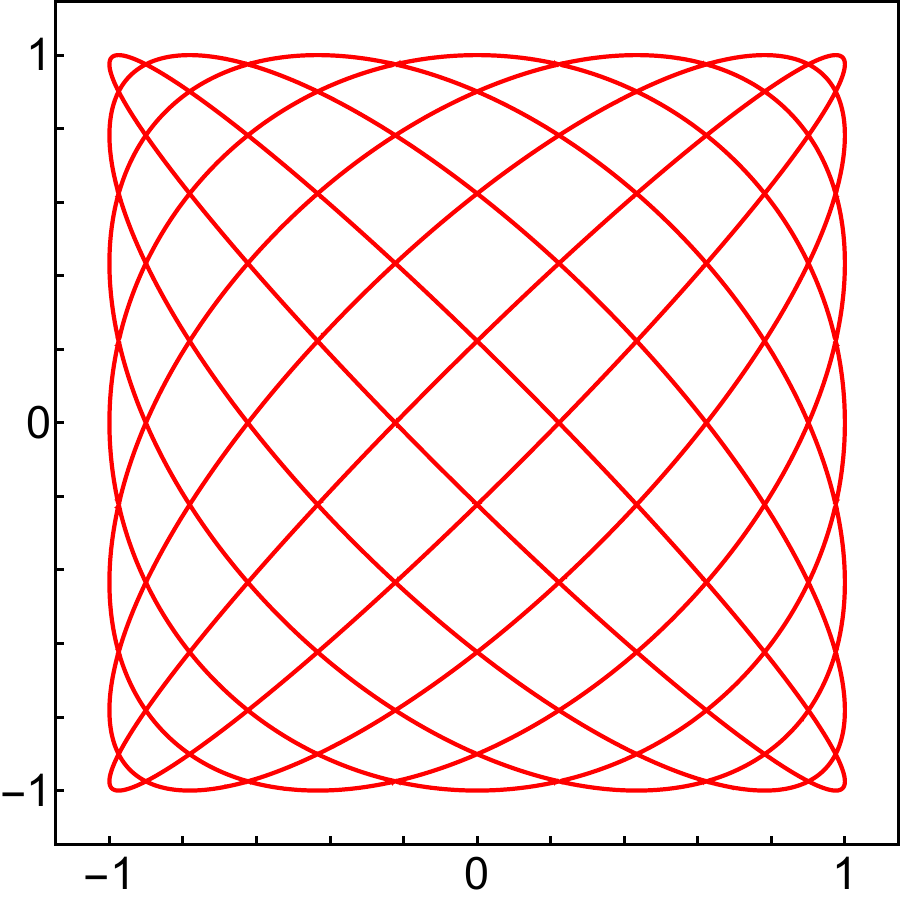}
    \includegraphics[scale=.35]{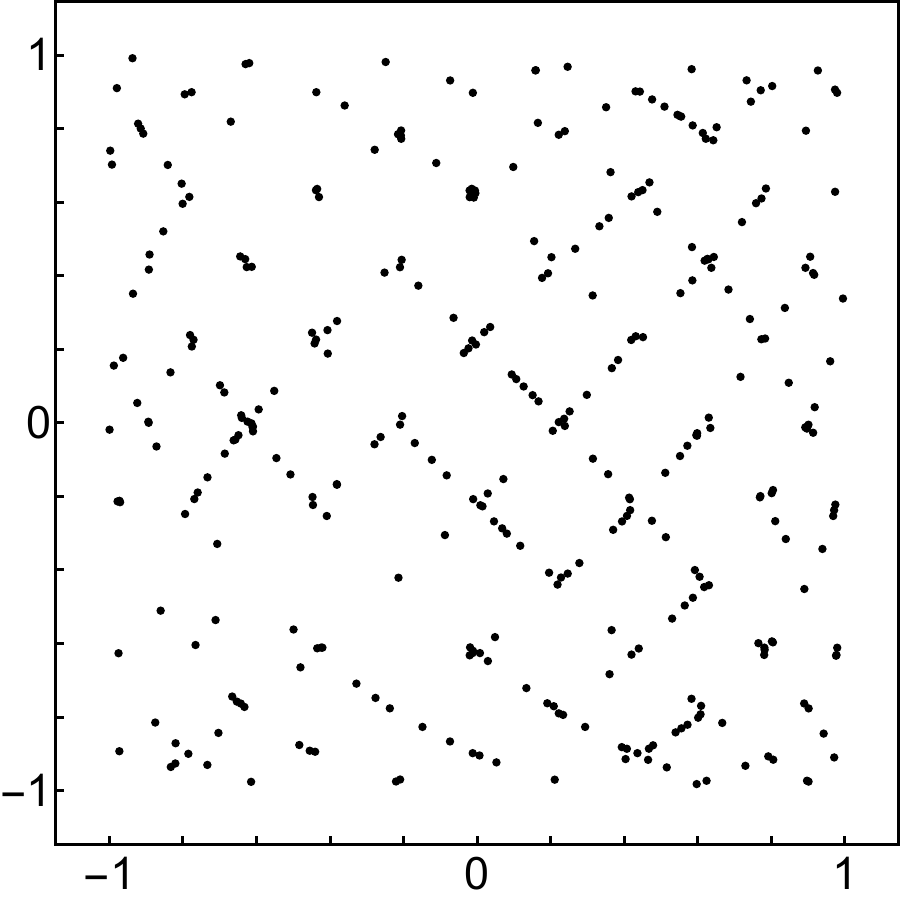}
    \includegraphics[scale=.35]{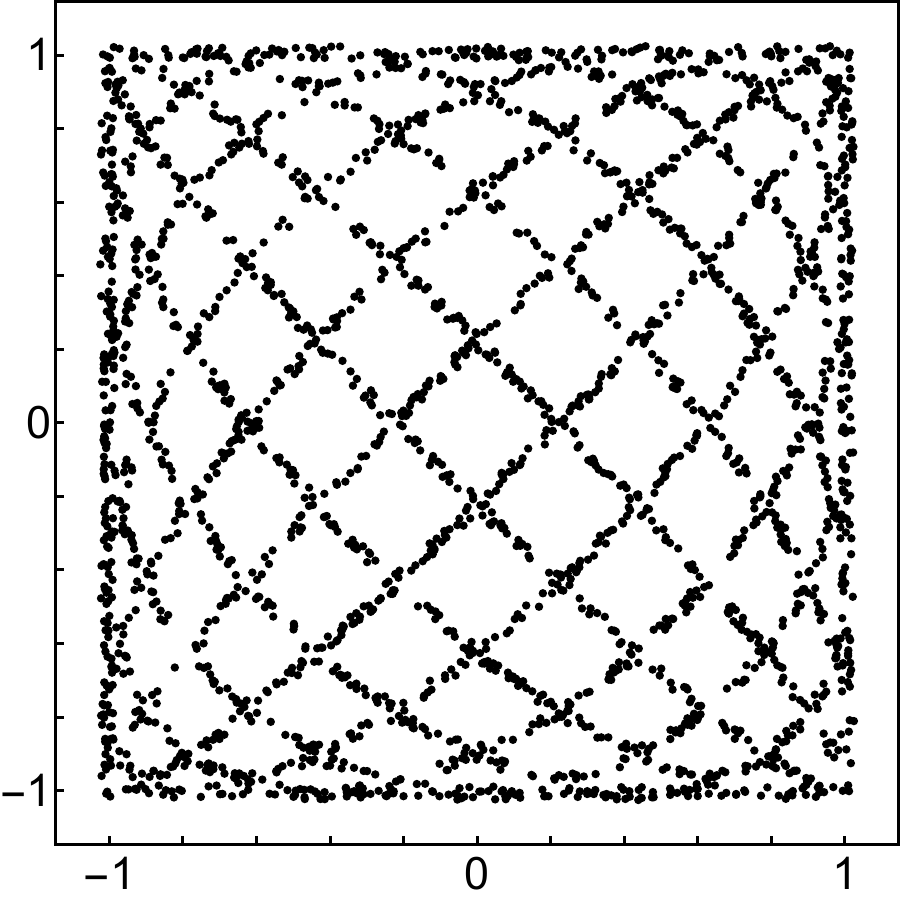}
\end{center}
\caption{The rejection sampling algorithm on the induced variety distribution
	$\Unif{-\ep}{\ep}$ on the degree-18 Lissajous polynomial's variety with
	$\ep = .15$. (Left) the variety, (middle) the induced distribution
	plugging in $g$, corresponding to randomly plugging in points and
	checking where the polynomial is $|g| \leq \ep$, and (right) the same
	inserting $\bar{g}$ instead of $g$. The acceptance rate changes from
	3.4\% to 28.3\% by using $\bar{g}$. $N = 10,000$.}
\label{fig:rejection-lissajous}
\end{figure}

Using this simple rejection scheme on the induced variety distribution formed by
plugging $\bar{g}$ into the PDF of a $\Unif{-\ep}{\ep}$ PDF for small $\ep$
provides an interesting, intuitive procedure. The PDF of the uniform
distribution is always proportional to a constant over its support and zero
elsewhere, so $\tilde{p}(x, y|g,\ep) = 1[-\ep \leq \bar{g} \leq \ep]$. If we use
the same proposal $q(\ve{x})$ that is uniform over the square $\bm{\mc{X}}$,
notice that we can skip the uniform sampling of the auxiliary variable $u$
entirely, since $\tilde{p}$ is either 0 or 1. In this case, the algorithm
reduces to this: 1) sample $x, y \sim \Unif{-1.5}{1.5}$; 2) evaluate $\bar{g}(x,
y)$; and 3) retain $(x,y)$ if $|\bar{g}(x, y)| \leq \ep$. This is very similar
to the naive technician's procedure that, when looking for where $g$ is 0,
simply evaluates $g$ at a collection of random points and retains the ones that
are close to zero, but with one distinction: the present procedure uses the
normalized version $\bar{g}$, not $g$, and that constitutes a significant
distinction. Figure~\ref{fig:rejection-lissajous} illustrates this distinction
on the Lissajous polynomials \citep{merino2003lissajous}.

Figures~\ref{fig:rejection} and \ref{fig:rejection-lissajous} are very suggest
that the method works very well, and it does, at least in low dimensions. It has
several advantages: it produces independent draws, it is almost trivial to
implement, and it is easy to monitor. But it suffers from a number of
disadvantages as well. First, one needs to know a priori some sense of what the
spatial extent is of the variety. It may be possible to use clever
transformations to alleviate this problem, but it still presents a challenge
that would likely need to be addressed on an individual basis. Second, and more
serious, is scaling. For varieties in higher dimensional spaces, the variety
distribution is likely to focus its mass near a very small region of the ambient
space. This region becomes vanishingly small as the dimension of the problem
increases. Naive uniform box-style samplers will be too inefficient to be of
practical value. Perhaps more efficient rejection sampling-style schemes can be
developed here, with better and perhaps even adaptive proposals, but we do not
consider them further. This exact same scenario is encountered routinely in
Bayesian statistics, so it is to their solutions we now turn.

\subsubsection{MCMC sampling}\label{sec:mcmc}

The major engine behind Bayesian sampling is Markov chain Monte Carlo (MCMC).
MCMC is a class of algorithms that enable one to sample from an un-normalized
distribution by constructing a Markov chain whose stationary distribution is the
target distribution. To generate observations from this distribution, the chain
is initialized at an arbitrary location in the support of the distribution and
proceeds by iterating through proposal and acceptance steps.  In the proposal
step, another point in the support of the distribution is generated using a
transition kernel, and in the acceptance step that proposed value is either
accepted or rejected with a probability designed in such a way that the
stationary distribution of the chain is the distribution of interest. In a
typical setting, this probability (the Metropolis probability) is the ratio of
the relative likelihoods of the proposed and current states. Consequently, the
normalization factors cancel and never need to be computed. MCMC is thus ideal
for situations where the distribution is only known up to a constant of
proportionality, like all the distributions described in
Sections~\ref{sec:algfams}~and~\ref{sec:induced}. Another way to think of MCMC
is as a stochastic mode-seeking algorithm, and this perspective in particular
may help to understand the use of Hamiltonian dynamics below.

\begin{figure}
\begin{center}
    \includegraphics[scale=.25]{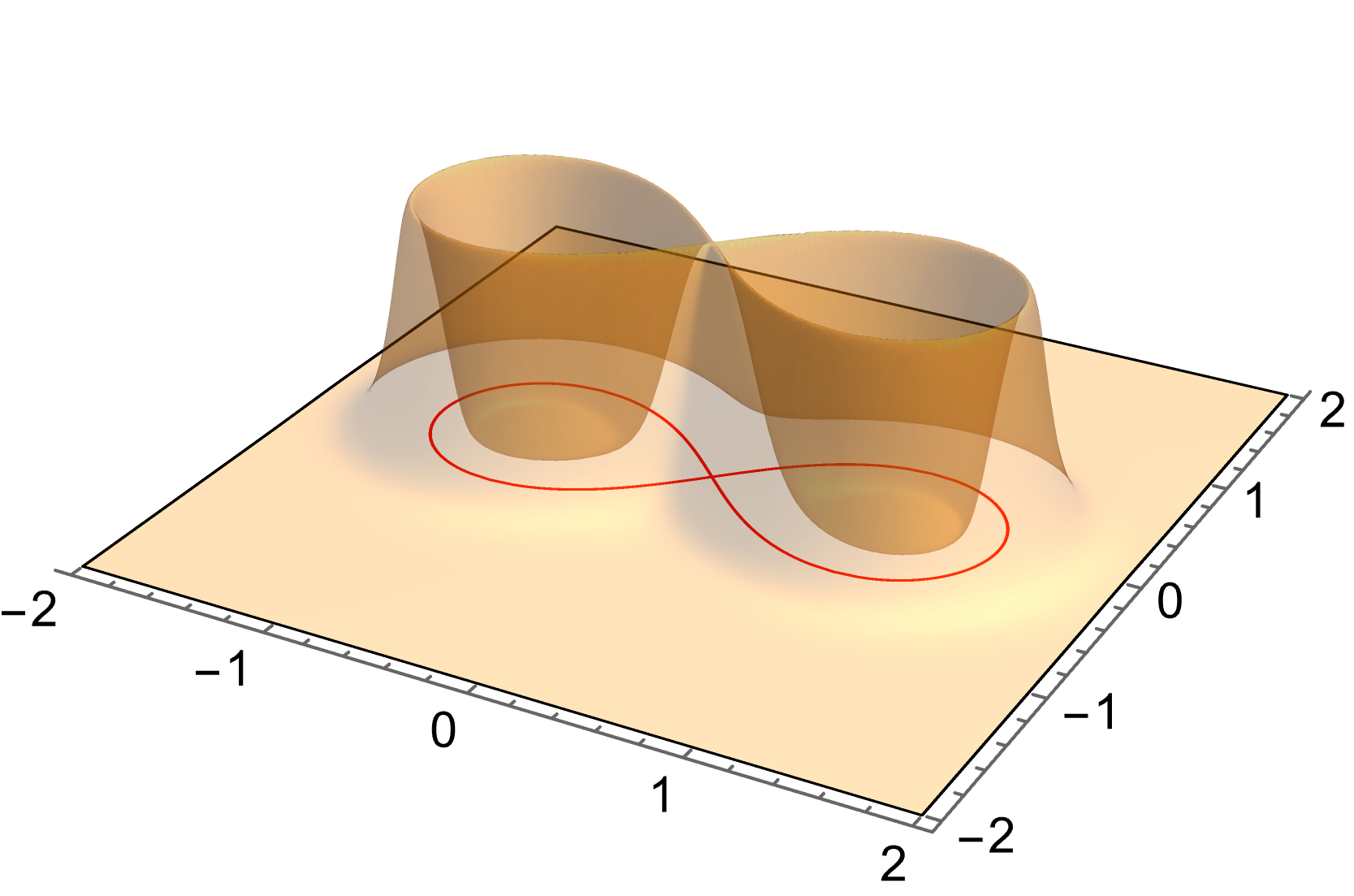}
    \includegraphics[scale=.25]{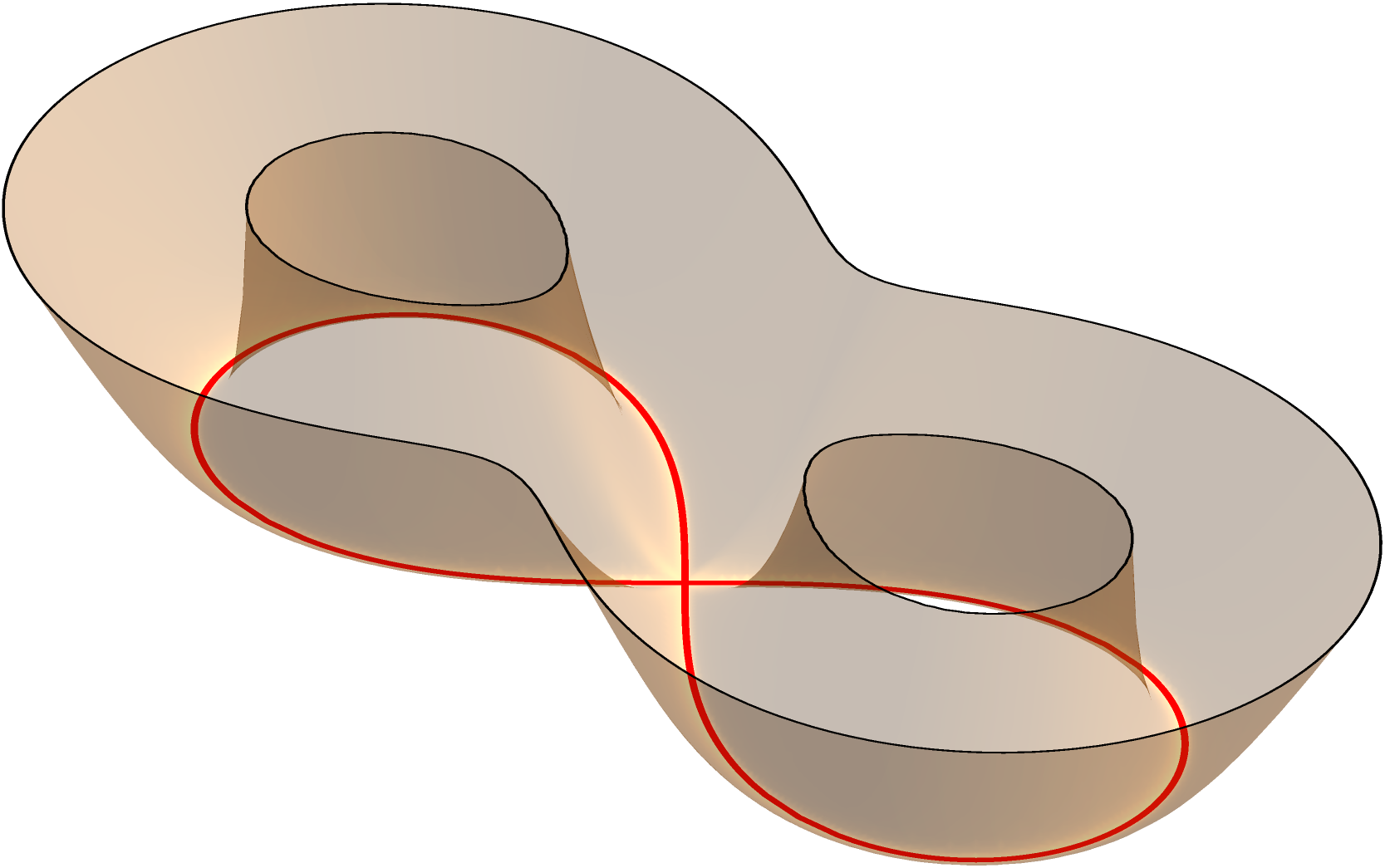}
    \includegraphics[scale=.25]{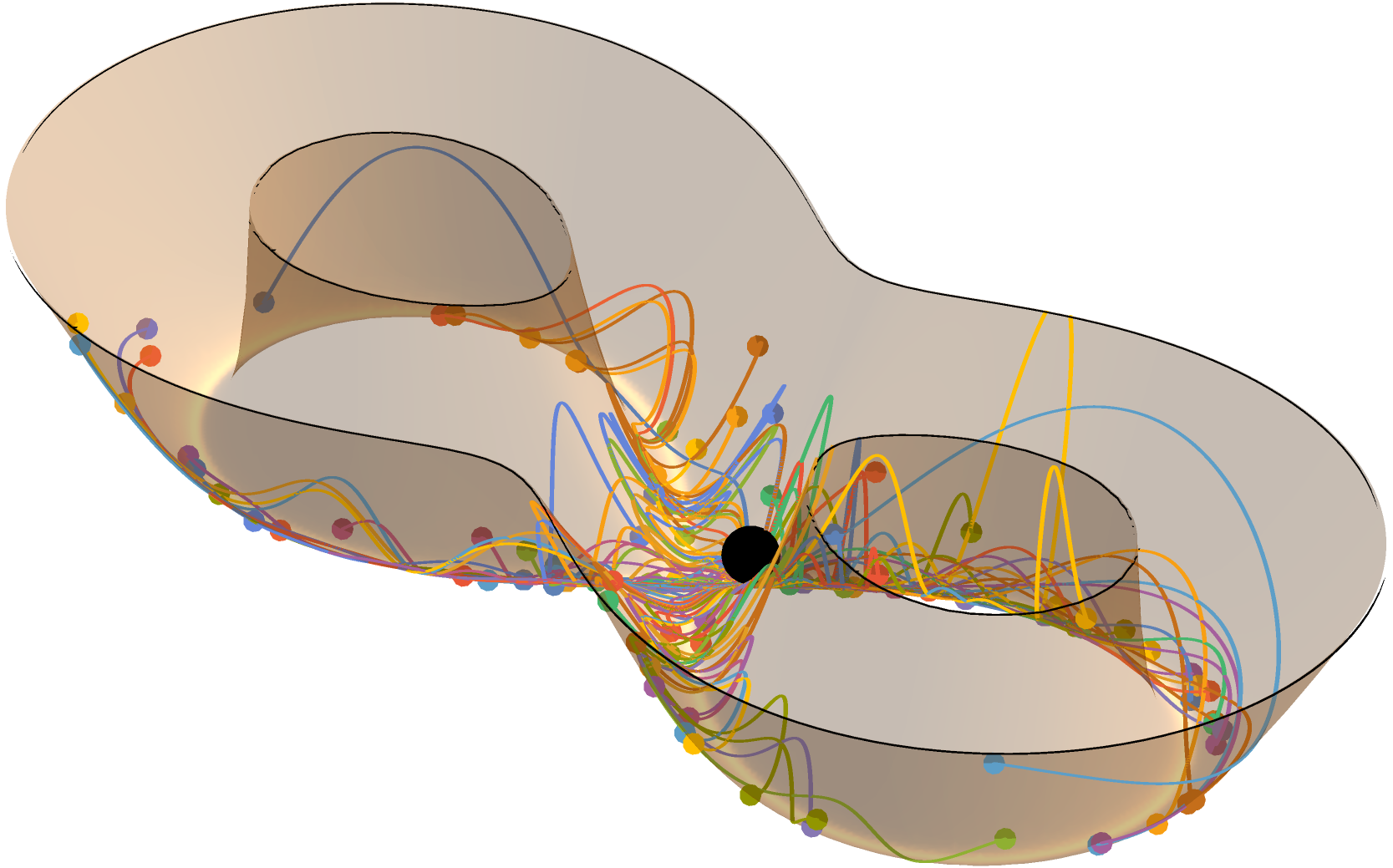} \\
    \vskip .1in
    \includegraphics[scale=.25]{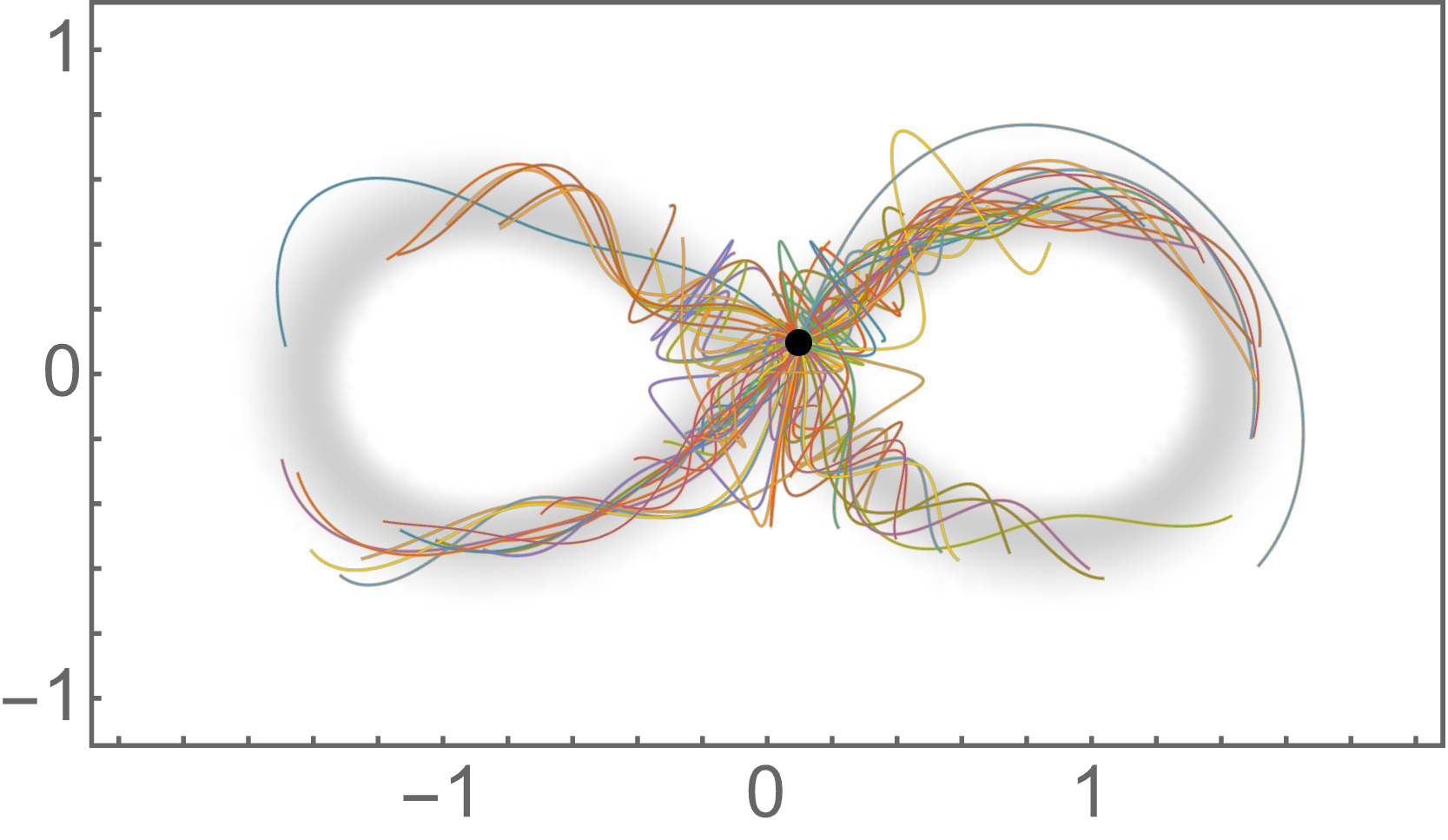}
    \includegraphics[scale=.25]{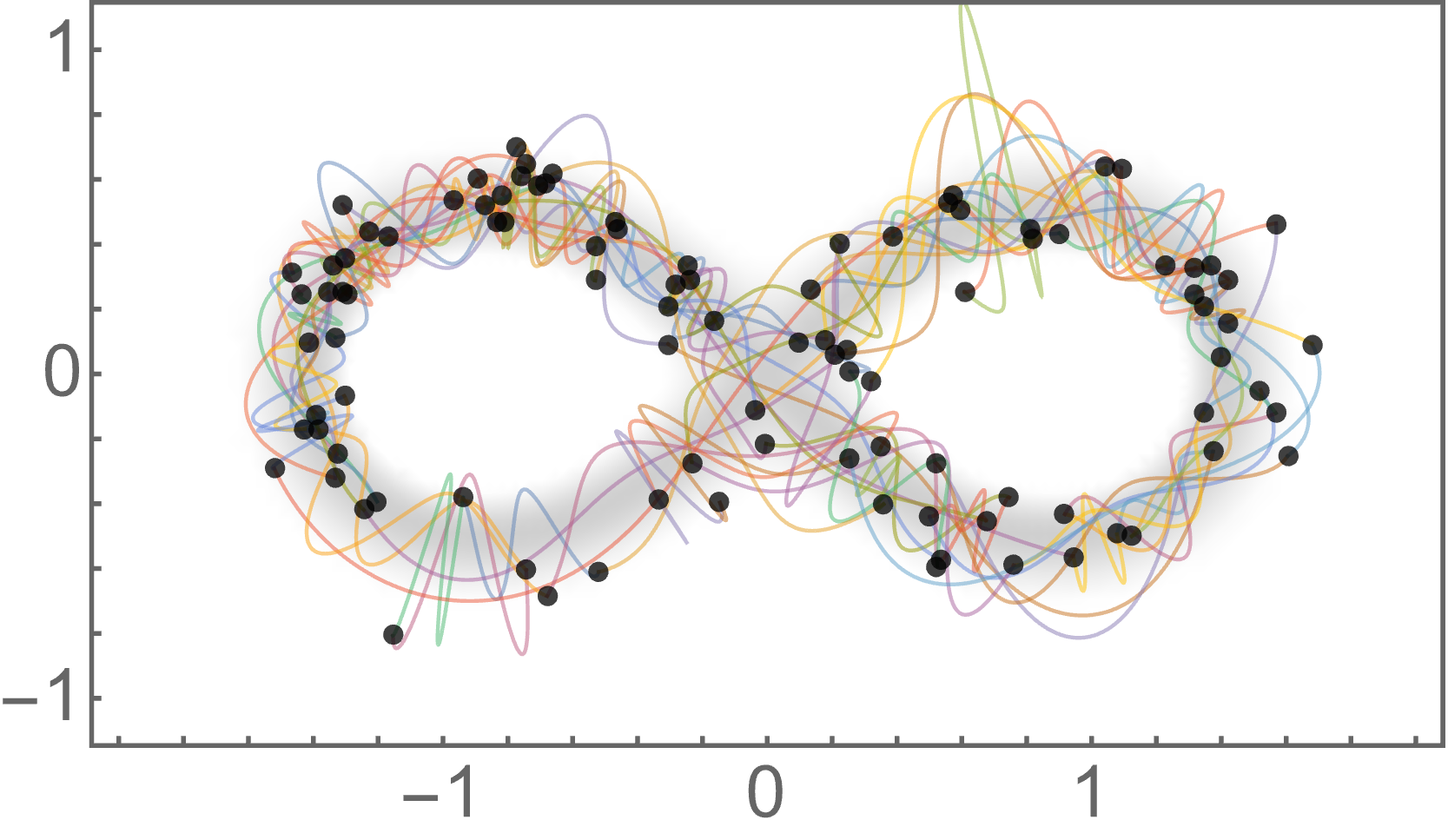}
    \includegraphics[scale=.25]{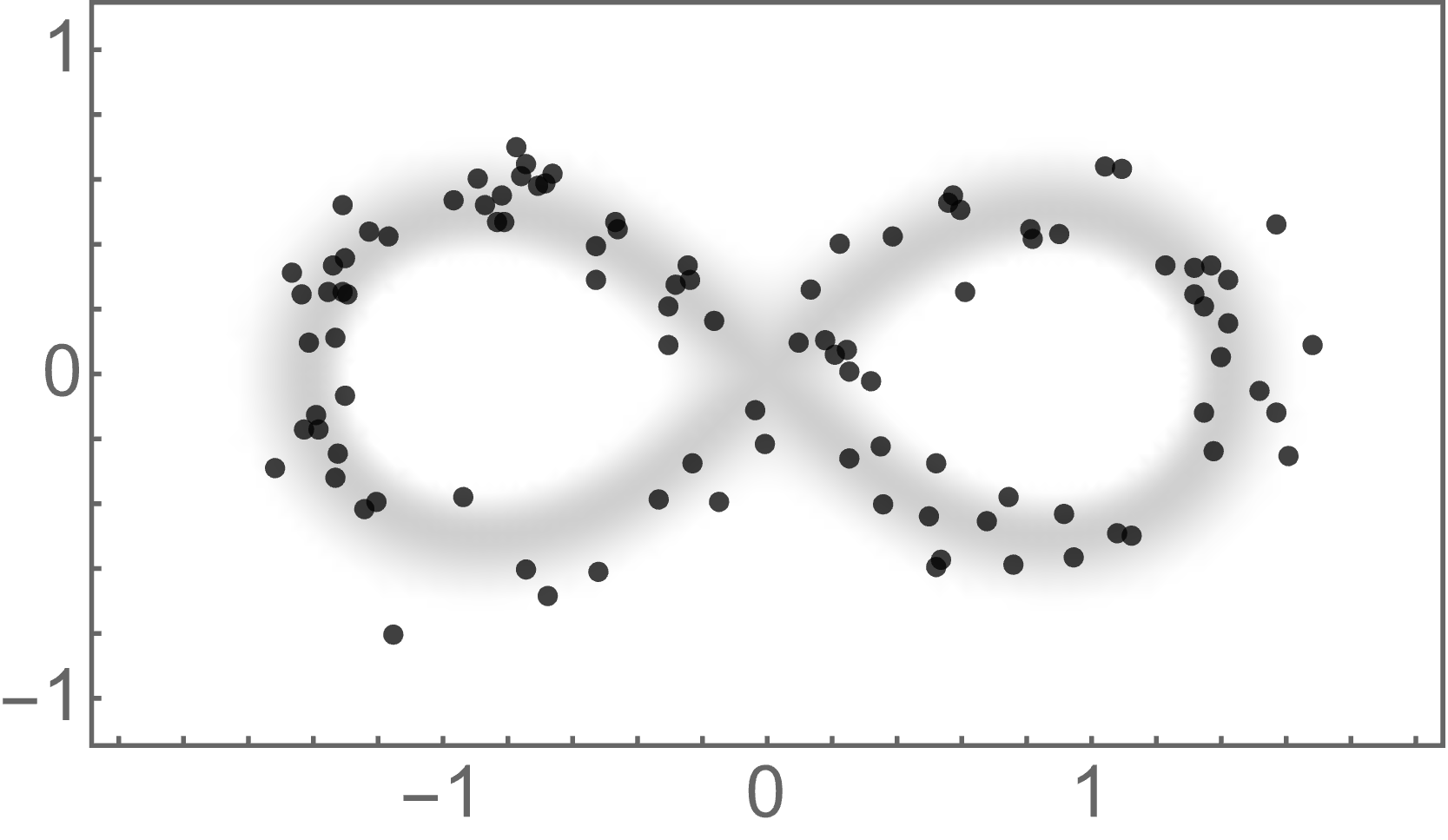}
\end{center}
\caption{HMC samples distributions by simulating Hamiltonian dynamics on $-\log
	\tilde{p}(\ve{x})$, which allows paths to efficiently explore the
	distribution. This display illustrates the process for the variety
	normal on the polynomial $g(x,y) = (x^2 + y^2)^2 - 2 (x^2 - y^2)$, the
	lemniscate of Bernoulli, $\si = \f{1}{10}$. Top: $\tilde{p}(\ve{x})$,
	$-\log \tilde{p}(\ve{x})$, and 100 different simulated paths originating
	from $(.1,.1)$. Bottom: The same 100 paths in the $(x,y)$ plane, and a
	real HMC simulation using head-to-tail simulations with and without the
	transition paths.}
\label{fig:hmc}
\end{figure}

There are many schemes commonly used for MCMC. In recent years Hamiltonian Monte
Carlo (HMC) has proved to be a robust general purpose strategy to sample from
continuous probability distributions where other MCMC algorithms, e.g.  Gibbs
sampling, fail \citep{duane1987hybrid, nealhandbook}. HMC simulates Hamiltonian
dynamics to generate proposal draws in high probability regions.  The simulation
is equivalent to monitoring the position of a puck imparted with a randomly
generated momentum and sliding on a frictionless surface whose height is based
on the negative log of the target density.  Regions of high probability density
are low regions on the surface, and so the puck tends to stay in them even as it
slides around the irregularly shaped surface. After some time, the puck is
stopped, and its location is used as the proposed value. Then the standard
Metropolis probability is computed and the draw is accepted or rejected in the
phase space and projected back down into the original space, much like rejection
sampling. Under general conditions, as long as the Hamiltonian dynamics are
simulated carefully this process leads to a Markov chain whose stationary
distribution is indeed the distribution of interest.  \cite{nealhandbook} is an
authoritative, accessible, and freely available introduction. The process is
illustrated in Figure~\ref{fig:hmc}.

A mathematical description of how the basic HMC algorithm works and how it
applies to the variety normal distribution is helpful to understand how nicely
suited it is to that task. In general, when performing HMC one constructs a
Hamiltonian $H(\ve{x},\ve{p}) : \R^{n} \times \R^{n} \to \R$ that satisfies
Hamilton's equations
$$
\f{dx_{i}}{dt} = \f{\pa}{\pa p_{i}}H(\ve{x},\ve{p}) \quad \mbox{and} \quad 
\f{dp_{i}}{dt} = -\f{\pa}{\pa x_{i}}H(\ve{x},\ve{p}).
$$
Here $\ve{x}$ represents the candidate draw from the distribution and $\ve{p}$ a
collection of auxiliary variables analogous to the single auxiliary variable $u$
in rejection sampling.  One typically assumes $H(\ve{x},\ve{p}) = U(\ve{x}) +
K(\ve{p})$. In the physical analogy, $\ve{x}$ is the position of the puck,
$\ve{p}$ is its momentum, $U(\ve{x})$ is its potential energy, and $K(\ve{p})$
is its kinetic energy; all of these vary in time.  The potential energy is
related to the (potentially un-normalized) target PDF via $U(\ve{x}) = -\log
\tilde{p}(\ve{x})$, and the momentum distribution is typically set to a
multivariate normal so that $K(\ve{p}) = \ve{p}'\bm{M}^{-1}\ve{p}/2$, where
$\bm{M}$, the ``mass matrix,'' is a tuning parameter that gauges the variability
of the imparted momentum, its magnitude and direction. It is usually estimated
in an introductory warmup stage of running the chains, often as a scaled
identity matrix, in an effort to optimize the efficiency of the sampling.  In
the case of the variety normal with $n \geq m$ (at least as many variables as
equations), $\ma{J}_{\ve{x}}^{+} =
\ma{J}_{\ve{x}}'(\ma{J}_{\ve{x}}\ma{J}_{\ve{x}}')^{-1}$, and the random vector
$\ve{X} \sim \MVpNorm{n}{\ve{g}}{\bm{\Si}}$ has PDF $p(\ve{x}|\ve{g},\bm{\Si})
\propto
\exp\set{-\f{1}{2}\bar{\ve{g}}(\ve{x}|\ma{B})'\bm{\Si}^{-1}\bar{\ve{g}}(\ve{x}|\ma{B})}$,
so
\begin{equation*}
U(\ve{x}) 
\ = \ \f{1}{2}\bar{\ve{g}}(\ve{x}|\ma{B})'\bm{\Si}^{-1}\bar{\ve{g}}(\ve{x}|\ma{B}) 
\ = \ \f{1}{2}\ve{g}(\ve{x}|\ma{B})'\Big((\ma{J}_{\ve{x}}^{+})'\bm{\Si}^{-1}\ma{J}_{\ve{x}}^{+}\Big)\ve{g}(\ve{x}|\ma{B})\big).
\end{equation*}
If $\bm{\Si} = \sisq \ma{I}_{n}$ as would typically be the case when exploring
varieties, from (\ref{eq:varty-normal-simple-not-over}) this further reduces to
\begin{equation}
U(\ve{x}) 
\ = \ \f{1}{2\sisq}\ve{g}(\ve{x}|\ma{B})'(\ma{J}_{\ve{x}}\ma{J}_{\ve{x}}')^{-1}\ve{g}(\ve{x}|\ma{B}).
\end{equation}
If $\ma{M} = \sisq_{m}\ma{I}_{n}$, Hamilton's equations reduce to
\begin{eqnarray}
\f{dx_{i}}{dt} 
&=& 
\f{\pa}{\pa p_{i}} K(\ve{p})
\ \ = \ \ \f{\pa}{\pa p_{i}} \ve{p}'\bm{M}^{-1}\ve{p}/2
\ \ = \ \ \f{p_{i}}{\sisq_{m}} \\
\f{dp_{i}}{dt} 
&=& 
-\f{\pa}{\pa x_{i}} U(\ve{x})
\ \ = \ \ -\f{\pa}{\pa x_{i}} (-\log \tilde{p}(\ve{x}))
\ \ = \ \ -\f{1}{2\sisq}\f{\pa}{\pa x_{i}} \set{\ve{g}(\ve{x}|\ma{B})'(\ma{J}_{\ve{x}}\ma{J}_{\ve{x}}')^{-1}\ve{g}(\ve{x}|\ma{B})}.
\end{eqnarray}
Of course, since $\ve{g} \in \R[\ve{x}]^{m}$ is polynomial, the computation of
$\ma{J}_{\ve{x}}$ is trivial. While in practical implementations of HMC such as
\pl{Stan} automatic differentiation is used because Jacobians virtually never
present in closed form, for the variety normal distribution this is the typical
case. 

As a general purpose algorithm, HMC has a number of parameters that require
tuning for the sampler to work. For example, considerable effort has been spent
answering questions that speak to how long the Hamiltonian dynamics should be
simulated and how time should be discretized. Since this work is not focused on
novel or optimially efficient samplers, we simply recommend the defaults of the
implementation used. In our case, this is the Bayesian software \pl{Stan}, which
implements the No-U-Turn sampler (NUTS) algorithm, a HMC variant that seeks to
tune the parameters governing the Hamiltonian dynamics by leveraging the
time-reversability of the process and avoiding random walk behavior
\citep{hoffman2014no}. 

While implementing HMC from scratch may seem daunting, a simple trick allows us
to sidestep the process and use state of the art implementations from the
Bayesian community. We illustrate this process with the variety normal
distribution, but similar manipulations enable its use for others we well. The
trick is to recognize the PDF of the variety normal distribution as the
posterior distribution of an overparameterized Bayesian model.  Recall that
Bayes' theorem provides that in a statistical modeling scenario where the data
model is $p(\ve{y}|\ve{\th})$ and the prior is $p(\ve{\th})$, the posterior
distribution is proportional to the product of the likelihood and the prior,
notationally $p(\ve{\th}|\ve{y}) \propto p(\ve{y}|\ve{\th})p(\ve{\th})$.
Assuming an improper flat prior $p(\ve{\th}) = 1$, this is $p(\ve{\th}|\ve{y})
\propto p(\ve{y}|\ve{\th})$, which is well known to produce proper posteriors
under suitable conditions, notably in the univariate normal case with $\sisq$
known \citep{berger1993statistical}. The key is to recognize the difference in
notational roles. In the case of the algebraic distributions of interest, the
coefficients $\ve{\be}$ are known and the values of $\ve{x}$ are desired, which
is precisely the opposite of the typical Bayesian scenario. In particular, the
role of the data $\ve{y}$ is played by the coefficient vector $\ve{\be}$, which
is known, and that of the parameter $\ve{\th}$ is played by the indeterminate
vector $\ve{x}$.  Moreover, observing that
$\bar{\ve{g}}(\ve{x}|\ma{B})'\bm{\Si}^{-1}\bar{\ve{g}}(\ve{x}|\ma{B}) =
\big(\ve{0}_{n}-\bar{\ve{g}}(\ve{x}|\ma{B})\big)'\bm{\Si}^{-1}\big(\ve{0}_{n}-\bar{\ve{g}}(\ve{x}|\ma{B})\big)$,
the variety normal PDF can be thought of as the posterior distribution over the
``parameter'' space $\R^{n}$ of a data model that is multivariate normal with
mean $\bar{\ve{g}}(\ve{x}|\bm{\be})$ and known covariance $\bm{\Si}$, a flat
prior on the variates $\ve{x}$, and a single observation of $\ve{0}_{n}$.

\pl{Stan} is a free and open-source state of the art probabilistic programming
language (PPL) that is both a specification language and a sampling engine that
implements HMC \citep{stan, rstan}. From a theoretical vantage, its engine
implements both NUTS and other HMC variants, including a robust automatic
differentiation suite.  From an applied perspective, it is cross-platform and
has interfaces in most major computing environments, including \pl{R},
\pl{Python}, \pl{Julia}, \pl{Matlab}, and others, in addition to a Linux command
line version. \pl{Stan} largely functions as a markup language for Bayesian
models that is translated into \pl{C++} and compiled before sampling. For
example, an implementation used to sample from the multivariety normal
distribution defined by the equations $x^{2} + y^{2} + z^{2} = 1$ and $z = x +
y$, for example, might be:

\begin{verbatim}
data { real<lower=0> si; }
      
parameters { real x; real y; real z; } 
        
transformed parameters {
  vector[2] g = [x^2+y^2+z^2-1, -x-y+z]';
  matrix[2,3] J = [
      [2*x, 2*y, 2*z], 
      [-1, -1, 1]
    ];
} 
        
model { target += normal_lpdf(0.00 | J' * ((J*J') \ g), si); }
\end{verbatim}

Since the $\bm{\be}$ (or $\ma{B}$) coefficients that define the polynomial(s)
function as data from the Bayesian perspective, in addition to creating
individual samplers for individual variety normal distributions, \pl{Stan} can
be used to create a single compiled binary that samples from a suitably small
class of variety normal distributions. For example, here is a \pl{Stan} program
to sample from the variety normal distribution on a quadratic polynomial in the
indeterminates $x$ and $y$ that can be compiled once and then simply provided
different coefficients, which this framework views as data, at run-time. It
limits its scope to the window $[-5,5]^2$:

\begin{verbatim}
data {
  real<lower=0> si;
  real b0; real bx; real by;
  real bx2; real bxy; real by2;
}

parameters {
  real<lower=-5,upper=5> x;
  real<lower=-5,upper=5> y;
}

model {
  real g = b0 + bx*x + bx2*x^2 + bxy*x*y + by*y + by2*y^2;
  real dgx = bx + 2*bx2*x + bxy*y;
  real dgy = by + 2*by2*y + bxy*x;
  real ndg = sqrt(dgx^2 + dgy^2);
  target += normal_lpdf(0.00 | g/ndg, si);
}
\end{verbatim}

\noindent Of course, it bears repeating that not all combinations of parameters
yield varieties over the reals, so we may need to check if an apparent component
corresponds to actual roots.

\subsection{Endgames}\label{sec:endgames}

In some cases points on the variety may be preferred to points near it. In the
current situation, if $\si$ is sufficiently small, the resulting points from the
sampler will be close to the variety, so a sensible approach to generating
points on the variety is is to project the points onto it. Mathematically, if
$\ve{x}_{0}$ is a point near the variety generated by the sampler, this
corresponds to solving the $\ell_{2}$ optimization problem 
\begin{equation}
\ve{x}^{*} = \argmin_{\ve{x} \in \varty{V}(\ve{g})} \lpnorm{2}{\ve{x}-\ve{x}_{0}}
= \argmin_{\ve{x} \in \varty{V}(\ve{g})} \lpnorm{2}{\ve{x}-\ve{x}_{0}}^{2}.
\end{equation}
The solution to this problem is unique for almost all $\ve{x}_{0} \in \R^{n}$. 

One may consider two basic naive strategies to compute $\ve{x}^{*}$ for which
there are many variants, both theoretical and implementation-wise. First, one
may consider running any nonlinear optimization algorithm on $h(\ve{x}) =
\ve{g}(\ve{x})'\ve{g}(\ve{x}) = \sum g_{i}(\ve{x})^{2}$ initialized at
$\ve{x}_{0}$, with the intuition being that if $\ve{x}_{0}$ is already close to
being a root of $\ve{g}$, perhaps the algorithm will move $\ve{x}_{0}$ to
$\ve{x}^{*}$, or at least something close.  For example, if $\ve{x}_{0} =
(x_{0},y_{y}) = (1,1)$ is a point drawn near the variety of $g =  x^{2} + y^{2}
- 1$ (the unit circle in $\R^{2}$), $g(x_{0},y_{0}) \neq 0$, but initializing
gradient descent or Newton's method (say) on the polynomial $g^{2}$ will find a
corresponding root of $g$, and one might expect that root to be close to the
true projection $\ve{x}^{*} = (x^{*}, y^{*}) = (\sqrt{2}/2, \sqrt{2}/2)$. The
resulting point will definitely be on the variety, the question is whether it is
in fact $\ve{x}^{*}$.

\begin{figure}[h!]
\begin{center}
    \includegraphics[scale=1]{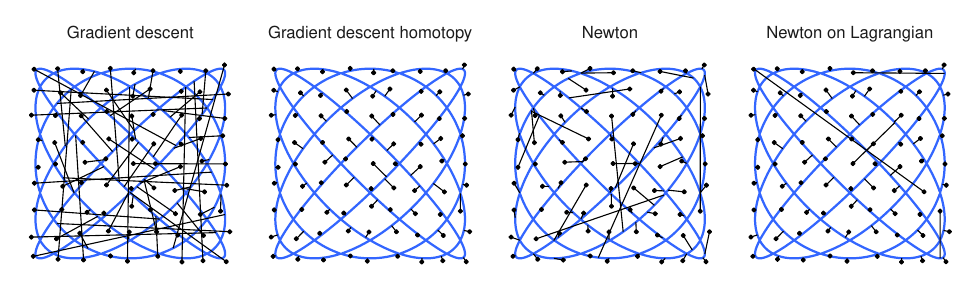}
\end{center}
\caption{Points sampled from variety distributions can be projected onto them
	using endgames, special homotopy-based solvers that exploit the
	algebraic structure of the optimization problem. Naive strategies to
	move points to the variety can produce very unreliable results. In this
	image a grid of points is projected onto the degree-18 Lissajous
	polynomial.}
\label{fig:projections}
\end{figure}

Alternatively, one may introduce Lagrange multipliers $\bm{\la} \in \R^{m}$ for
each of the $g_{i}(\ve{x})$'s, form the Lagrangian $\mc{L}(\ve{x}, \bm{\la}) =
\lpnorm{2}{\ve{x}-\ve{x}_{0}}^{2} + \bm{\la}'\ve{g}(\ve{x})$, take the gradient
with respect to both variables, and solve
$\nabla_{\ve{x},\bm{\la}}\mc{L}(\ve{x},\bm{\la}) = \ve{0}_{n+m}$.
Unfortunately, since $\mc{L}(\ve{x}, \bm{\la}) \in \R[\ve{x},\bm{\la}]$ is
polynomial, this system is itself a system of nonlinear polynomial equations
requiring solving. Again a naive strategy may be employed to solve it, but the
strategy again will not exploit the very highly structured algebraic nature of
the problem. Worse, it easily produces incorrect results, as illustrated in
Figure~\ref{fig:projections}.

The numerical algebraic geometry community has developed powerful methods based
on homotopy continuation to numerically solve systems of polynomial equations
\citep{sommese2005numerical, bates2013numerically}. These methods create similar
systems of equations with known solutions and carefully numerically track a
homotopy that deforms the solutions of one system into solutions of another.
This is done through a sequence of prediction-correction steps, usually Euler
steps on the Davidenko equation followed by Newton iterations. Near the end of
the path, when the solutions being tracked are near the roots of the target
system, the algorithms change their behavior in order to avoid numerical
problems caused by potential algebraic pathologies. For example, Newton's method
is well known to not achieve quadratic convergence near singular roots. These
finishing algorithms are called endgames. They only track one path, unlike
homotopy continuation in general which tracks one for each potential solution.
In the current situation, if $\si$ is sufficiently small, the resulting points
from the sampler will be sufficiently close to the variety to employ endgames to
move the points to the variety.

Recent advances in numerical algebraic geometry have adapted these endgames into
homotopy-based critical point methods for the purpose of projecting points
suitably close to a variety onto it \citep{griffinhauenstein2015,bdhs2019}.
Describing the critical points using Fritz John conditions yields the homotopy
\begin{equation}
  \ve{H}(\ve{x},\ve{\la};t) = 
  \twovec{ \ve{g}(\ve{x}) - t \ve{g}(\ve{x}_{0}) }{
\la_{0}(\ve{x}-\ve{x}_{0}) + \sum_{i=1}^{m} \la_{i} \nabla g_{i}(\ve{x})} = \ve{0}_{m+n}.
\end{equation}
where $\ve{\la}\in \P^{m}$, projective space of dimension $m$.  One starts the
homotopy at $t = 1$ and start point $(\ve{x}, \bm{\la}) = (\ve{x}_{0},
[1,0,\ldots,0])$ and aims to track to $t = 0$.  In our case, this is most easily
done using the single polynomial $h(\ve{x}) = \ve{g}(\ve{x})'\ve{g}(\ve{x}) =
\sum_{i=1}^{m}g_{i}(\ve{x})^{2}$ so that $m = 1$, and the computations can be
done over an affine patch of $\P^{1}$, yielding the homotopy
\begin{equation}
  \ve{H}^{a}(\ve{x},\ve{\la};t) = 
  \threevec{ h(\ve{x}) - t h(\ve{x}_{0}) }
  {\la_{0}(\ve{x}-\ve{x}_{0}) + \la_{1} \nabla h(\ve{x})}
  {\la_{0} + \al_{1}\la_{1} - \al_{0}} = \ve{0}_{n+2},
\end{equation}
where $(\al_{0}, \al_{1})$ can be selected randomly from the standard normal
distribution.

\begin{figure}
\begin{center}
    \includegraphics[scale=.25]{hmc-6.png}
    \includegraphics[scale=.25]{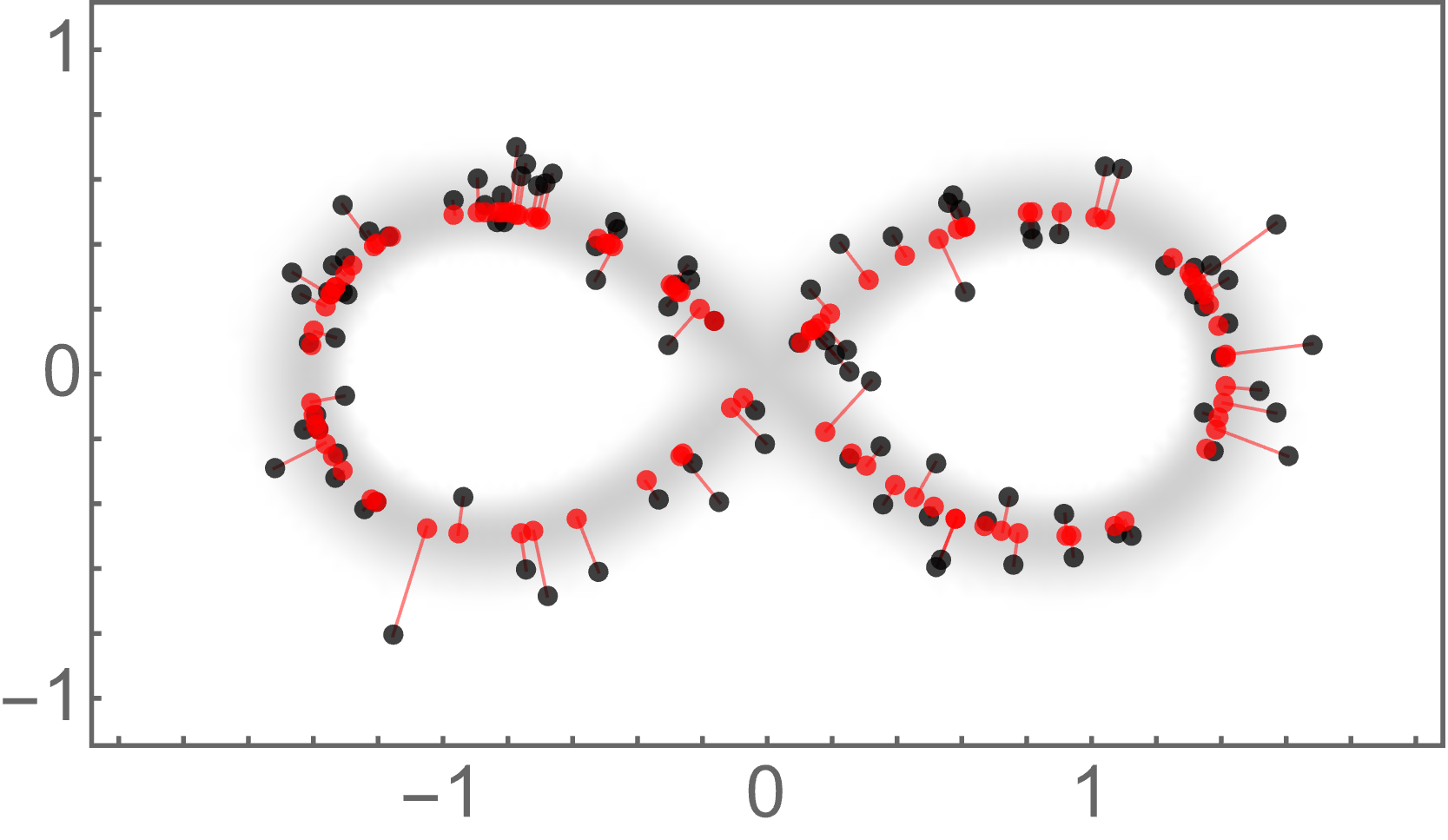}
    \includegraphics[scale=.25]{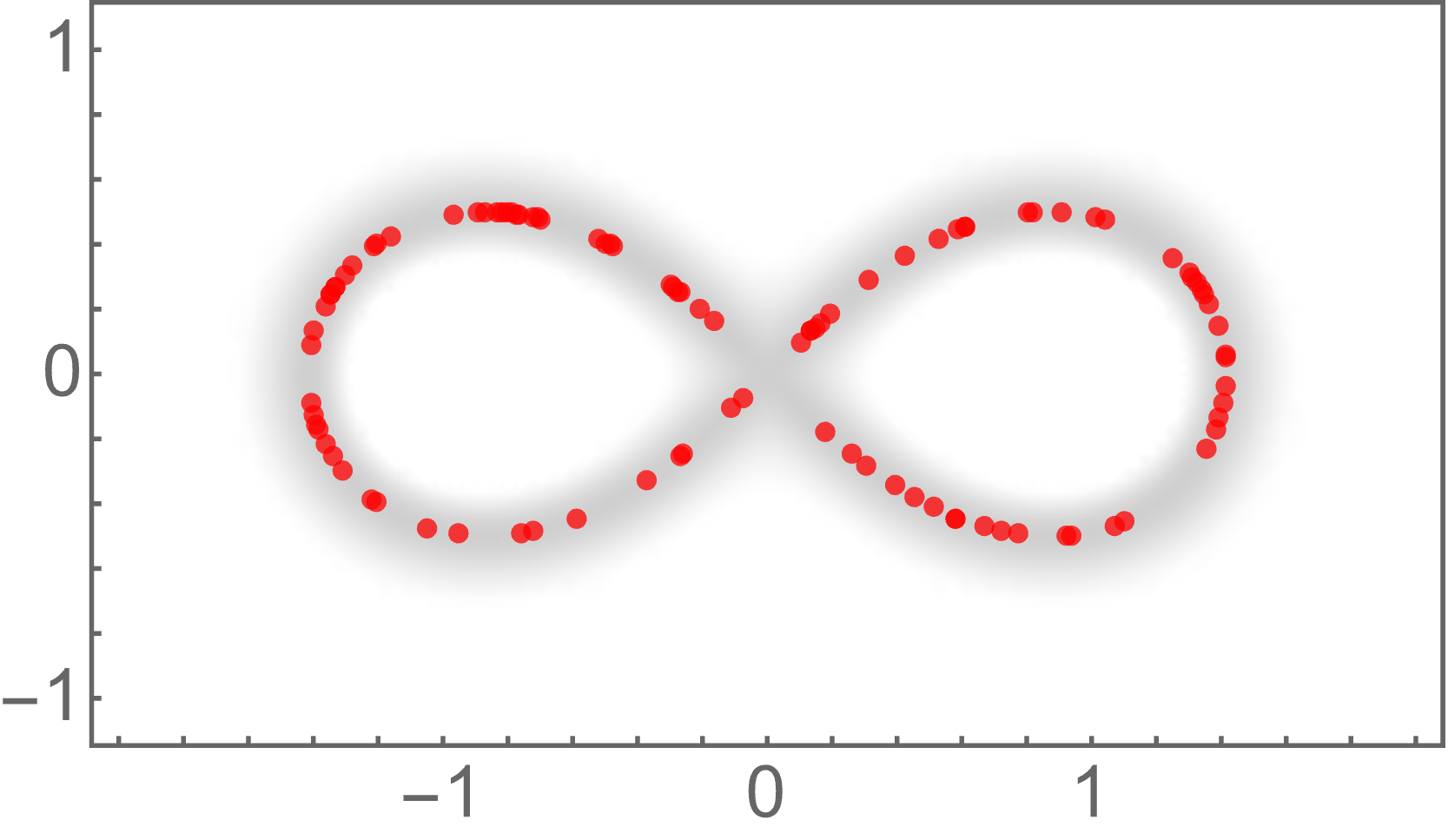}
\end{center}
\caption{100 draws from the lemniscate distribution in Figure~\ref{fig:hmc} and
	their projections via endgames.}
\label{fig:procedure}
\end{figure}

Summing up, we arrive at the following straightforward procedure: to obtain
points on the variety given only its defining polynomial(s), sample from a
corresponding variety distribution (usually the variety normal) and project the
points onto the variety using endgames. If the variety is low-dimensional,
rejection sampling is likely sufficient; otherwise, HMC via \pl{Stan} works
well.  An illustration of this is included in Figure~\ref{fig:procedure}.

\begin{figure}[h!]
\begin{center}
    \includegraphics[scale=.235]{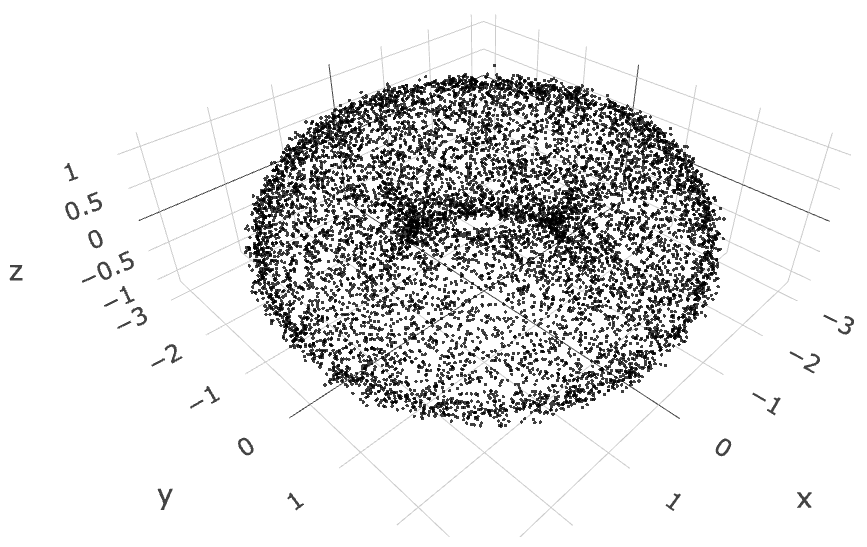}
    \includegraphics[scale=.235]{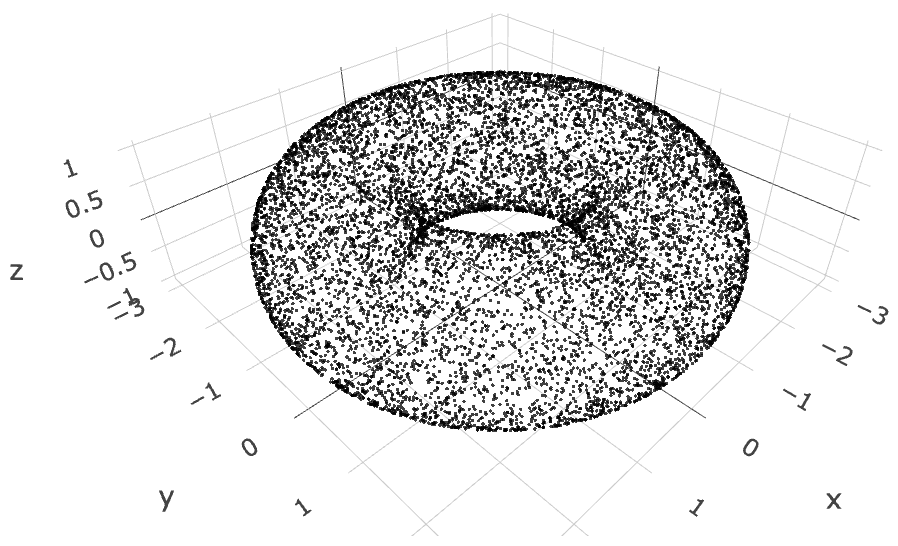} \\
    \includegraphics[scale=.15]{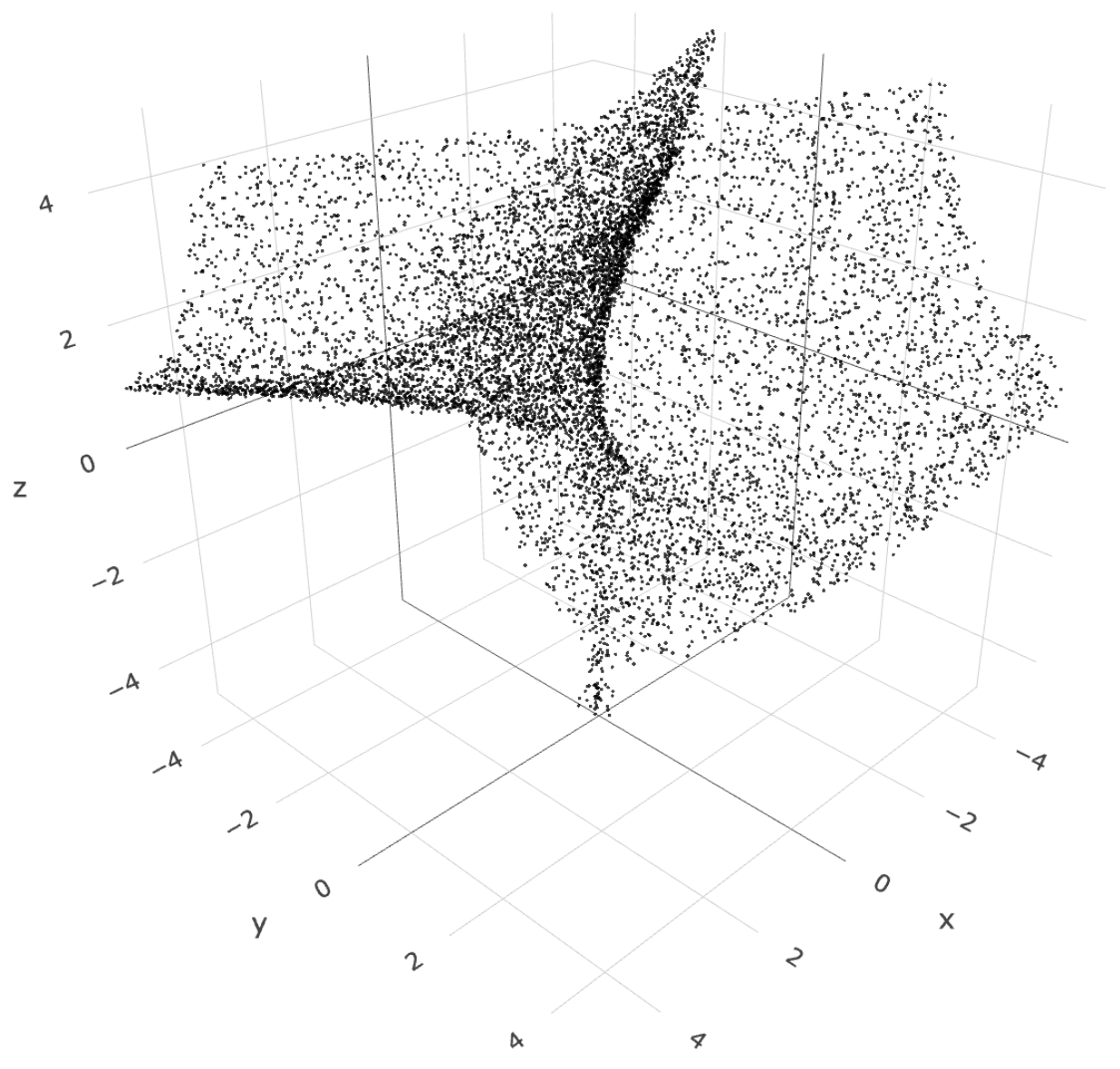}
    \includegraphics[scale=.14]{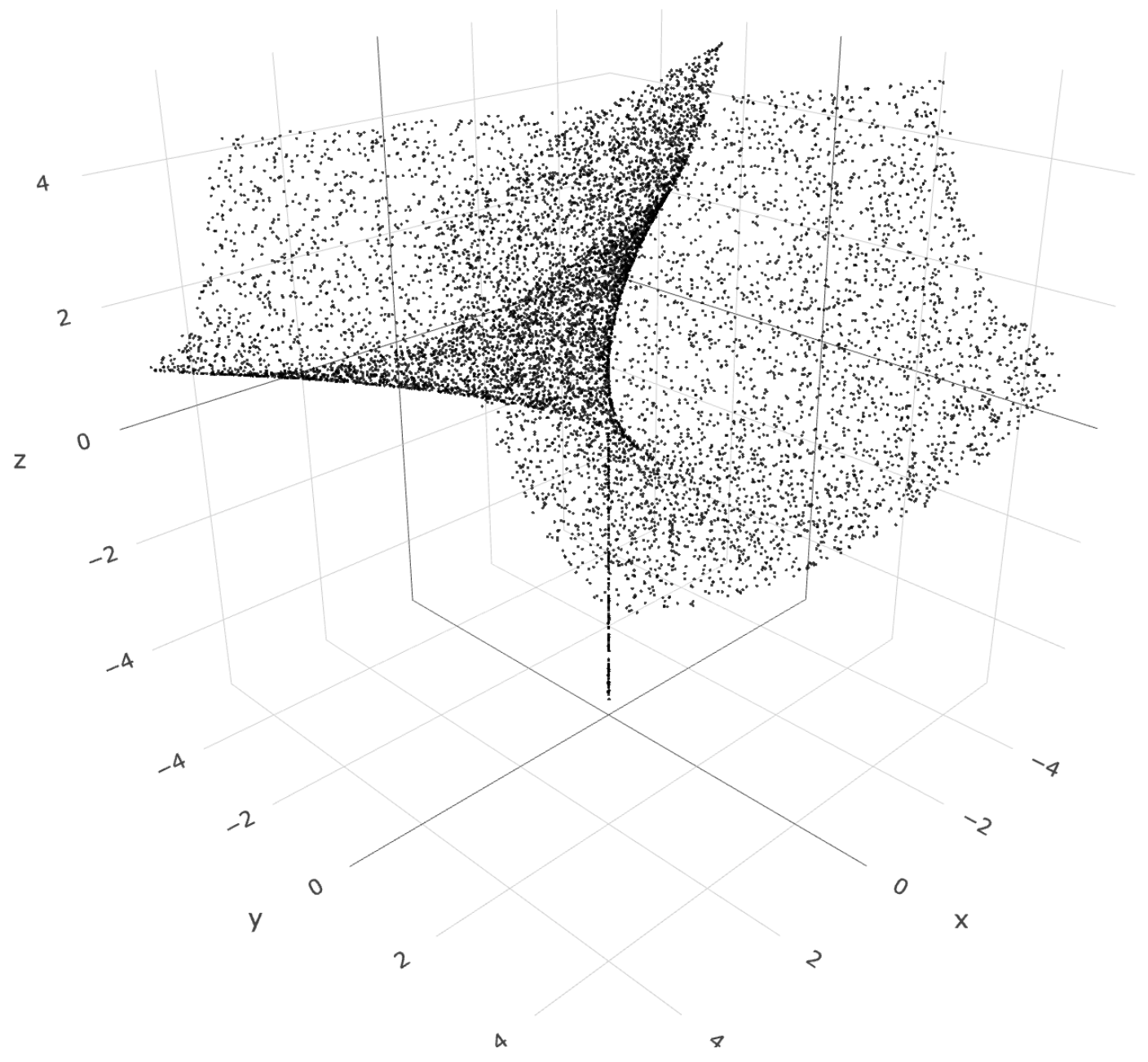} 
\end{center}
\caption{10,000 draws from the torus $g = (x^2 + y^2 + z^2 + R^2 - r^2)^2 - 4
	R^2 (x^2 + y^2)$ with $R = 2$ and $r = 1$ (top) and Whitney umbrella
	(bottom) variety normal distributions with $\sisq = .25$ as drawn (left)
	and projected with endgames (right). Notice that draws are obtained on
	the handle and the canope of the umbrella.}
\label{fig:projected}
\end{figure}

\section{Applications}\label{sec:applications}
In this section we present a few applications of the methods presented in the
previous sections. While we envision a broad array of potential applications, we
only present three here: independence models and optimization, a model of
synchronization arising from a dynamical system, and the generation of test
datasets for the evaluation of various pattern recognition and machine learning
tasks. In all examples we showcase the general idea of how the methods might be
used and leave any optimization or comparisons to existing methods as future
work.  All computations were performed in \pl{R}, often using the packages
\pkg{mpoly} and \pkg{algstat} \citep{r, mpoly, algstat}.

\subsection{Independence and optimization}\label{sec:indep}

A foundational perspective of algebraic statistics is that probabilistic
independence is naturally encoded implicitly. This applies in several settings,
but the most studied have been conditional independence models for multiway
contingency table analysis and Gaussian graphical models
\citep{sullivant2018algebraic, drton2008lectures}. In the latter setting, the
covariance matrix is positive definite (a semialgebraic positivity constraint on
the principal minors of the matrix), marginal independence is encoded as
0-covariance constraints, and conditional independence is encoded as 0
constraints on the inverse covariance matrix. The samplers from this work enable
the generation of actual distributions satisfying these constraints quite
easily.

\begin{figure}[h!]
\begin{center}
\includegraphics[scale=.30]{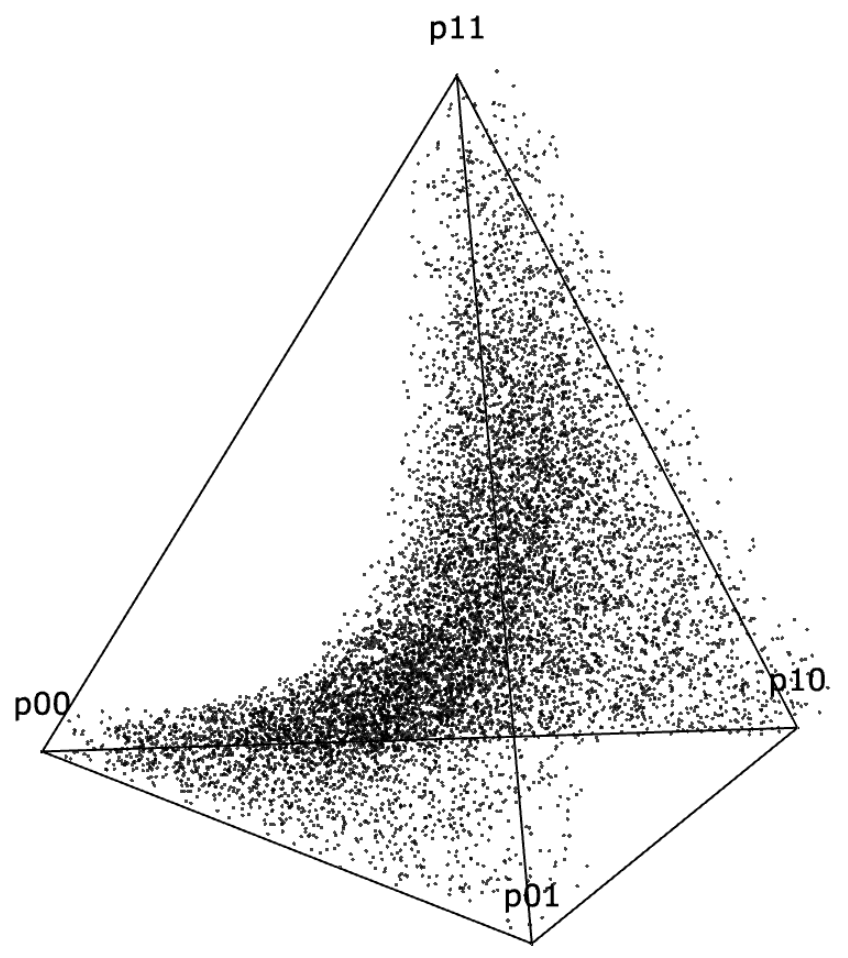}
\includegraphics[scale=.30]{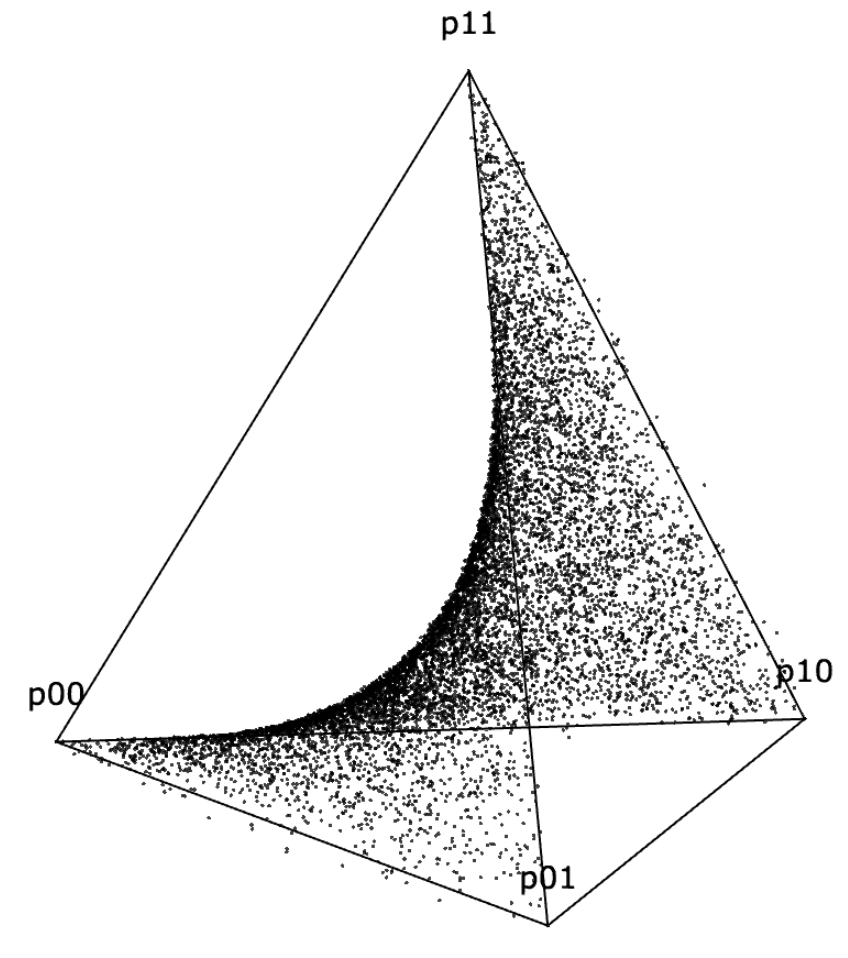}
\includegraphics[scale=.30]{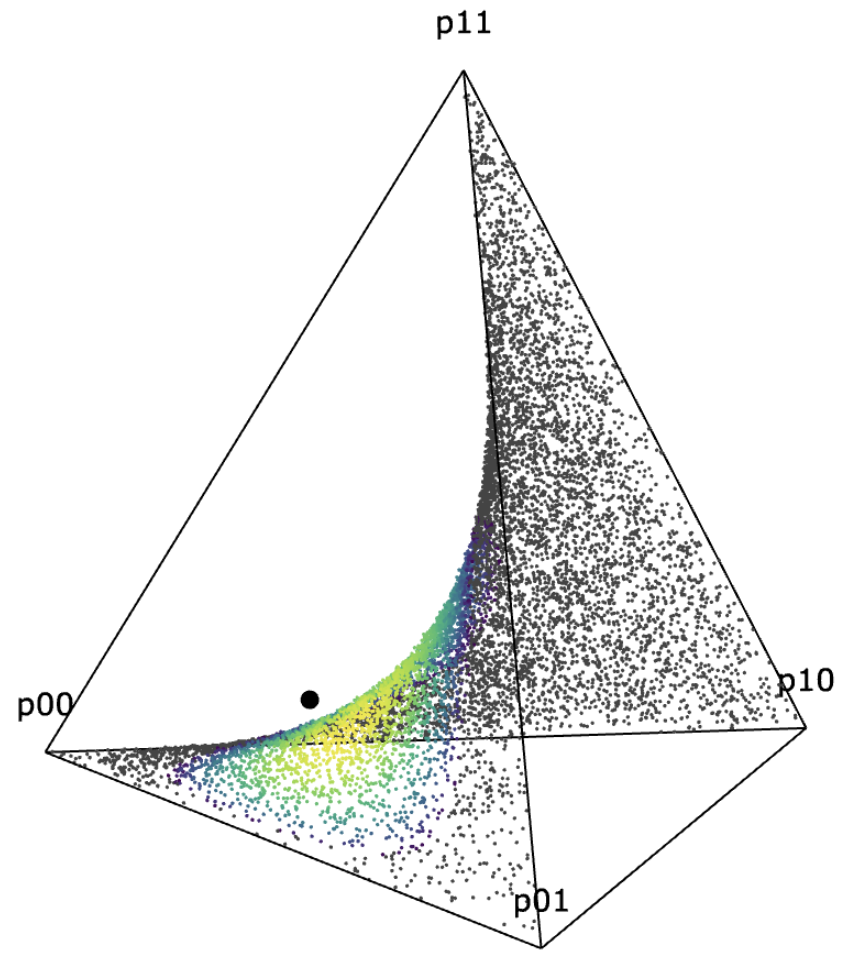}
\includegraphics[scale=.40]{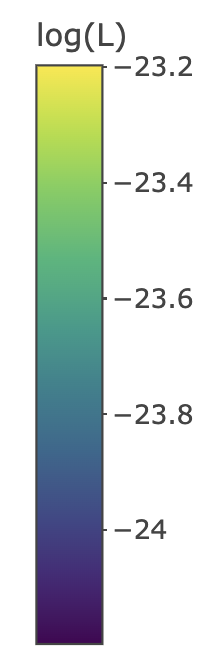}
\end{center}
\caption{Independence models on contingency tables are varieties in the
	probability simplex that can be explored using variety distributions.
	Here 10,000 such distributions were sampled (left) and projected
	(middle) using rejection sampling.  The projection took some points
	outside the simplex, but this could be corrected. For those inside, the
	log-likelihood is calculated with respect to the empirical distribution
	$(8, 8, 1, 3)$, illustrating how the sampler can be used for
	optimization (right).}
\label{fig:indep}
\end{figure}

A simple example of an algebraic statistical model on a contingency table is
instructive. The most basic manifestation of independence is in a $2 \times 2$
table representing two binary random variables $X$ and $Y$. In this setting
$p_{ij} = \PRRV{X = i, Y = j}$ for $i, j = 0, 1$. By definition, independence
demands four quadratic polynomial constraints $p_{ij} = p_{i+}p_{+j}$, where a
$+$ in the subscript indicates summing the indeterminates over the index.  These
are taken together with the constraint $p_{++} = 1$ and the semialgebraic
constraints $p_{ij} \geq 0$.  As is well known, the four independence
constraints can be summarized into one: $p_{00}p_{11} = p_{01}p_{10}$. Ignoring
the semialgebraic constraints, the model is given by $\ve{g} = [p_{00}p_{11} -
p_{01}p_{10}, p_{00} + p_{01} + p_{10} + p_{11} - 1] \in \R[\ve{p}]$. 10,000
draws from the corresponding variety normal distribution with $\si = .025$,
obtained via rejection sampling over the unit cube, is quite efficient. Results
are shown in barycentric coordinates in Figure~\ref{fig:indep}.

As in most settings, the statistical problem of estimation corresponds to an
optimization problem. The feasiblity region of the optimization problem in these
settings is a variety and the objective function depends on the type of
estimation desired \citep{kahle2011minimum, read2012goodness}. In the case of
the $2 \times 2$ contingency table, the empirical distribution of a dataset, the
relative frequency distribution, is another point in the simplex, and its
closest point on the independence surface constitutes an estimate for the model.
Different distance measures describe different estimation strategies, e.g.
maximum likelihood or minimum chi-square, which may or may not be algebraic in
some suitable sense, e.g. polynomial, rational, etc. Numerically, solutions to
the optimization problem may be obtained via a variety of different approaches,
from general routines such as Newton-Raphson, iteratively reweighted least
squares, or Nelder-Mead to more specialized routines such as semidefinite
programming solutions or sequences thereof \citep{anjos2011handbook}. In
principle, variety sampling can provide a Monte Carlo scheme for such
optimization problems by simply evaluating the objective function at the points
sampled or (preferably) after projection. In general this is likely more useful
for gaining a sense of the objective function over the variety, and not
necessarily optimization itself.

As a final component of this application, one can imagine using the sampler to
conduct Monte Carlo inference via inversion \citep{kahle2011minimum}. The
empirical distribution given by the relative frequency distribution of the
dataset exhibits a well-known central limit theorem (CLT) used throughout
categorical data analysis \citep{agresti2012categorical}. This CLT can be used
to construct an ellipsoidal confidence region around the empirical distribution.
In the same way that the elementary single proportion CLT interval $\hat{p} \pm
z_{\al/2}\sqrt{\hat{p}(1-\hat{p})/n}$ can be inverted to produce an
asymptotically correct hypothesis test for the hypotheses $H_{0} : p \in
\Th_{0}$ vs. $H_{1} : p \in \Th \setminus \Th_{1}$ by simply checking for
intersection with $\Th_{0}$, this ellipsoidal region can be inverted into an
inferential procedure to assess the algebraic statistical model, and precisely
the same way. If enough draws are taken from the implicitly described model,
operating under the reasonable assumption that they are sufficiently uniform on
it, one simply has to check whether any of the points sampled falls within this
ellipsoid, a trivial computation. If they do not, the procedure rejects in favor
of the alternative $H_{1}$ in the same way that the elementary interval does if
the intersection of $\Th_{0}$ and the interval is empty.

\subsection{Exploring solutions of the Kuramoto model}\label{sec:Kuramoto}

Another area of intense interest in recent years has been solution sets of
systems of differential equations that are polynomial and whose varieties
correspond to equilibrium states. Such is the case in chemical reaction
networks, for example \citep{gatermann2001counting, craciun2009toric}. 

Such varieties can exhibit very interesting behavior. A commonly occurring
situation for varieties arising in applications is that they decompose into
components of different dimensions.  Variety samplers can be applied to these
situations, too, as demonstrated by computing the steady states of a dynamical
system used to model synchronous behavior, namely the Kuramoto model
\citep{Kuramoto1975}.  Following an algebraic geometrization of the Kuramoto
model~\citep{KuramotoAG}, we demonstrate by solving the following system derived
from $N=5$ oscillators:
$$\hbox{\small $
\begin{array}{cc}
5s_{1} - 5c_{1}s_{2} + 5c_{2}s_{1} - 5c_{1}s_{3} + 5c_{3}s_{1} - 5c_{1}s_{4} + 5c_{4}s_{1}, &
5s_{2} + 5c_{1}s_{2} - 5c_{2}s_{1} - 5c_{2}s_{3} + 5c_{3}s_{2} - 5c_{2}s_{4} + 5c_{4}s_{2}, \\
5s_{3} + 5c_{1}s_{3} - 5c_{3}s_{1} + 5c_{2}s_{3} - 5c_{3}s_{2} - 5c_{3}s_{4} + 5c_{4}s_{3}, & 
5s_{4} + 5c_{1}s_{4} - 5c_{4}s_{1} + 5c_{2}s_{4} - 5c_{4}s_{2} + 5c_{3}s_{4} - 5c_{4}s_{3},\\
c_{1}^2 + s_{1}^2 - 1, &
c_{2}^2 + s_{2}^2 - 1, \\
c_{3}^2 + s_{3}^2 - 1, &
c_{4}^2 + s_{4}^2 - 1.
\end{array}$}
$$
The top four describe the interaction between the oscillators, while the bottom
four restrict the oscillators to the unit circle.  To remove rotational
symmetry, the $5^{\rm th}$ oscillator is fixed with $s_{5} = 0$ and $c_{5} = 1$.

\begin{figure}[h!]
\begin{center}
\includegraphics[scale=.25]{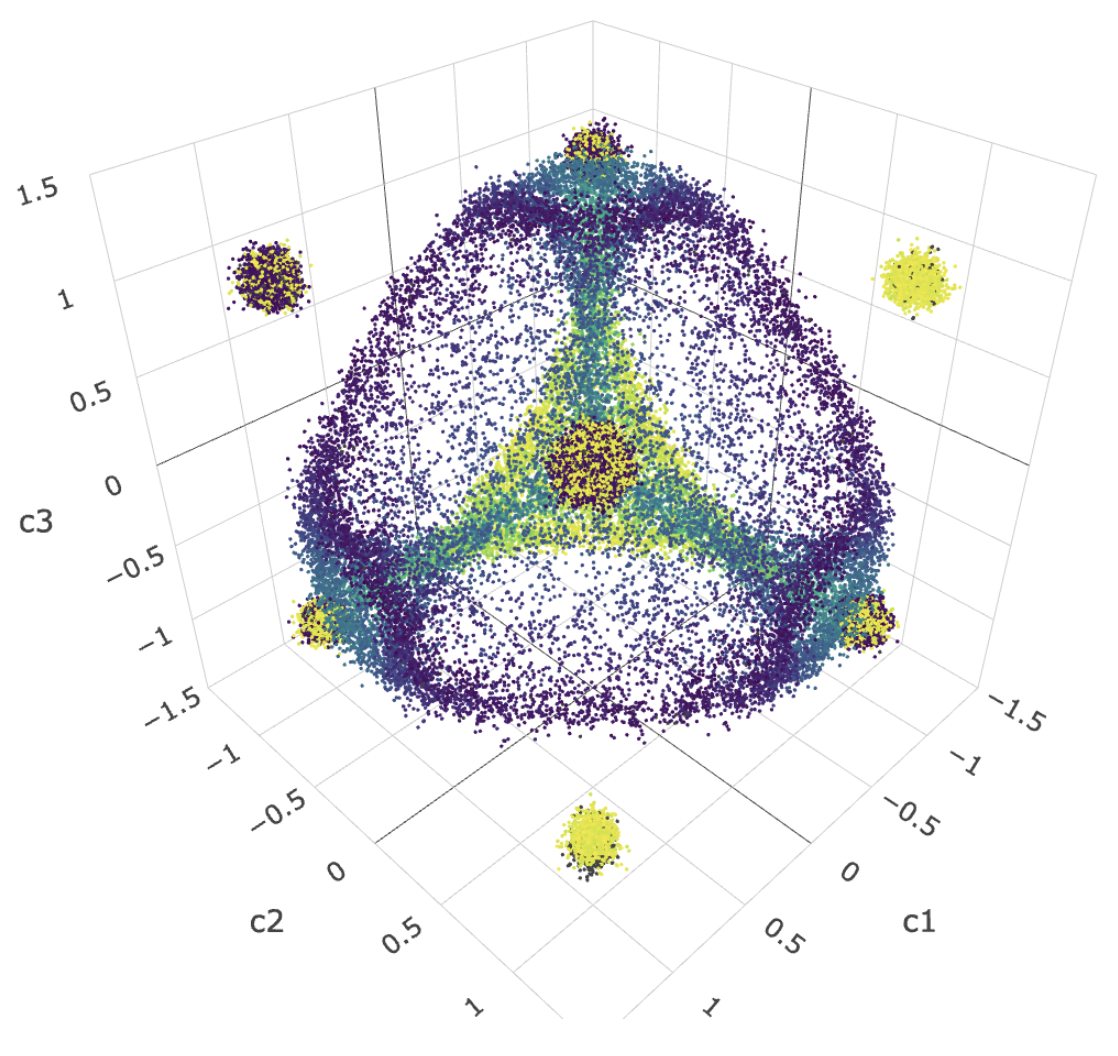}
\includegraphics[scale=.25]{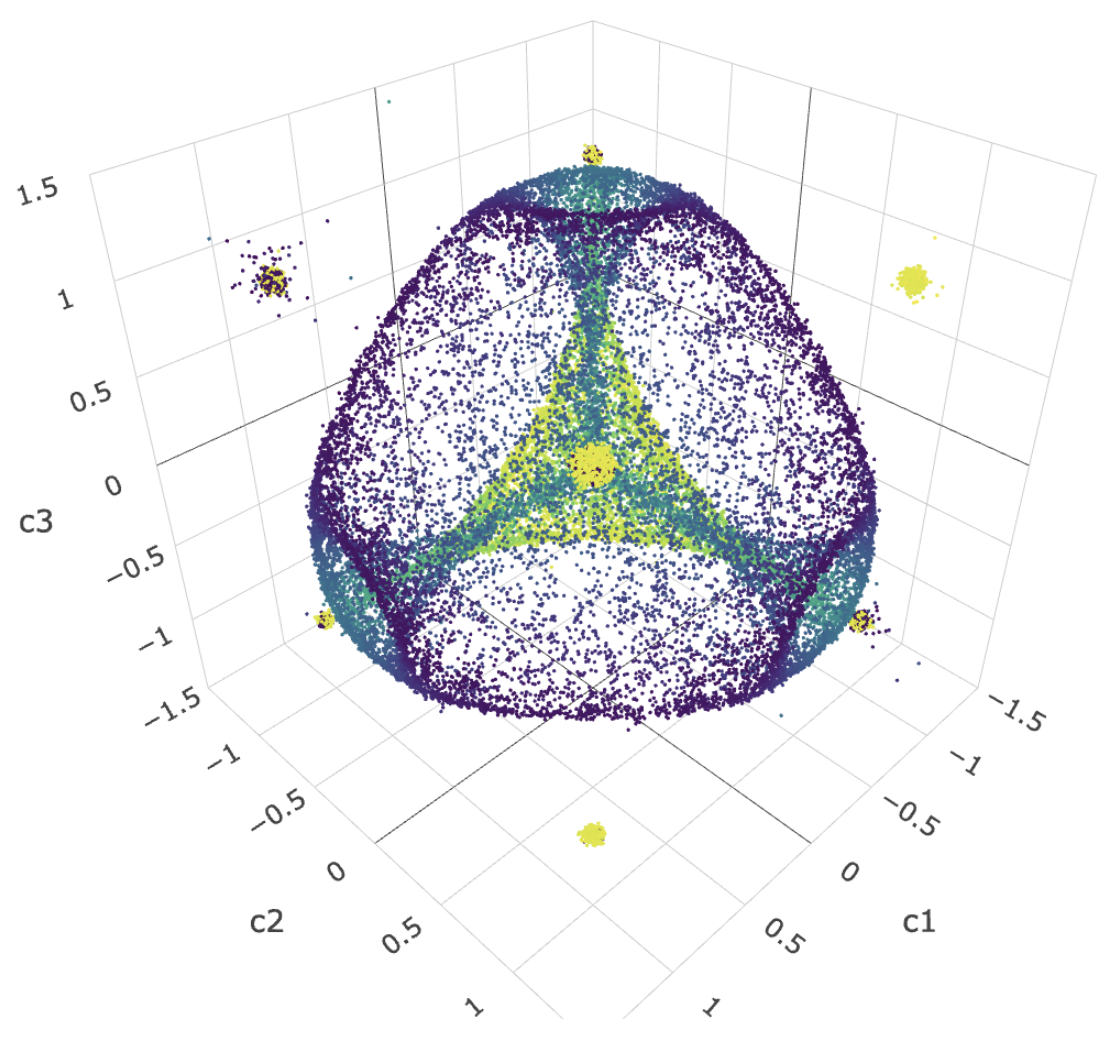}
\includegraphics[scale=.25]{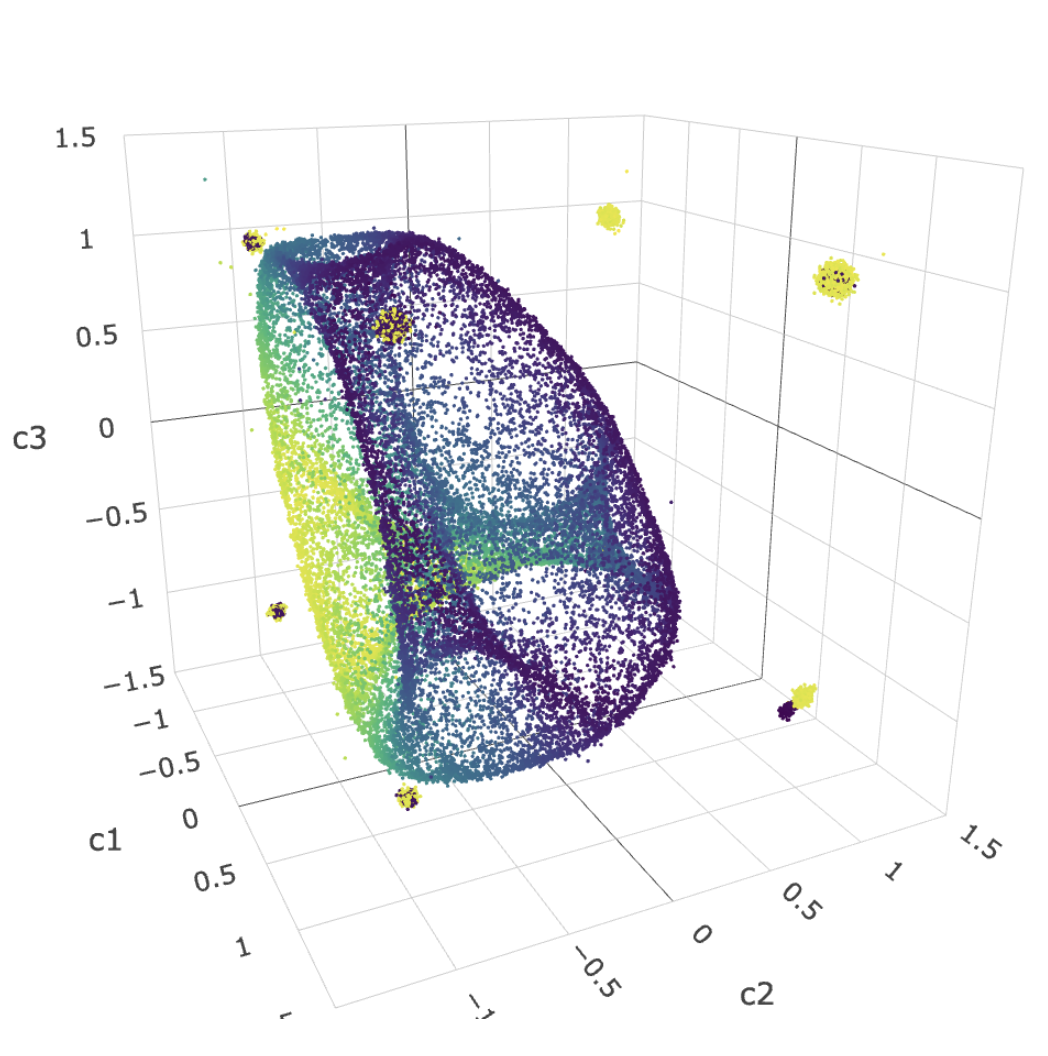}
\includegraphics[scale=.375]{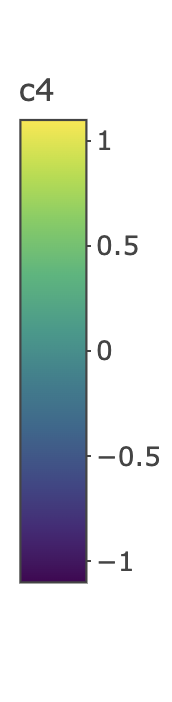} \\
\includegraphics[scale=.25]{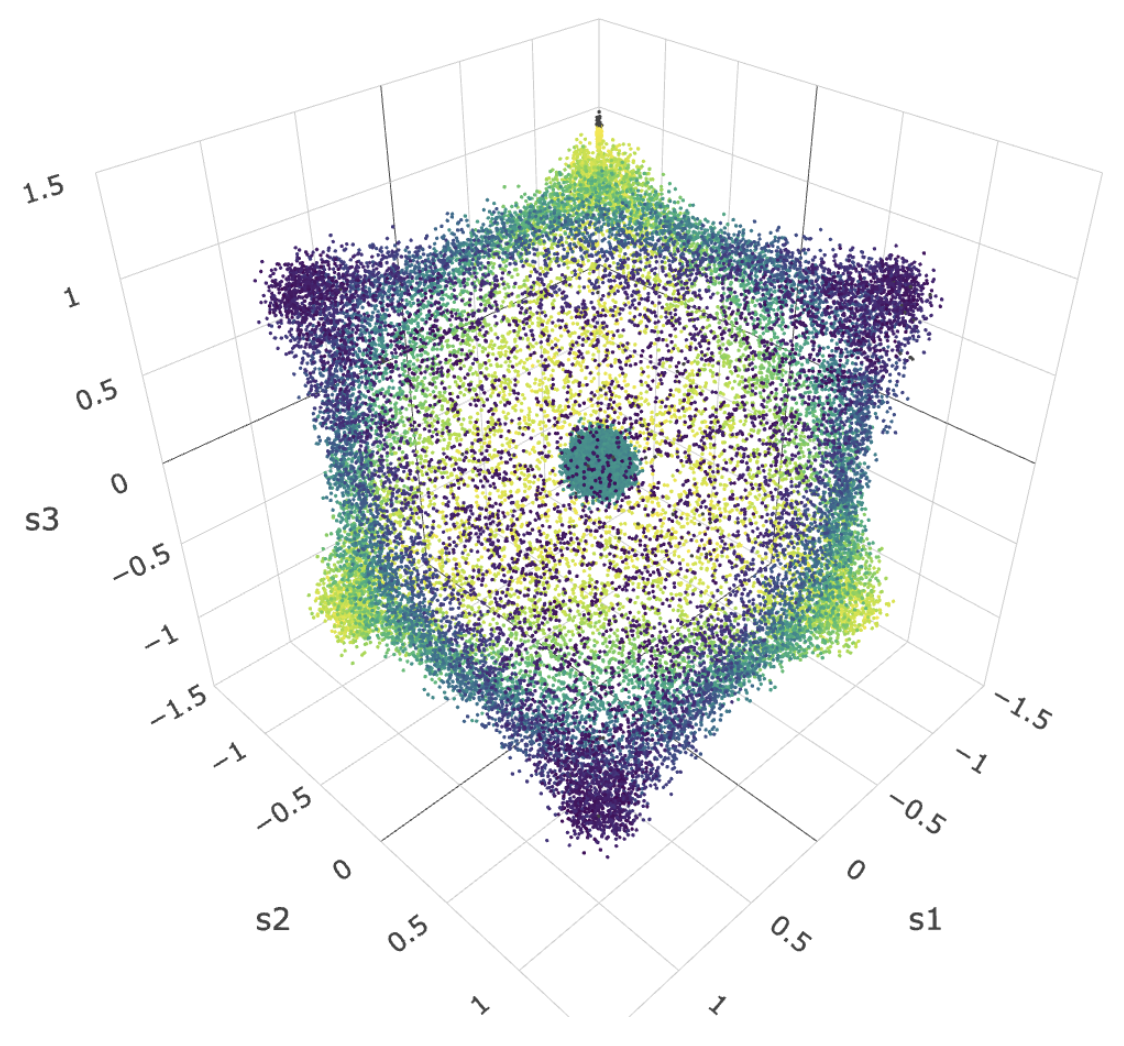}
\includegraphics[scale=.25]{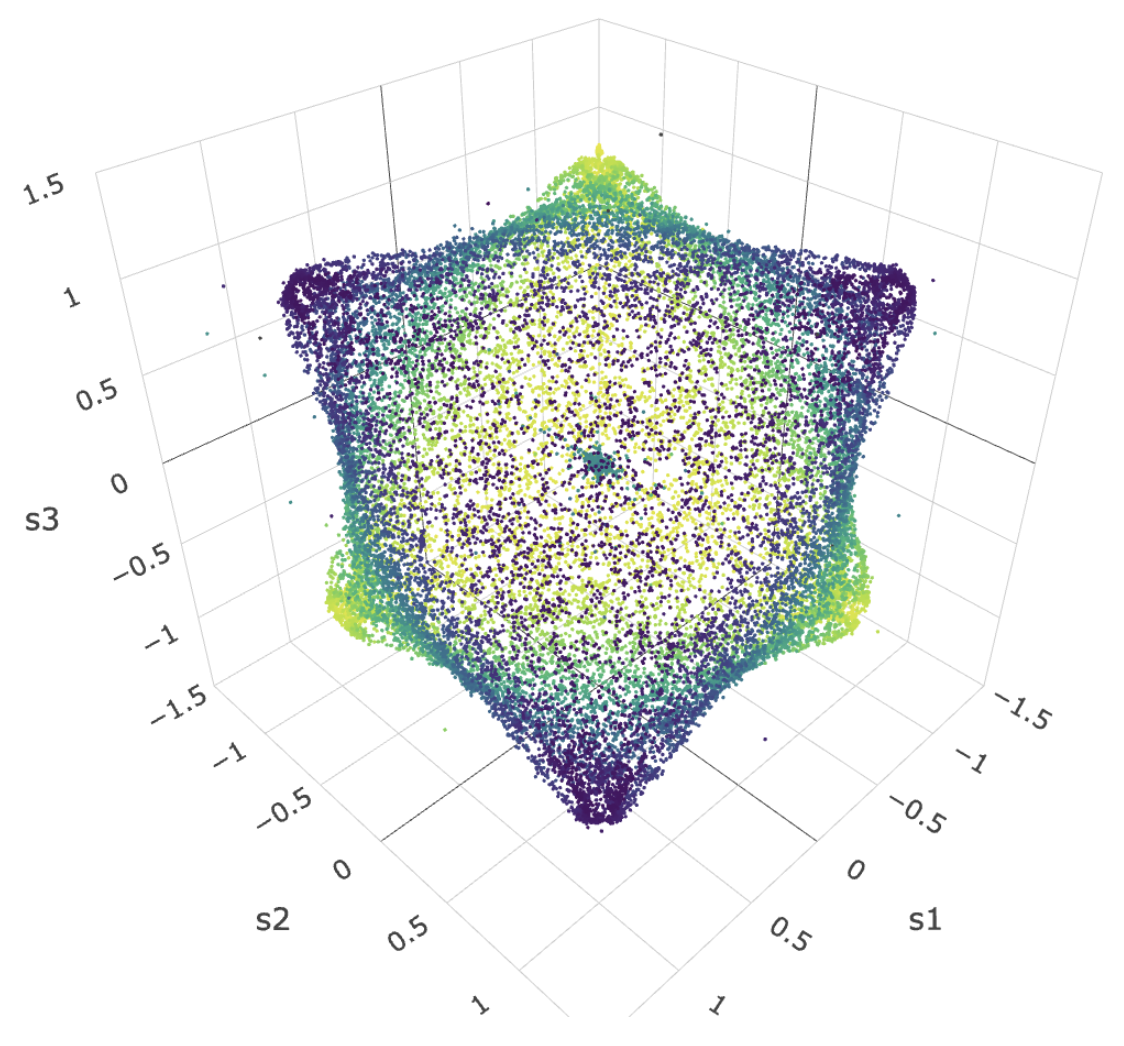}
\includegraphics[scale=.25]{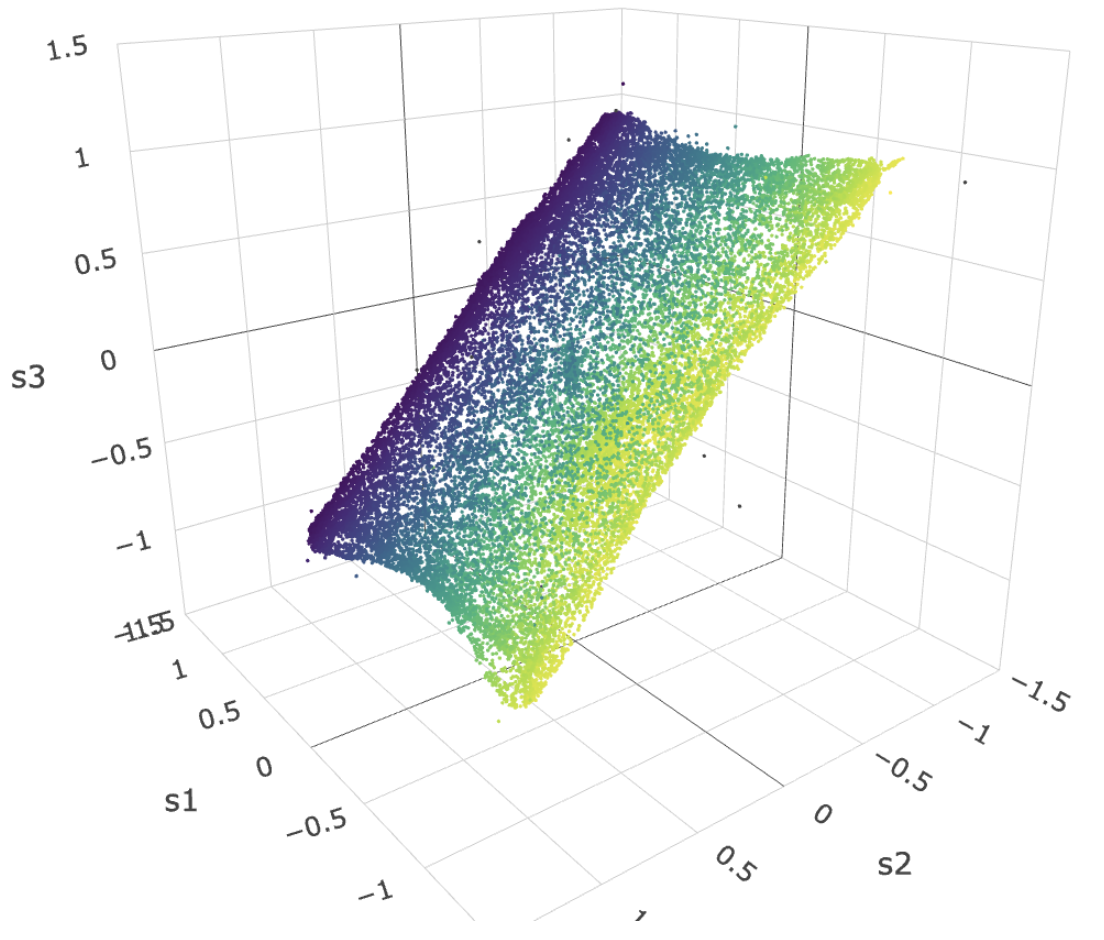}
\includegraphics[scale=.375]{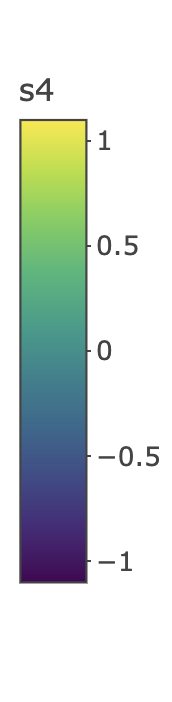}
\end{center}
\caption{HMC with \pl{Stan} can be effective at sampling intricate variety
	distributions in high dimensional settings. Here 64,000 draws from the
	variety normal on the $N=5$ Kuramoto model in 8 dimensions with $\si =
	.05$ were drawn with 64 HMC chains in parallel.  These are illustrated
	in projection onto $(c_{1}, c_{2}, c_{3})$ (top) and $(s_{1}, s_{2},
	s_{3})$ (bottom). From left to right: raw draws, projections, and
	projections from a different perspective.}
\label{fig:kuramoto}
\end{figure}

The system is known to have a variety in $\R^{8}$ that consists of a
$2$-dimensional component and $2^4 = 16$ points such that $s_{i} = 0$ and
$c_{i}^2 = 1$ for $i = 1,\dots,4$. Disregarding these known solutions, to
understand this space of possible equilibria we sampled from the variety normal
on the system and viewed the points on projection interactively.  We note in
passing that these points representations of varieties are easy to project since
projection simply corresponds to projecting points in $\R^{8}$ onto a linear
space, and moreover projection onto coordinate axes is achieved by simply
selecting component values.  The results are illustrated in
Figure~\ref{fig:kuramoto} and demonstrate the power of HMC to scale variety
sampling into higher dimensional spaces where rejection sampling, gridding
solutions, and graphical algorithms such as marching cubes are prohibitively
inefficient.

A few practical observations can be gleaned from this example. 
\begin{enumerate}
\item As in all major MCMC schemes, HMC gets stuck in local modes when $\sisq$
	is sufficiently small.  Variety components correspond to modes of the
		variety normal. Thus, some of the 64 chains migrated to and
		explored the 2d component while others went to different
		isolated solutions. As in Bayesian sampling, initializing many
		chains at random points, described in the Bayesian literature as
		``overdispersed'' \citep{lunn2012bugs}, is prudent in an effort
		do discover all components. 

In general there is a trade-off with respect to $\sisq$ in the variety normal.
		The larger it is, the easier the sampling is---that is, samplers
		are much faster---and the less likely chains are to get stuck in
		disconnected components, but the fuzzier the variety appears.
		The smaller it is, the harder the points are to sample, but the
		projections of the gradient descent homotopies are more
		reliable. Concrete guidelines on selecting an ideal $\sisq$
		seems like a hard problem; however, in our limited experience
		with small-scale varieties getting a decent $\sisq$ has been
		quite easy.

\item The chains ran with quite different efficiencies. While most chains
	finished within a second, some took a few minutes. Presumably, the
		chains corresponding to isolated points, which are effectively
		multivariate normals at those points, sampled very quickly,
		whereas the 2d component was explored more slowly.

\item The projections did not work perfectly, as illustrated by the stray points
	falling off of the isolated solutions. This was likely due to a naive
		implementation of the gradient descent homotopy but warrants
		future investigation.

\end{enumerate}

We suspect that the samplers perform well in much higher dimensional settings as
well, as Bayesian sampling via HMC has been applied successfully in scenarios
with thousands of parameters (and more). We have not experimented with these
scenarios. It should be noted that the implementation we suggest in the previous
section considers each of the Jacobian elements as parameters also.

\subsection{Algebraic pattern recognition and topological data analysis}\label{sec:apr}

In addition to providing the ability to explore varieties, the methods described
in this work also admit inverse problems that generalize interpolation and
regression with interesting applied math and statistical flavors. These problems
typically present with a dataset (point cloud) that exhibits an algebraic
structure and asks the analyst to recover that structure. For example, if in an
exploratory data analysis one encounters a 2d scatterplot of points as in
Figure~\ref{fig:procedure}, left or right, the task may be to recover the
defining polynomial or some aspect thereof.

As one might suspect, these kinds of questions are not new. In the context of
computer graphics, in a highly cited article \cite{zhang1997parameter} discusses
estimation methods to recover conic sections from noisy data of the form $\mc{D}
= \set{(x_{1},y_{1}), \ldots, (x_{N},y_{N})}$. While geometrically
sophisticated, the general nature of the solution is statistically routine:
create a parameterized family of parametrically described conic sections and
estimate the (first set of) parameters statistically. Of course, one may ask the
question more generally: given $\mc{D}$, recover the algebraic curve from which
the points were observed with noise. This problem is much more subtle, since (1)
algebraic curves need not have parametric representations and (2) the model
parameter space is not like those traditionally encountered in statistical
applications.  For example, two distinct polynomials may describe the same
curve.  The question can be stated more generally still: given a dataset
$\mc{D}$ of $N$ points in $\R^{n}$, recover the variety from which they were
observed, with or without error. The same may be considered for points in a
semi-algebraic set.  A similar pursuit motivated the nonlinear generalization of
principal components analysis in \cite{vidal2005generalized}, which has the same
order of citations and has been developed into a book-length treatment
\citep{gpca-book}.

Variety distributions and samplers provide a Monte Carlo framework for
experimenting with candidate solutions to these kinds of pattern recognition
problems. Many areas of current interest---deep learning, manifold learning, and
topological data analysis for example---seek to understand nuanced structure in
point clouds; however, strategies for generating example cases to test these
procedures seems to be absent from the communities and, as a consequence, one
sees the same examples repeated in the literature. The distributions and
samplers provided in this work seem to be excellent for this purpose. Moreover,
they avoid the too-often used solution to the problem of meshing a parameter
space, transporting the points to the manifold (say) via a parameterization, and
adding noise. In addition to not having many such parameterized objects, the
resulting clouds exhibit various oddities, from being too symmetric, in the
sense of not exhibiting gaps one might expect from a randomly drawn collection
of points, to being too biased to particular regions of the object of interest.
These features may in fact be present in real problems, but they feel arbitrary
and artefactual from a Monte Carlo perspective.

\begin{figure}[h!]
\begin{center}
\includegraphics[scale=.80]{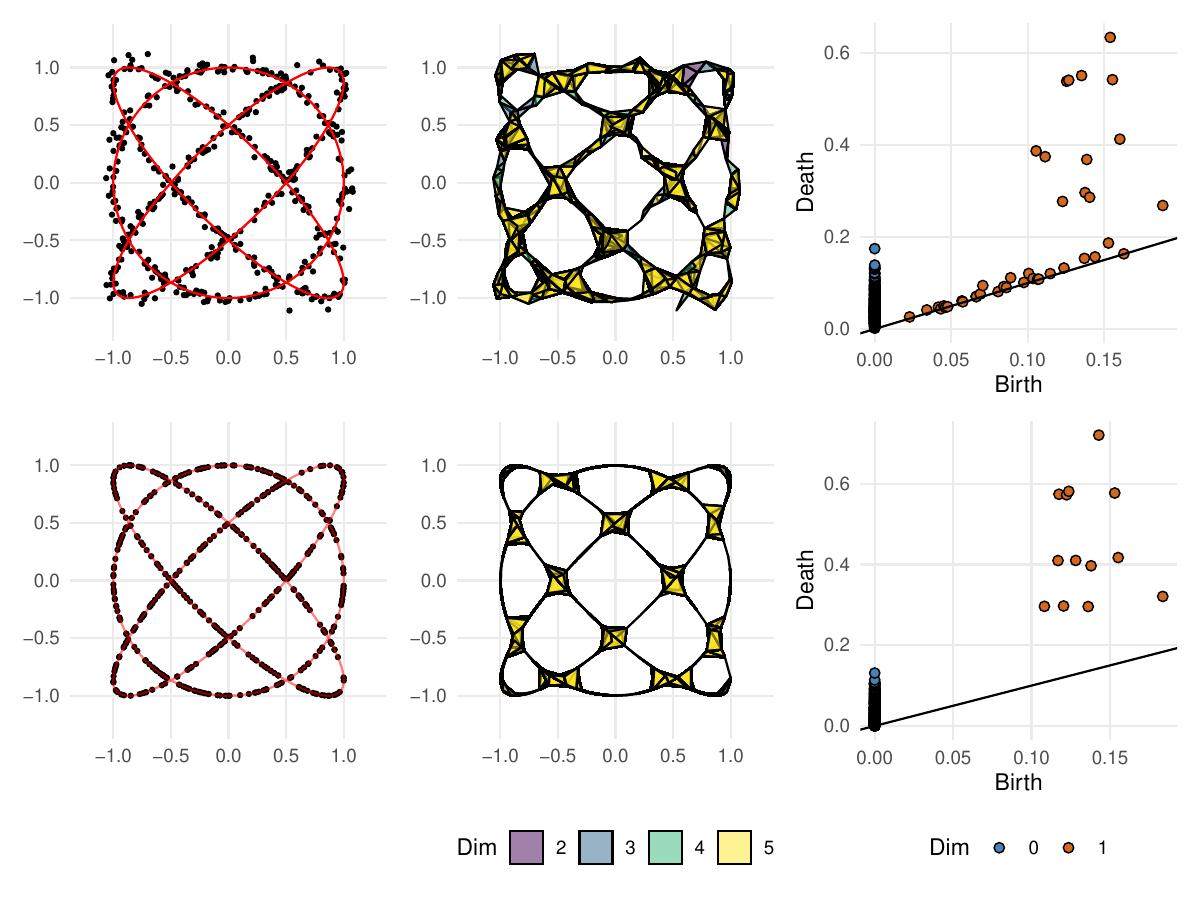}
\end{center}
\caption{Variety distributions can be used for pattern recognition experiments,
	here TDA. A 2d point cloud, simplicial complex, and persistence diagram
	are illustrated using 500 draws from the degree-6 Lissajous variety
	normal distribution $\Norm{9 x^2  -  24 x^4  +  16 x^6  +  9 y^2  -  24
	y^4  +  16 y^6  -  1}{\si = .025}$, raw (top) and projected (bottom).
	Note that the 13 orange points in the persistence diagram correspond to
	the 13 cells bounded by the varieties; their horizontal alignment
	corresponds to their size being the same, showing the symmetry of the
	$13 = 4 + 4 + 4 + 1$ cells. Ordinarily only the persistence diagram
	would be visualized.}
\label{fig:tda}
\end{figure}

As a specific example, variety distributions seem particularly useful for
experimenting with topological data analysis (TDA), which seeks to learn the
topological structure of a point cloud \citep{carlsson2009topology}.  As
varieties often have interesting topological features, sampling them provides
for the creation of a rich collection of synthetic datasets with which to assess
TDA procedures.  One of the most well-known TDA procedures, persistent homology
(PH), seeks to discover topological features of point clouds of data that
manifest at many different scales, e.g. holes of different dimensions.
Figure~\ref{fig:tda} shows an example of the PH of draws from the degree-6
Lissajous variety normal distribution sampled via rejection sampling. PH is
computed via \pl{R}'s \pkg{ripserr} package, which uses the TDA software
\pl{Ripser} and plotted using \pkg{TDAvis} soon to be incorporated into
\pkg{ggtda} \citep{ripserr, bauer2021ripser, TDAvis, ggtda}.  The procedure is
able to discover that the variety bounds 13 cells and to some extent is even
able to speak to the relative size of those cells.  This suggests that the Monte
Carlo methods can feed back into (real) algebraic geometry as a tool for
experimentation, possibly even providing statements about topological features
(or even algebraic or geometric ones) that hold with a certain probability, akin
to confidence regions.

\section{Discussion}\label{sec:discussion}

This article has introduced new methods to stochastically explore real algebraic
and semi-algebraic sets. There are many directions to go from here; many have
been suggested in other sections. In terms of the distributions in particular, a
few directions would be interesting: a clearer view of when a polynomial will
admit a normalizable HVN or VN; better ways to reduce the skewness for overly
large dispersion $\sisq$, or even useful ways to select $\sisq$ for a given
$\ve{g}$; better treatment of semi-algebraic distributions that don't
overconcentrate on the boundaries; consideration of distributions over $\C^{n}$
(and applications); adjustments that reduce the effects of multiplicity on
variability (e.g.  $g$ vs $g^{2}$, $g^{3}$, etc.); and further investigations
into which variety distributions might have special properties akin to the
variety normal or variety uniform. Each of these would be fascinating and of
practical utility. Similar improvements would be useful on the sampling front,
including easy-to-use implementations of HMC routines for users inexperienced in
that area or different strategies altogether. A thorough investigation of the
samplers with large scale, or even varying scale, and complex (real) varieties
would be interesting as well. We look forward to all these in the future.

In closing, one may reasonably offer the critique that most of the discussion
provided in this work applies not only to varieties, but level sets of functions
in general.  The argument is similar to numerical continuation in the context of
solving systems of algebraic equations. While it is true that the methods might
work for exploring level sets of reasonably nice functions, polynomials are the
primary class of such functions, mostly because the mathematics in general is
more tractable with polynomials and some guarantees can be offered. For
instance, it is possible to speak to normalizability to some extent over
polynomials, whereas arbitrary functions $g$ seem more challening.  In
Section~\ref{sec:sampling}, the computations for sampling varieties are nice,
unlike level sets in general, since differentiation of polynomials is nice. To
check for zeros, the gradient descent homotopies are able to certify
projections, theory of which depends on the underlying algebraic structure.
These, combined with our own interest, has lead us to consider first and
foremost zero level sets (varieties) of polynomials; however the same strategies
can certainly be entertained for general functions $\ve{g}$.

\bibliographystyle{plainnat}
\bibliography{__0-bibliography}

\end{document}